\documentclass[a4paper,12pt]{article}

\usepackage{geometry} 
\usepackage{graphicx}
\usepackage{lineno}
\usepackage{comment}
\usepackage{tabularx}
\usepackage{subcaption}
\usepackage{multirow}
\usepackage{tabularx}
\usepackage{xcolor}
\usepackage{tikz}
\usepackage{authblk}
\usepackage{hyperref}
\usepackage{amsmath}

\newcommand{\keywords}[1]{
  \begin{flushleft}
    \textbf{Keywords:} #1
  \end{flushleft}
}

\usepackage[style=numeric-comp, sorting=none]{biblatex}
\renewbibmacro*{in:}{}
\title{\textbf{Characterization and Novel Application of Power Over Fiber for Electronics in a Harsh Environment}}

\author[a]{M. A. Arroyave}
\author[b]{B. Behera}
\author[a]{F. Cavanna}
\author[a]{A. Feld}
\author[c]{F. Guo}
\author[b]{A. Heindel}
\author[c]{C. K. Jung}
\author[a]{K. Koch}
\author[b]{D. Leon Silverio}
\author[b,1,$\ddag$]{D. A. Martinez Caicedo}
\author[c]{C. McGrew}
\author[a]{A. Paudel}
\author[a,1,$*$]{W. Pellico} 
\author[a]{R. Rivera}
\author[b]{J. Rodríguez Rondon}
\author[e]{S. Sacerdoti}
\author[a]{P. Shanahan}
\author[c,1, $\dagger$]{W. Shi}
\author[b]{D. Torres Muñoz}
\author[d]{D. Totani}
\author[b]{C. Uy}
\author[b]{C. Vermeulen}
\author[e]{H. Vieira de Souza}

\affil[a]{Fermi National Accelerator Laboratory, Batavia, IL, 60510, USA}
\affil[b]{South Dakota School of Mines and Technology, Rapid City, SD 57701, USA}
\affil[c]{Stony Brook University, SUNY, Stony Brook, New York 11794, USA}
\affil[d]{University of California Santa Barbara, Santa Barbara, CA 93106, USA}
\affil[e]{Universit\'e Paris Cit\'e, CNRS, Astroparticule et Cosmologie, Paris, France}

\date{}

\begin{document}

\maketitle

\abstract{ Power-over-Fiber (PoF) technology has been used extensively in settings where high voltages require isolation from ground. In a novel application of PoF, power is provided to photon detector modules located on a surface at $\sim$ 300 kV with respect to ground in the planned DUNE experiment. In cryogenic environments, PoF offers a reliable means of power transmission, leveraging optical fibers to transfer power with minimal system degradation. PoF technology excels in maintaining low noise levels when delivering power to sensitive electronic systems operating in extreme temperatures and high voltage environments. This paper presents the R$\&$D effort of PoF in extreme conditions and underscores its capacity to revolutionize power delivery and management in critical applications, offering a dependable solution with low noise, optimal efficiency, and superior isolation.}


\begin{tikzpicture}[remember picture,overlay]
    \node[anchor=north east, align=right, font=\small, yshift=-10mm, xshift=-5mm] at (current page.north east) {FERMILAB-PUB-24-0265-AD-ETD-LBNF};
\end{tikzpicture}

\footnotetext[0]{This manuscript has been authored by Fermi Research Alliance, LLC under Contract No. DE-AC02-07CH11359 with the U.S. Department of Energy, Office of Science, Office of High Energy Physics. Additionally, this material is based upon work supported by the U.S. Department of Energy, Office of Science, Office of High Energy Physics under Award Number DE-SC0024450. Moreover, this work is supported by subcontract No. 687391 between Fermi Research Alliance, LLC and South Dakota School of Mines and Technology.}

\footnotetext[1]{\textbf{Corresponding authors:}%
\begin{itemize}
    \setlength\itemsep{0pt}
    \item[$*$] \texttt{pellico@fnal.gov}
    \item[$\ddag$] \texttt{David.MartinezCaicedo@sdsmt.edu}
    \item[$\dagger$] \texttt{wei.shi.1@stonybrook.edu}
\end{itemize}
}

\section{Introduction}
\label{sec:intro}

\begin{figure}[h!]
    \centering
    \includegraphics[width=0.9\textwidth]{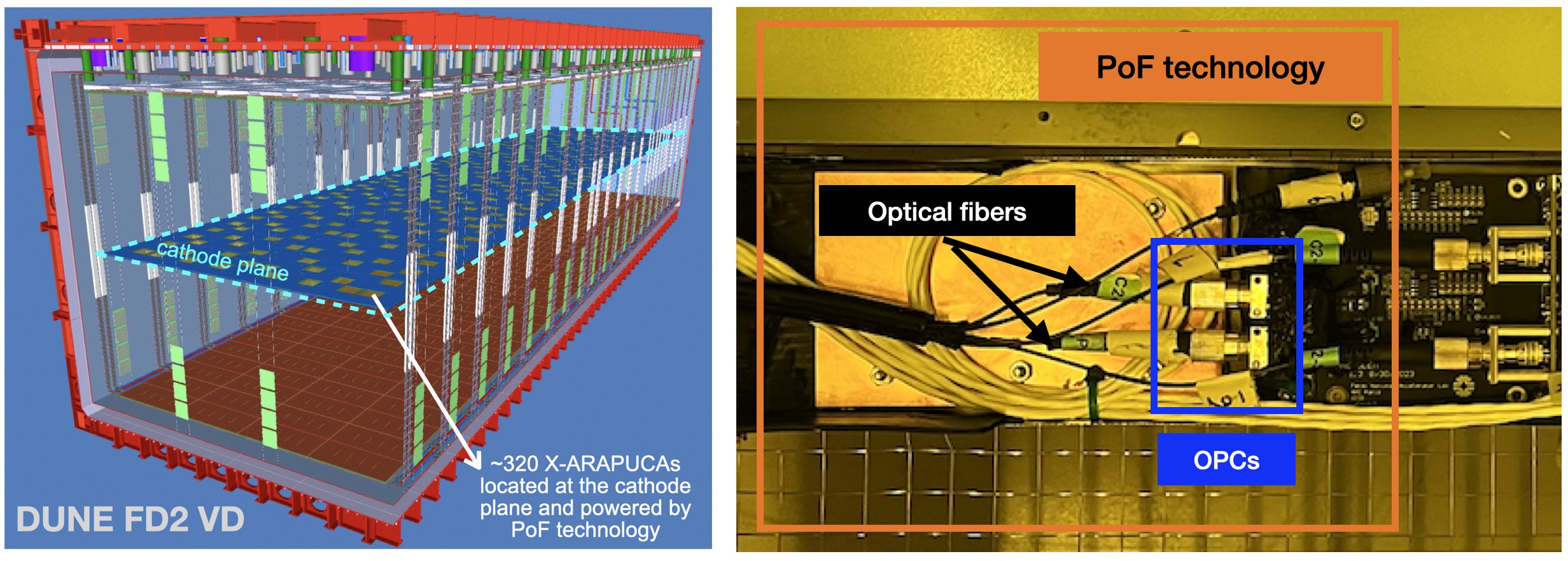}
    \captionsetup{width=0.9\textwidth}
    \caption{Left: Schematic of the DUNE vertical drift concept featuring the X-ARAPUCAs PDS located on the cathode plane. The figure is obtained and adapted from the DUNE FD2 TDR \cite{DUNE:2023nqi}. Right: A cold readout electronic board, powered by PoF, designed for a DUNE single photo-detector module prototype located on a HV cathode~\cite{DUNE:2023nqi}.} 
    \label{fig:Broadcom_laser_PoFBoard}
\end{figure}

The DUNE experiment will soon deploy large liquid argon (LAr) time projection chambers (TPC) to detect neutrino interactions and other particle physics phenomena~\cite{DUNE:2020jqi, DUNE:2020ypp, DUNE:2020lwj, DUNE:2020txw, DUNE:2023nqi}. In addition to the particle tracking provided by the TPC, photon detectors in the cryostat will leverage the high scintillation light yield of LAr to provide crucial timing and additional calorimetric information. The development of the DUNE vertical drift (VD) TPC configuration with its largely unobstructed geometry presents an opportunity to provide excellent photo-detection coverage using $\sim$ 320 photo-detectors known as X-ARAPUCAs~\cite{Machado} placed in the cathode plane (see Figure \ref{fig:Broadcom_laser_PoFBoard}, left) and read them out by DUNE VD cold electronics motherboard (DCEM)~\cite{DUNE:2023nqi}, provided that the X-ARAPUCAs can be electrically isolated for safe operation on high-voltage surfaces of the TPC. This requirement is met for the Photon Detector System (PDS) of the DUNE VD detector using power-over-fiber (PoF) in a novel application both in particle detectors for high energy physics and in a cryogenic environment. 

PoF transmits laser power over (non-conductive) lightweight, non-conductive optical fiber to a remote photovoltaic receiver (also called an optical power converter, OPC), allowing the operation of remote sensors or electronic devices. Existing PoF technologies are commonly employed for voltage isolation between source and receiver and embedded electronics in high voltage or high noise environments~\cite{pofreview, PoFapp1, PoFapp2, PoFapp3}. However, none of the commercially available technologies are rated to operate in cryogenic environments such as liquid argon (LAr) or liquid nitrogen (LN2). 

In the PoF application shown in Figure~\ref{fig:Broadcom_laser_PoFBoard} (right), multimode fibers with a core radius of 62.5 $\mu$m are utilized to transmit the laser diode power. The fibers have a black polyvinylidene fluoride (PVDF) jacket and are FC terminated. Sheathing opacity prevents light from leaking out of the fiber and minimizes background light noise to the PDS (Section~\ref{sec_fiber_jacket}). The electronics power is obtained by stacking two gallium arsenide (GaAs) OPCs connected in parallel on a cold readout board as shown in Figure~\ref{fig:Broadcom_laser_PoFBoard} (right). The GaAs OPCs convert the optical power to electrical power at an efficiency that depends upon the load matching to laser power and to a lesser extent upon the temperature. The GaAs OPC maximum efficiency can reach $\sim$ 50 - 55\%, which is significantly higher than the silicon OPC option that can reach between $\sim$ 20 - 40\%~\cite{DUNE:2023nqi}. This innovative PoF solution provides three major benefits: (1) voltage isolation, (2) noise immunity, and (3) spark-free operation. A summary of the developed PoF system capability is shown in Table~\ref{tab:PoF_parameters}.

\begin{table}[h!]
    \centering
    \begin{tabular}{l|l}
    Parameter                                    & Value                          \\\hline
    808nm Broadcom laser power per unit          & tunable up to 2 W               \\
    Max OPC Efficiency (Optical-to-Electrical)             & $\sim$51\% ($\sim$45\% at room temperature) \\
    Output voltage per OPC                       & 3-38 V \footnotemark        \\
    Output current per OPC                       & up to 100 mA                    \\
    Output power per OPC                         & up to 600 mW                    \\
    \end{tabular}
    \caption{PoF system capability at LN2 temperature.}
    \label{tab:PoF_parameters}
\end{table}

\footnotetext{The OPC base voltage is determined by OPC design. The number of semiconductor gaps/layers sets the base voltage.}

The remainder of this paper is structured as follows. Section~\ref{sec_opc} summarizes the performance of optical power converters (OPC) in warm (room temperature) and cryogenic environments. Section~\ref{sec_fiber} summarizes R\&D of the optical fibers used in the PoF system. Section~\ref{sec_laser} describes the laser transmitter unit operation within the PoF system.  Section \ref{sec_pd} describes an example of the electrical power delivered by an OPC within a PoF system. Section~\ref{sec_dune} describes the characterization and mitigation of the light leakage from a PoF system application alongside photodetectors, where photon background is a relevant concern. Section~\ref{sec_dune_application} details the practical application of PoF technology within the HV environment of the CERN Cold Box Test Facility. Finally, conclusions are provided in Section~\ref{sec_con}. 

\section{Optical Power Converter (OPC)}
\label{sec_opc}

OPCs are semiconductor devices used to convert optical power into an electrical power. OPCs are typically categorized as either single junction or multi junction devices, with the majority of modern OPCs falling into the multi junction category. These multi junction designs, such as the vertical epitaxial heterostructure architecture (VEHSA) design from Broadcom~\cite{photonics9080579}, are composed of many semiconductor sub cell layers \cite{opc_overview}. The material used for the absorbing layer of these sub cells is primarily dependent on the wavelength of the laser being used. Presently, 808 nm lasers are among the most widely available commercially. At this wavelength, the most efficient OPCs utilize GaAs absorbing layers. For higher spectral ranges, such as 960-990 nm, Silicone is employed as the absorbing layer, while at 1500-1600 nm, InGaAs lattice-matched to InP is used \cite{opc_overview}. Similar to the wavelength used, the maximum power capabilities of these devices are constantly evolving and can vary significantly. Generally, optical power converters can be categorized into three power levels: ``regular-power" devices with input power below 1 W, ``medium-power" devices with input power below 6 W, and ``high-power" devices with input power below 50 W \cite{opc_overview}. These OPC devices are typically tested and certified for an operating temperature range of -40°C to 85°C, which is significantly above the cryogenic range. This section describes the procedure and results from the cryogenic characterization of GaAs Broadcom OPCs~\cite{photonics9080579}.

\subsection{OPC Characterization at Room and Cryogenic Temperatures}
The OPC characterization measures the current-voltage (I-V) and power resistance (P-R) curves using different loads and input powers from the laser box. The output power of the OPCs (P = I $\times$ V) was obtained using the I-V curves. Later, the maximum efficiency of the OPC is obtained by calculating the ratio between the maximum output power provided by the OPC ($P^{OPC}_{\text{max output}}$) and the input power provided by the laser box ($P^{laser}_{\text{input}}$), as described in Equation \ref{eg:OPCeff}. 

\begin{figure}[h!]
    \centering
    \includegraphics[width=0.9\textwidth]{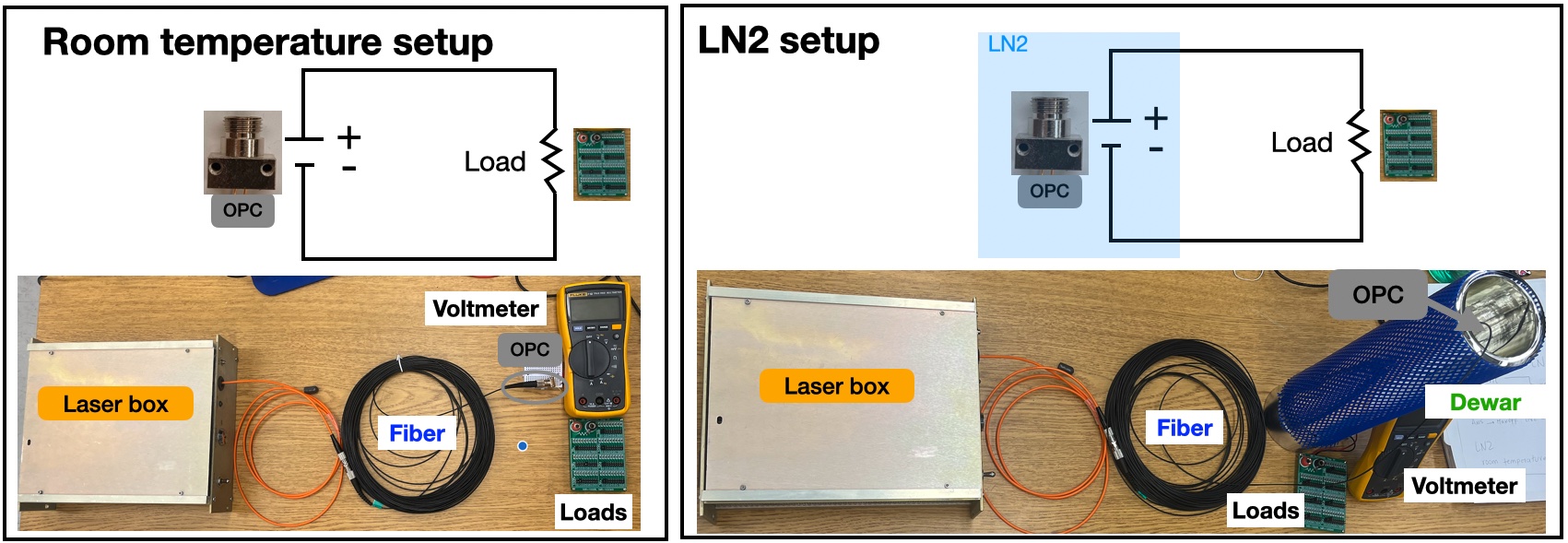}
    \captionsetup{width=0.9\textwidth}
    \caption{The test stand designed and built to characterize the OPCs at room temperature (left) and in LN2 temperature (right). \label{fig:opc_test_setup_cryo}}
\end{figure}

\begin{equation}
    \text{Efficiency} = \frac{P^{OPC}_{\text{max output}} (W)}{P^{Laser}_{\text{input}} (W)}
    \label{eg:OPCeff}
\end{equation}

Figure \ref{fig:opc_test_setup_cryo} shows the test stand designed and built to characterize the OPCs at room and LN2 temperatures. ``Room temperature" refers to the typical ambient temperature of the environment, typically around 20 to 25°C, while ``LN2" refers to liquid nitrogen, with a temperature around -196°C. The setup includes a single-channel laser box with a tunable input power up to $\sim$ 2 W connected to an optical fiber with a 62.5 $\mu$m core, which is then connected to a single OPC. The output voltage of the OPC is measured using a voltmeter across a range of loads controlled by a variable load board, which enabled the load to be adjusted from 50 $\Omega$ to 140 $\Omega$ in 1 $\Omega$ increments. First, the test was conducted at room temperature (Figure~\ref{fig:opc_test_setup_cryo}, left), and was later repeated at LN2 temperatures (Figure \ref{fig:opc_test_setup_cryo}, right). During the LN2 measurements, the OPCs were submerged $\sim$ 5 cm in LN2 to ensure that the measurements were not affected by any power loss due to thermal contraction in the fiber jacket, as described in Section~\ref{sec_fiber_jacket}.

\begin{figure}[h!]
    \centering
    \includegraphics[width=.46\textwidth]{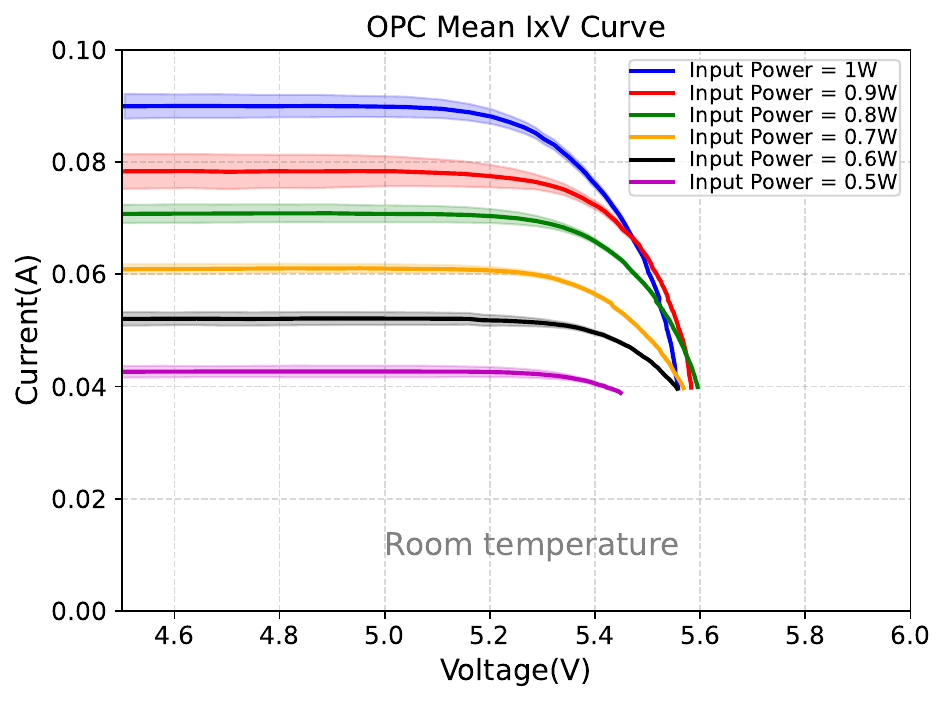}
    \includegraphics[width=.46\textwidth]{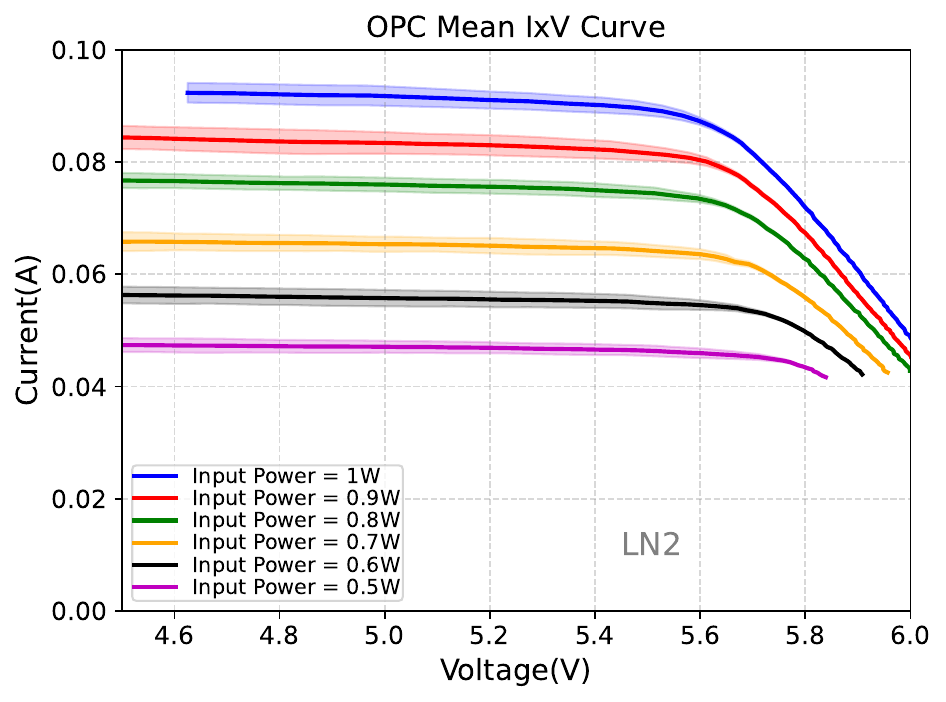}\\
    \includegraphics[width=.46\textwidth]{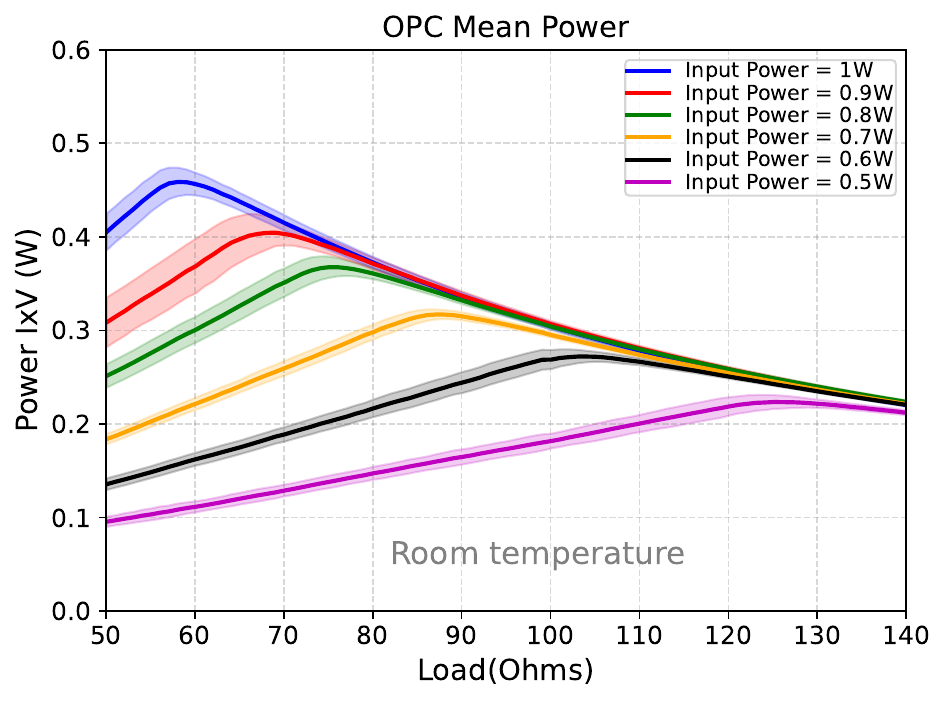}
    \includegraphics[width=.46\textwidth]{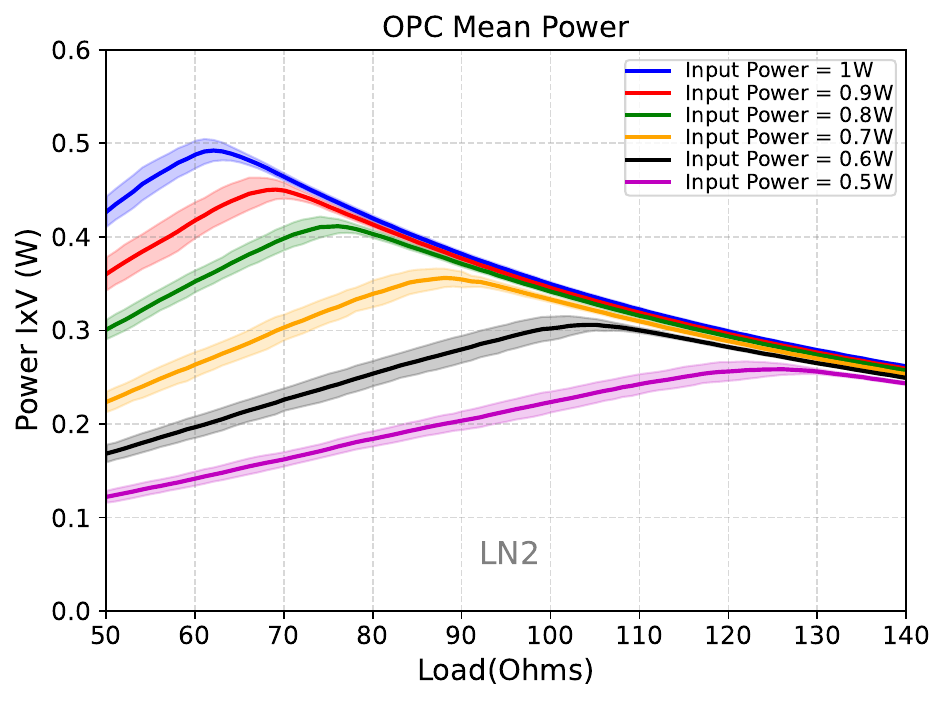}
    \captionsetup{width=0.9\textwidth}
    \caption{I-V curves (upper panel) and P-R curves (lower panel) for 12 OPCs (shaded color bands represent standard deviation) at room temperature (left column) and cryogenic temperatures (right column).\label{fig:opc-IV_PR_air_ln}}
\end{figure}

Figure \ref{fig:opc-IV_PR_air_ln} displays the I-V curves (upper panel) and P-R curves (lower panel) including the standard deviation (shaded color bands) obtained for the 12 OPCs at room temperature (left column) and LN2 temperatures (right column). For each OPC a total of six I-V and P-R curves were measured, with each colored curve corresponding to a different input power provided by the laser box, ranging from 0.5 W up to 1.0 W in steps of 100 mW. The I-V curves obtained at room temperature and LN2 for the OPCs exhibited similar shapes, with the measured OPC output voltage in LN2 being  higher. The P-R curves showed a slight upward shift, indicating that higher output powers on the OPCs are achieved in LN2 when compared to room temperature measurements. Additionally, the P-R curves showed that the output power of the OPCs decreases after the maximum OPC output power is achieved  as the input load is increased, both at room temperature and in LN2 conditions.

\begin{figure}[h!]
    \centering
    \includegraphics[width=.46\textwidth]{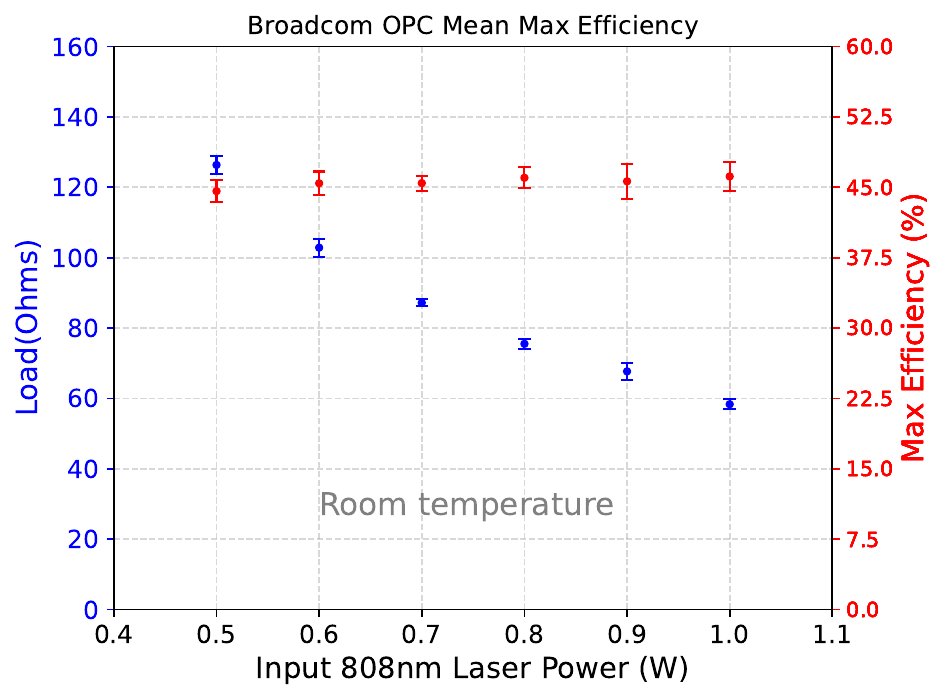}
    \includegraphics[width=.46\textwidth]{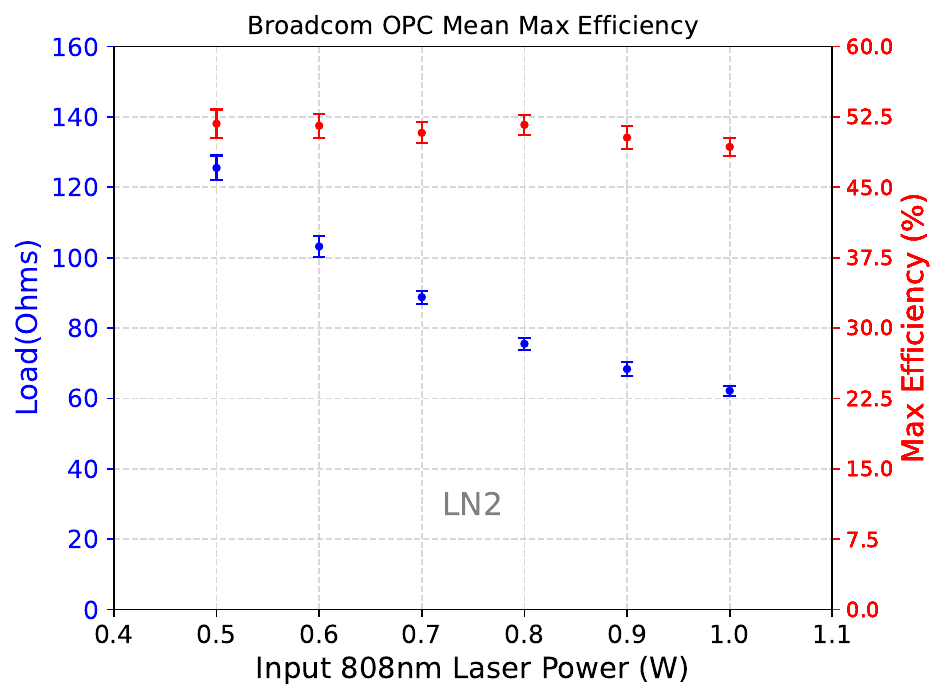}
    \captionsetup{width=0.9\textwidth}
    \caption{The input power provided by the laser box (X-axis), maximum efficiency obtained in the OPCs (Y-axis in red), and the corresponding load where the maximum OPC output power was achieved (Y-axis in blue) for both room temperature (left) and LN2 measurements (right). \label{fig:opc-maxeff_air_ln2}}
\end{figure}

\begin{table}[h!]
    \centering
    \smallskip
    \begin{tabular}{c|c|c|c}
    Input Power (W) & Environment Condition & Load (ohms) &  Max Efficiency (\%) \\
    \hline
    \multirow{2}{*}{0.5}  & room temperature & 126.35 $\pm$ 2.59 & 44.59 $\pm$ 1.19 \\
                          & LN2              & 125.52 $\pm$ 3.52 & 51.78 $\pm$ 1.51 \\ \hline
    \multirow{2}{*}{0.6}  & room temperature & 102.85 $\pm$ 2.55 & 45.42 $\pm$ 1.25 \\
                          & LN2              & 103.18 $\pm$ 2.90 & 51.56 $\pm$ 1.29 \\ \hline
    \multirow{2}{*}{0.7}  & room temperature & 87.18 $\pm$ 1.04  & 45.44 $\pm$ 0.80 \\
                          & LN2              & 88.77 $\pm$ 1.89  & 50.80 $\pm$ 1.11 \\ \hline
    \multirow{2}{*}{0.8}  & room temperature & 75.52 $\pm$ 1.50  & 46.01 $\pm$ 1.12 \\
                          & LN2              & 75.52 $\pm$ 1.71  & 51.66 $\pm$ 1.05 \\ \hline
    \multirow{2}{*}{0.9}  & room temperature & 67.68 $\pm$ 2.40  & 45.63 $\pm$ 1.86 \\
                          & LN2              & 68.35 $\pm$ 1.92  & 50.29 $\pm$ 1.20 \\ \hline
    \multirow{2}{*}{1.0}  & room temperature & 58.35 $\pm$ 1.48  & 46.15 $\pm$ 1.55 \\
                          & LN2              & 62.18 $\pm$ 1.44  & 49.31 $\pm$ 0.97 \\ 
    \end{tabular}
    \captionsetup{width=0.9\textwidth}
    \caption{The results obtained for the 12 OPCs maximum efficiency achieved for the six input powers and the corresponding loads. A maximum efficiency of $\sim$51\% was achieved for the 12 OPCs measured in LN2, while $\sim$45\% was achieved for the room temperature measurements. }
    \label{tab:MaxEff_12OPCs}
\end{table}

Figure~\ref{fig:opc-maxeff_air_ln2} shows the input power provided by the laser box (X-axis), maximum efficiency obtained in the OPCs (Y-axis in red), and the corresponding load where the maximum OPC output power was achieved (Y-axis in blue). The error bars in the maximum efficiency and the loads corresponds to the the standard deviation calculated for the 12 OPCs measurements. The OPCs maximum efficiency does not change significantly, and the load decreases as the input power provided by the laser box increases for both room temperature (Figure~\ref{fig:opc-maxeff_air_ln2}, left) and LN2 measurements (Figure~\ref{fig:opc-maxeff_air_ln2}, right). Table~\ref{tab:MaxEff_12OPCs}, summarizes the results obtained for the 12 OPCs maximum efficiency achieved for the six input powers and the corresponding loads. A maximum efficiency of $\sim$51\% was achieved for the 12 OPCs measured in LN2, while $\sim$45\% was achieved for the room temperature measurements.

Modifications to the OPC improved the cryogenic efficiency with a slight negative impact to the warm efficiency. Additionally, efficiency is dependent upon the matching of the laser light wavelength and the OPC semiconductor gap energy. The cryogenic temperature will cause a physical change in the material that would require a corresponding laser wavelength change to optimize efficiency. However, price considerations often require use of off-the-shelf lasers at the cost of efficiency.

\subsection{Long Term Test at Cryogenic Temperatures}
\label{subsec_longterm}

\begin{figure}[htbp!]
    \centering
    \includegraphics[width=.7\textwidth]{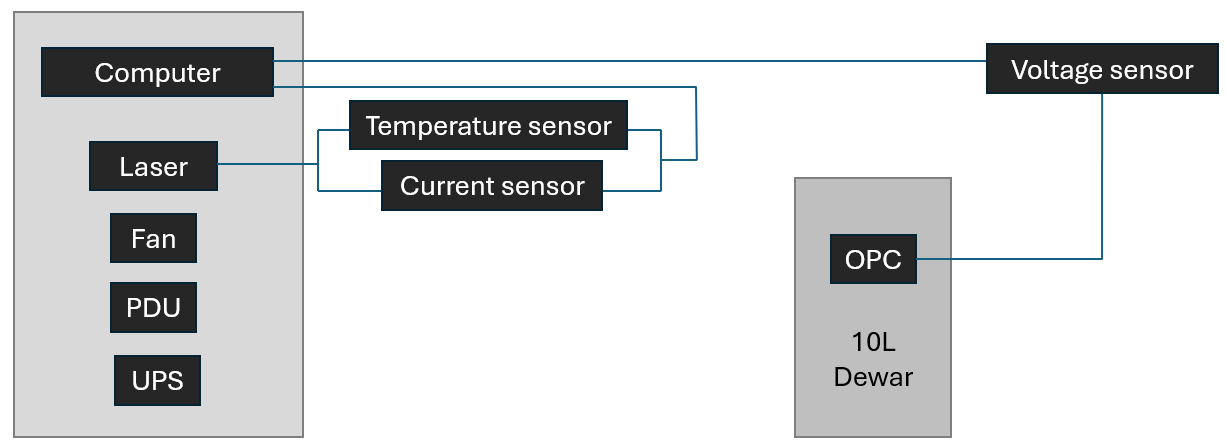}
    \captionsetup{width=0.9\textwidth}
    \caption{A schematic view of the long term test stand with an OPC submerged in a 10L Dewar.\label{fig:long_term_single_OPC_stand}}
\end{figure}

\begin{figure}[htbp!]
    \centering
    \includegraphics[width=.7\textwidth]{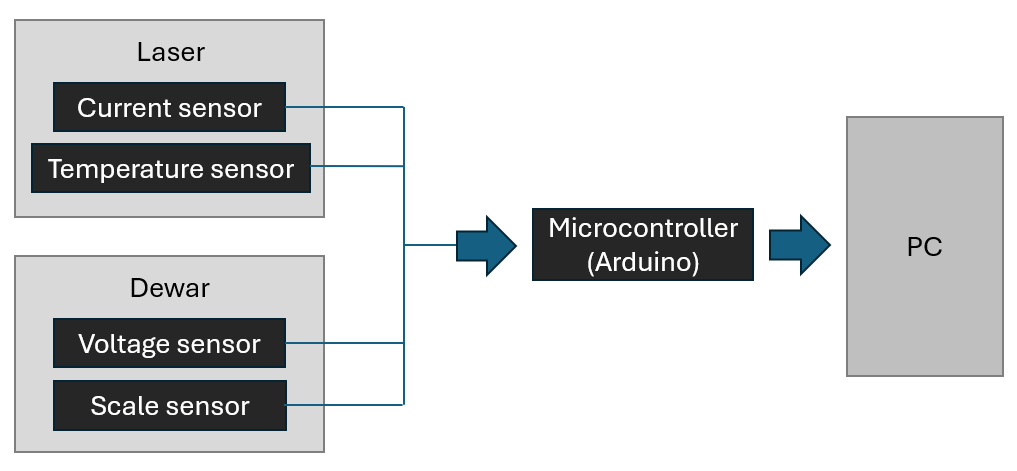}
    \captionsetup{width=0.9\textwidth}
    \caption{The voltage of the OPC and the current and temperature of the laser is monitored by various sensors using an Arduino Uno microcontroller. Additionally, the data from each sensor was recorded using a dedicated computer. \label{fig:long_term_sensor_flow}}
\end{figure}

To assess the feasibility of using OPCs under cryogenic conditions for extended time periods, a long-term test setup was designed, constructed, and conducted. Figure \ref{fig:long_term_single_OPC_stand} shows the long-term test setup which includes a 10-liter Dewar and a scale to continuously monitor Dewar weight. An optical fiber was used to connect the laser to the OPC submerged within the Dewar. The voltage across the OPC was continuously monitored using a voltage sensor connected to an Arduino Uno for data acquisition. Additionally, sensors were deployed to monitor laser current and temperature to readout the laser's performance and overall test stability over extended durations. The sensitivity of each sensor used in the long term test setup is determined by the resolution of the analogue-to-digital conversion (ADC) levels within the Arduino Uno micro-controller. The schematic illustration of each sensor's connection to the Arduino Uno is shown in Figure~\ref{fig:long_term_sensor_flow}. Data from various sensors are captured using an Arduino Uno, which operates with a 10-bit resolution providing 1024 discreet levels. Each sensor's sensitivity is derived from the Arduino Uno's ADC precision, involving two discrete levels for each measurement. The voltage sensor can measure up to 25 V and has a resolution of $\pm$50 mV. The current sensor is a Hall effect sensor that can record up to 30 A and has a resolution of $\pm$30 mA. Lastly, the temperature sensor is a 1 k$\Omega$  platinum resistance temperature detector (RTD) that can measure between -200 $^{\circ}$C and 200 $^{\circ}$C, and has a resolution of $\pm$0.8$^{\circ}$C. To safeguard against potential power interruptions, an uninterruptible power supply (UPS) was employed to ensure continuous operation throughout the months-long test period. 

\begin{figure}[htbp!]
    \centering
    \includegraphics[width=.43\textwidth]{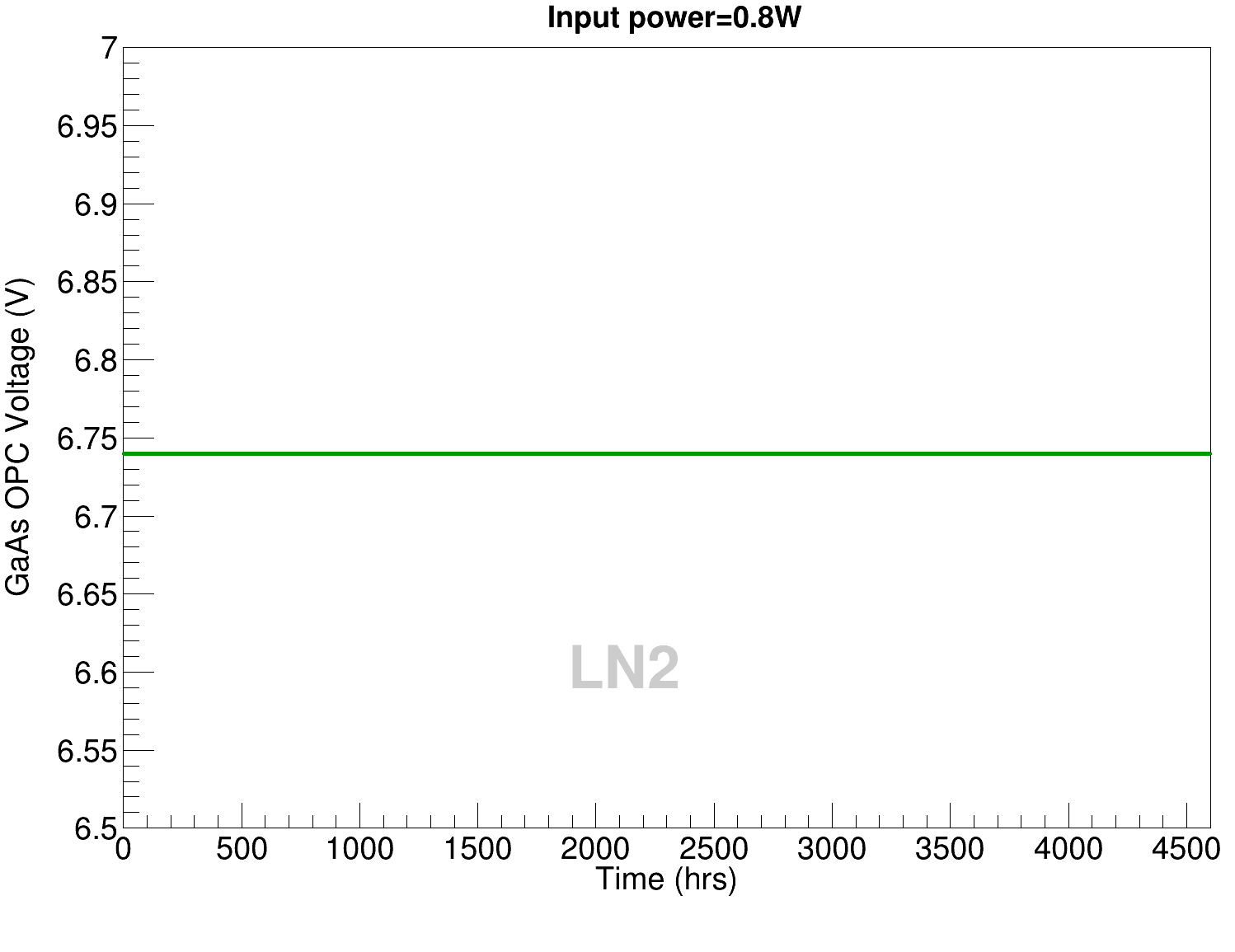}
    \qquad
    \includegraphics[width=.43\textwidth]{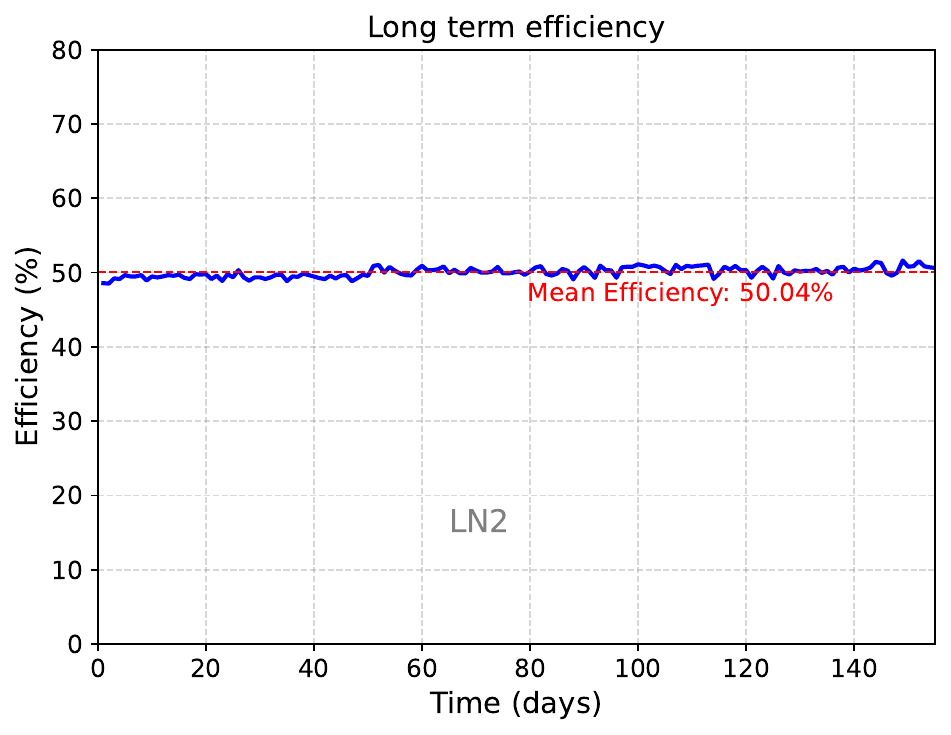}
    \captionsetup{width=0.9\textwidth}
    \caption{The OPC output voltage has remained stable at 6.74 V throughout the 4500-hour monitoring period, equivalent to 6.1 months (left), and the OPC efficiency over the last 155 days (right) of data collection shows a mean efficiency of 50.04\%. \label{fig:opcv-eff}}
\end{figure}

The monitoring of the output OPC voltage has shown a steady value of 6.74 V with an input power of 0.8 W provided by the laser box over a testing period of 4500 hours, equivalent to 6.1 months of data collection (Figure~\ref{fig:opcv-eff}, left). Additionally, a voltmeter was used to measure the voltage under an 80 $\Omega$ load to monitor the efficiency of the OPC, calculated using Equation~\ref{opc_eff}, during the last 155 days of the data collection. Figure~\ref{fig:opcv-eff} (right) illustrates that the OPC efficiency has not changed significantly (less than 2.6\%) over the last 155 days of the data collection, with a mean efficiency of 50.04\%.

\begin{equation}
    \text{Efficiency}(\text{at 80}\Omega) = \frac{P^{OPC}_{\text{output at 80}\Omega} (W)}{P^{Laser}_{input} (W)} = \frac{V^{2}/{80\Omega}}{0.8W}
    \label{opc_eff}
\end{equation}

\section{Optical Fibers}
\label{sec_fiber}

\begin{figure}[htbp!]
    \centering
    \includegraphics[width=0.9\textwidth]{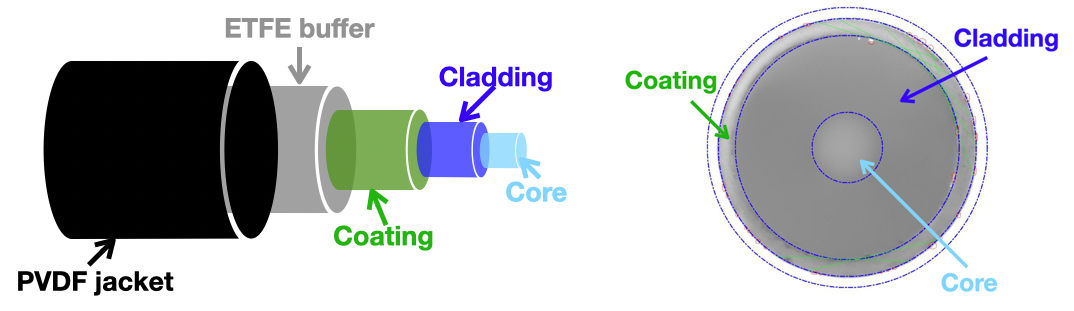}
    \captionsetup{width=0.9\textwidth}
    \caption{A schematic representation detailing the structure of an optical fiber (left). The cross-sectional view (right) shows the core, cladding, and coating layers, as observed through Thorlabs' optical fiber scope. The green lines outline scratches on the surface of the fiber end.
    \label{fig:FiberStructure}}
\end{figure}

Optical fibers find wide applications in underwater power transmission, telecommunications, rotor blade monitoring systems, current and voltage sensors in high-voltage environments, and many other areas discussed in literature~\cite{pofreview}. Optical fibers are used in cryogenic applications for tasks like detecting quenches in high-temperature superconductors~\cite{Li}, infrared instruments for space~\cite{lee2001properties}, and in quantum computing~\cite{lecocq2021control} experiments conducted at very low temperatures. There are two types of optical fibers commonly used: single-mode~\cite{TRICKER200337-1} and multimode~\cite{kao1966dielectric}. Several multimode fiber configurations were studied, but eventually a focus was placed on fibers with a core diameter of 62.5 $\mu$m. These fibers consist of a core, cladding, coating, buffer, and outer jacket, as depicted in Figure~\ref{fig:FiberStructure}. Table~\ref{tab:Optical-fiber-description}, details the specific fiber configuration, featuring a large cladding, coating, buffer, and a black PVDF jacket, which substantially minimizes light leakage along the optical fiber and offers non-contamination at cryogenic temperatures. More details about the PVDF jacket and its temperature-dependent properties are discussed in Section~\ref{sec_fiber_jacket}. 

\begin{table}[htbp]
    \centering
    \begin{tabular}{l|l}
    Item                    & Description \\ \hline
    Numerical Aperture (NA) & 0.27 \\ 
    Index Profile           & GI \\ 
    Core Diameter           & 62.5 $\pm$ 3 $\mu$m\\
    Cladding Diameter       & 200 $\pm$ 4 $\mu$m  \\ 
    Coating Diameter        & 230 $\pm$ 10 $\mu$m \\
    ETFE Buffer Diameter    & 500 $\pm$ 50 $\mu$m \\
    Black PVDF jacket       & 1.5 $\pm$ 0.1 mm \\ \hline
    Connector Type          & FC/PC ceramic \\
    Fiber Cable Length      & 40 $\pm$ 0.2 m 
    \end{tabular}
    \captionsetup{width=0.9\textwidth}
    \caption{Characteristics of MH GoPower Company Limited~\cite{MHGoPower} optical fibers includes a cladding, coating, buffer, black PVDF jacket, connector type and the cable length.}
    \label{tab:Optical-fiber-description}
\end{table}

The process of down-selecting optical fibers suitable for PoF at cryogenic temperatures took $\sim$ 3 years of research and testing across 5 types of fibers. Initial efforts on fiber discovery for cryogenic PoF in high energy physics (HEP) PDS looked at two aspects: Light leakage and physical degradation. Fibers with non black jackets were quickly determined to not be suitable. Jacket material was also explored extensively to understand cryogenic viability, including PVC and silicone jacketed fibers being tested. Table~\ref{tab:FiberType} provides a summary of the optical fiber types that underwent testing, detailing their specific cladding and jacket materials.

\begin{table}[htbp]
    \centering
    \smallskip
    \begin{tabular}{c|c|l}
    Fiber Core ($\mu$m) & Length (cm) & Jacket Material/Coating Material \\ \hline
    62.5 & 4000 & PVDF/Acrylate  \\ 
    62.5 & 2500 & without jacket/Acrylate\\ 
    105  & 200  & ETFE/Polyimide  \\
    105  & 100  & PFA/Acrylate \\ 
    105  & 100  & PFA/Polyimide 
    \end{tabular}
    \captionsetup{width=0.9\textwidth}
    \caption{List of different optical fibers tested with their varying specifications.}
    \label{tab:FiberType}
\end{table}

\subsection{Quality Assurance (QA/QC) Procedures}
\label{QA/QC}

\begin{figure}[htbp!]
    \centering
    \includegraphics[width=0.5\textwidth]{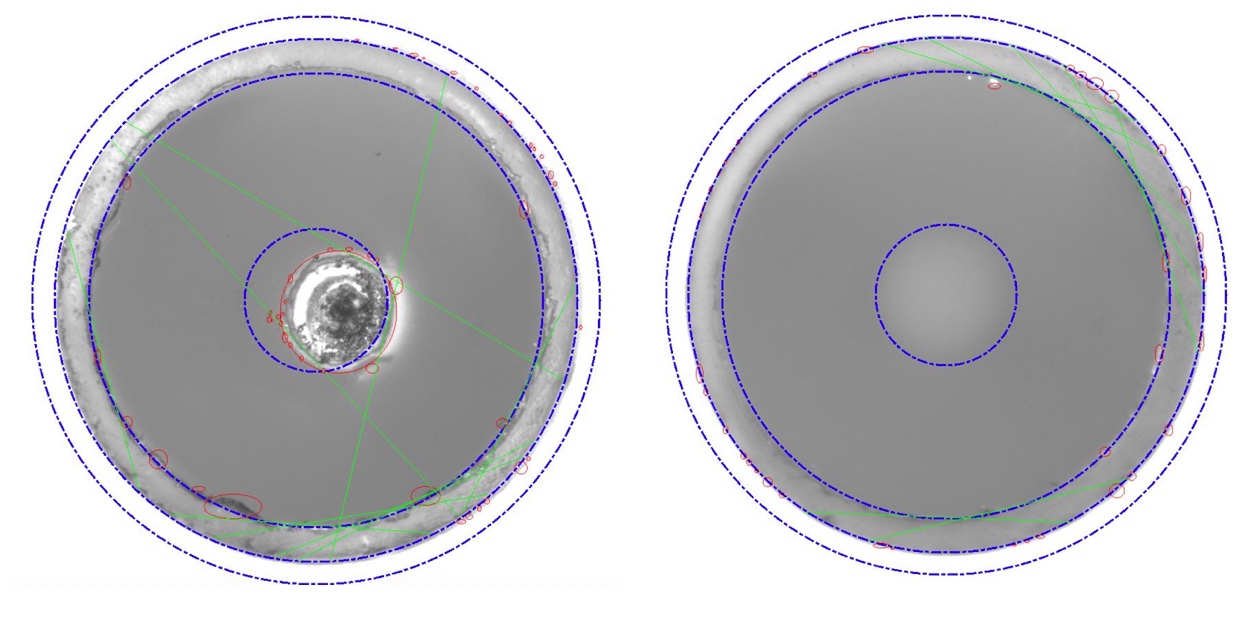}
    \captionsetup{width=0.9\textwidth}
    \caption{Left: Optical fiber end burned due to inadequate cleaning. Right: Optical fiber end that passed successful visual inspection, no damage is observed along the core and cladding. Optical fiber inspection was conducted using a Thorlabs' optical fiber scope.}
    \label{fig:fiber_visual_inspection}
\end{figure}

In order to ensure that optical fibers are suitable for their use in the PoF technology, a three step quality assurance procedure was developed:

\textit{1. Visual inspection:} A visual inspection is needed to look for any visible physical damage of the fiber, fiber ends (core and cladding) and jacket surface. Since PoF technology uses a class-4 laser \cite{LaserBroadcom} as an optical power source, damages like scratches, cracks, or chips can affect the light transmission resulting in power loss. Scratches and cracks in the optical fiber core not only impact power transmission, but it can also absorb a significant amount of light, leading to hot spots in certain areas that could potentially damage the fiber. Visual inspection makes it possible to detect contaminants such as dust, dirt or oils that may accumulate on the fiber ends during handling or cleaning. Figure \ref{fig:fiber_visual_inspection} (left) illustrates an example of a fiber end damaged by burns resulting from dust accumulation when the proper cleaning procedure was not followed.

\begin{figure}[htbp]
    \centering
    \includegraphics[width=0.45\textwidth]{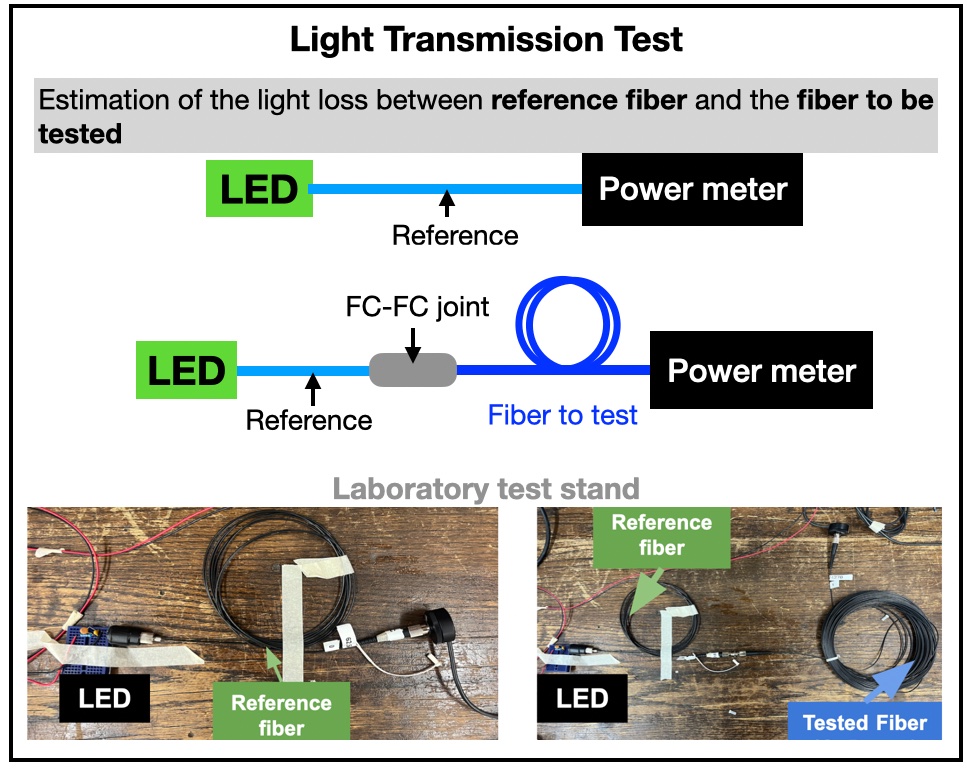}
    \includegraphics[width=0.45\textwidth]{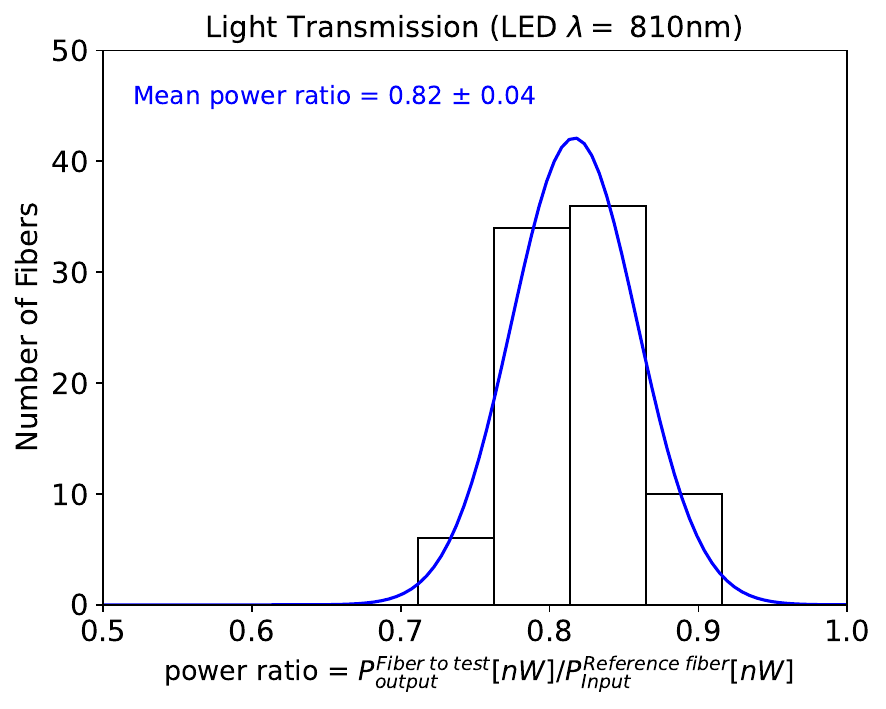}
    \captionsetup{width=0.9\textwidth}
    \caption{Light transmission test stand (left) with results of 87 optical fibers (right)}
    \label{fig:transparency}
\end{figure}

\textit{2. Light transmission:} This test serves to check the light propagation along the fiber as a function of the optical fiber's length in room temperature. Figure \ref{fig:transparency} (left) shows a basic light transmission setup using an LED of 810 nm and a photo-diode power sensor. The light transmission test quantifies the ratio of light power loss between a reference fiber and the fiber under examination as shown in Equation \ref{eq:ratio_light}. 
The ratio between the input power measured at the reference fiber and the output power measured for each fiber under test as shown in Figure \ref{fig:transparency} (right). A total of 87 fibers with a 62.5 $\mu$m core and 40-meter length were tested obtaining a mean value for the ratio of 0.82 $\pm$ 0.04.

\begin{equation}
   \text{ratio} = \frac{\text{P}^{\text{Fiber to test}}_{output} (nW)}{\text{P}^{\text{Reference fiber}}_{input} (nW)}
    \label{eq:ratio_light}
\end{equation}

Although this test verifies light transmission along the fiber, two of the eighty seven fibers exhibited significant power loss when used to transmit optical power O(100 mW) from a laser source, even though they were measured to have a power ratio greater than 80\%. Therefore, conducting an optical power test is necessary to determine the suitability of the optical fibers for PoF transmission. 

\begin{figure}[htbp]
    \centering
    \begin{subfigure}[b]{0.2\textwidth}
        \includegraphics[width=\textwidth]{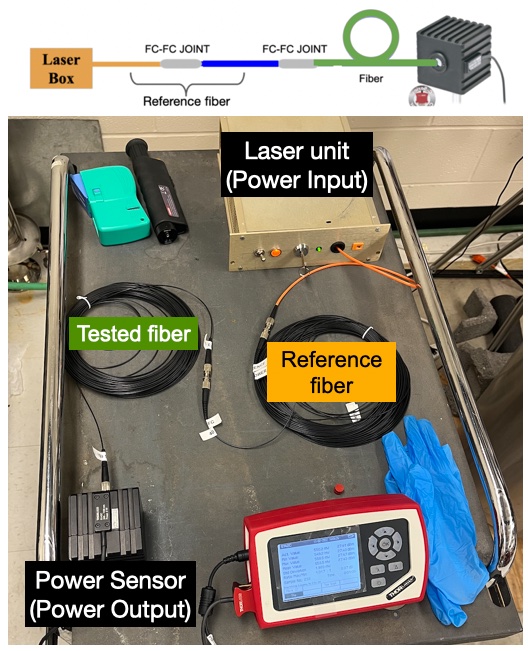}
        \caption{Optical power performance test stand}
        \label{fig:PowerTestStand}
    \end{subfigure}
    \begin{subfigure}[b]{0.37\textwidth}
        \includegraphics[width=\textwidth]{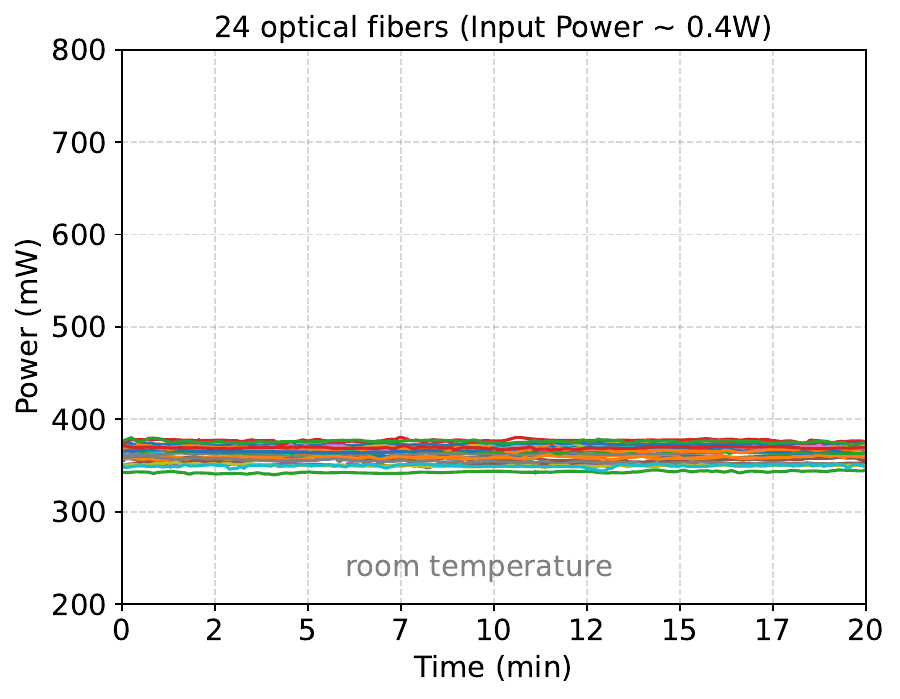}
        \caption{Optical power performance test at 0.4 W for 24 fibers}
        \label{fig:PowerTest_0.4W}
    \end{subfigure}
    \begin{subfigure}[b]{0.37\textwidth}
        \includegraphics[width=\textwidth]{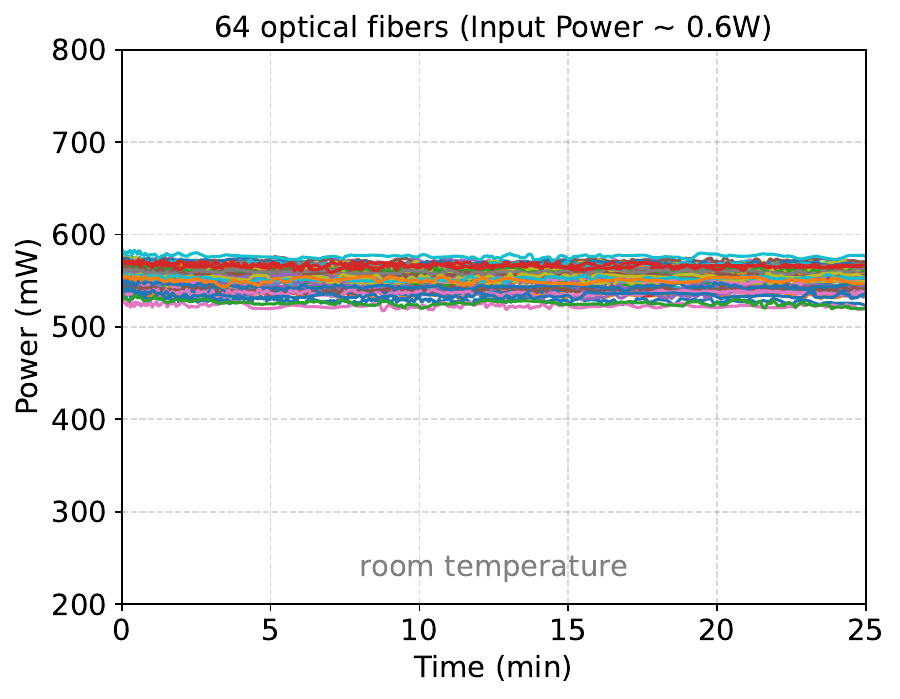}
        \caption{Optical power performance test at 0.6 W for 64 fibers}
        \label{fig:PowerTest_0.6W}
    \end{subfigure}
    \captionsetup{width=0.9\textwidth}
    \caption{Power test for optical fibers to monitor optical power performance using PoF technology at room temperature.}
    \label{fig:PowerTestAll}
\end{figure}

\textit{3. Optical Power Performance:} Since optical fibers are primarily used for signal transmission, this does not guarantee their suitability for optical power transmission on the order of hundreds of mWs. Conducting an optical power performance test at room temperature involves monitoring the optical power transmission over an extended period of time using an integrating sphere photo-diode power sensor, as shown in Figure \ref{fig:PowerTestStand}. For this study, two sets of fibers consisting of 24 and 64 fibers (62.5$\mu$m core and 40 m length) were tested for optical power performance under two laser input powers. Figure \ref{fig:PowerTest_0.4W} shows the power performance test conducted over 24 fibers at an input power of 0.4 W and Figure \ref{fig:PowerTest_0.6W} displays the power performance of 64 fibers for 25 minutes using a laser input power of 0.6 W. Each colored line in both figures corresponds to the power performance of an individual fiber. A mean power performance of 362 $\pm$ 7 mW at an input power of 0.4W and 553 $\pm$ 15 mW at 0.6W were measured, with a mean power loss of $\sim$38 mW and $\sim$47 mW, respectively. \\

The described three-step quality assurance procedure made it possible to measure a power loss of $\sim$ 10\% with respect to a laser power input of 0.4 W and $\sim$ 9\% with respect to a laser power input of 0.6 W at room temperature. However, applications in harsh conditions such as cryogenic temperatures present challenges not only for the optical fiber integrity, but also for optical power transmission (see Sections \ref{Sec:thermal}, \ref{Sec:loss_power_cryo}, \ref{bending} and \ref{sec_fiber_jacket}).

\subsection{Thermal Cycling Test}
\label{Sec:thermal}
 
\begin{figure}[h!]
    \centering
    \includegraphics[width=0.45\textwidth]{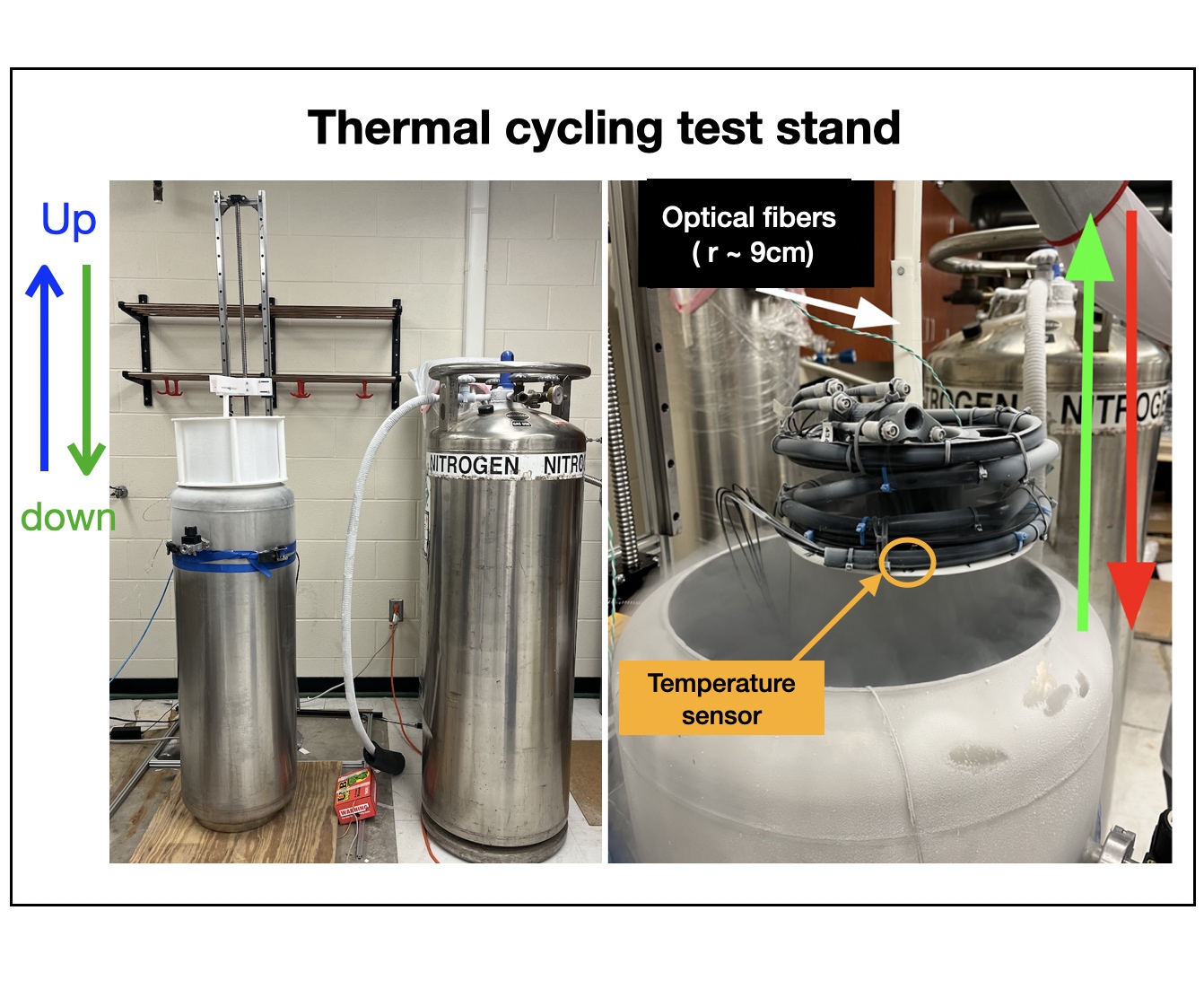}
    \includegraphics[width=0.45\textwidth]{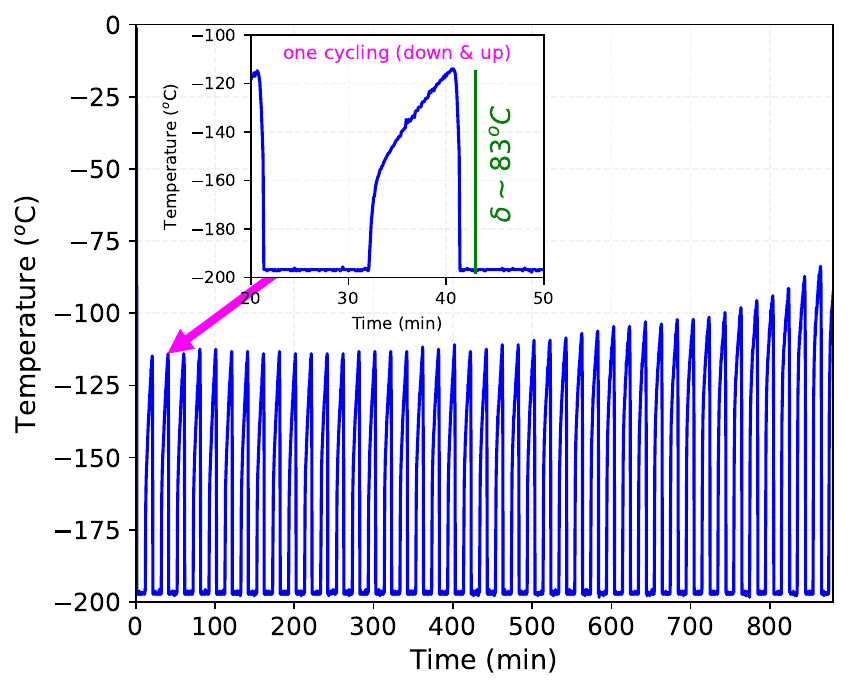}
    \captionsetup{width=0.9\textwidth}
    \caption{Left: A test stand for thermal cycling testing of a set of eight optical fibers, each with a 62.5 $\mu$m core and a length of 40 m. The fibers were wrapped to a spool of radius of $\sim$ 9 cm.
    Right: Example of the first 45 cycles recorded during the thermal cycling test for the set of 8 optical fibers. Each thermal cycle had a duration of 20 minutes, distributed as submerging the fibers in LN2 for 9 minutes, raising them for a period of 1 minute, holding the fibers at room temperature for 9 minutes, and then lowering the fibers back into the LN2 for 1 minute, where a temperature difference of $\sim$83$^{o}$C was recorded.}
    \label{fig:thermalcycling}
\end{figure}

A thermal cycling test makes it possible to evaluate the optical fibers under cryogenic conditions. This test subjects the optical fibers to controlled thermal stress through repetitive cycles of warming and cooling using liquid nitrogen. Figure~\ref{fig:thermalcycling} (left) shows the thermal stress test stand designed and constructed for this study, where optical fibers underwent thermal cycling. 

Each cycle had a duration of 20 minutes, distributed as  submerging the fibers in LN2 for 9 minutes, raising them out of the LN2 for a period of 1 minute, holding the fibers at room temperature for 9 minutes, and then lowering the fibers back into the LN2 for 1 minute (see Figure \ref{fig:thermalcycling}, right). The temperature profile of the cycle is monitored by a temperature sensor. 

Two sets of optical fibers are evaluated under thermal cycling tests at cryogenic conditions. One set with eight optical fibers undergoes a single thermal cycling run of $\sim$ 129 cycles, and another set with five optical fibers undergoes $\sim$ 72 cycles during three thermal cycling runs.

A total of eight optical fibers, each with a 62.5$\mu$m core and a length of 40 m were inserted into a black Polytetrafluoroethylene (PTFE) tube and wrapped around a spool with a radius of $\sim$ 9 cm, as depicted in Figure \ref{fig:thermalcycling} (left). The PTFE tube has a slit along its entire length, allowing the LN2 to enter and circulate within. Light transmission tests were conducted both before and after the thermal cycle test. Table~\ref{tab:Thermal_129} presents the ratio between the input and output power (see Section~\ref{QA/QC}) before and after conducting the thermal cycling test. These results showed that the ratio before and after conducting the 129 thermal cycles are similar and within the statistical errors of the measurement. 

\begin{table}[h!]
    \centering
    \smallskip
    \begin{tabular}{c|c}
    & ratio = P$_{output}$(nW)/P$_{input}$(nW) \\ \hline
    Before 129 cycles   & 0.78 $\pm$ 0.02 \\ 
    After 129 cycles    & 0.78 $\pm$ 0.01 \\
    \end{tabular}
    \captionsetup{width=0.9\textwidth}
    \caption{Light transmission tests conducted on a set of eight optical fibers before and after 129 thermal cycles. Details of the light transmission test are provided in Section \ref{QA/QC}.}
    \label{tab:Thermal_129}
\end{table}

\begin{figure}[h!]
    \centering
    \includegraphics[width=0.37\textwidth]{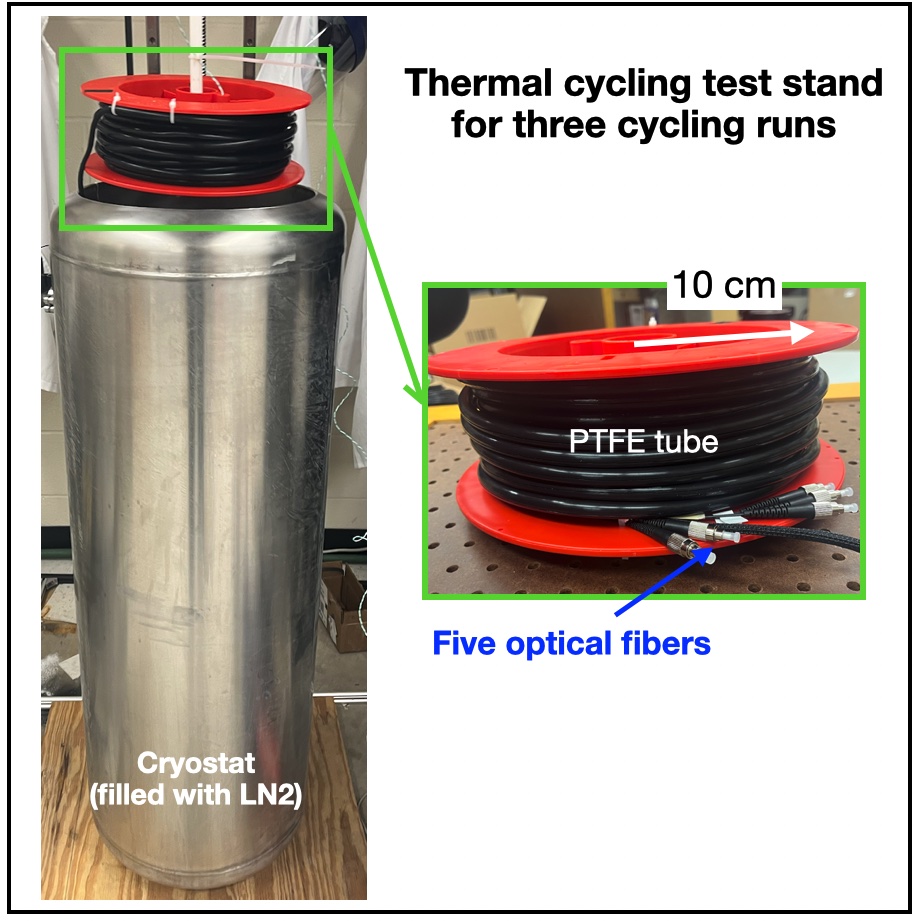}
    \includegraphics[width=0.55\textwidth]{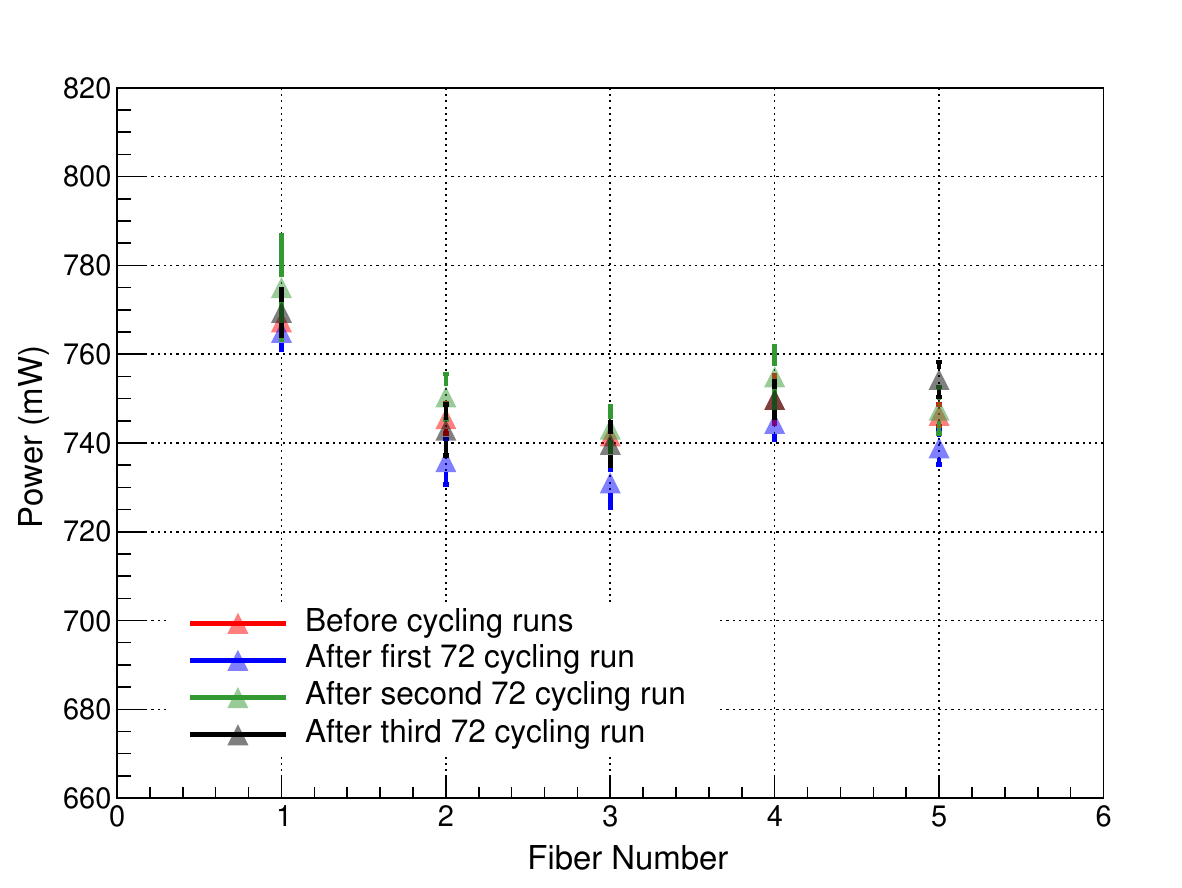}
    \captionsetup{width=0.9\textwidth}
    \caption{Left: Thermal cycling test stand for five optical fibers under three thermal cycle runs. Right: Optical power performance test for a set of five fibers before and after the three thermal cycles. Details of the optical power performance test procedure are provided in Section~\ref{QA/QC}.}
    \label{fig:MultipleThermalCycling_PowerTest}
\end{figure}

A total of five new optical fibers with a core diameter of 62.5 $\mu$m and a length of 40 m were also tested and encased within a protective polyester sleeve and then inserted into a black PTFE tube. Each optical fiber underwent three thermal cycling runs, each run consisting of $\sim$ 72 cycles, where the fibers were spooled at a radius of $\sim$ 10 cm (see Figure \ref{fig:MultipleThermalCycling_PowerTest}, left). Light transmission tests were conducted both before and after the thermal cycling tests. Table \ref{tab:MultipleThermalCycling}, summarizes the light transmission results over the three thermal cycling runs, and again no noticeable difference between the three measurements was observed. In addition, an optical power performance test  over  the five optical fibers was done before and after each thermal cycling run using a laser input power of 0.8 W. Figure~\ref{fig:MultipleThermalCycling_PowerTest} (right) shows the results of the optical power test measured for each fiber before and after each thermal cycling run. No significant deviation was found in the optical fiber performance after each test.

\begin{table}[h!]
    \centering
    \smallskip
    \begin{tabular}{l|c}
    & ratio = P$_{output}$(nW)/P$_{input}$(nW) \\ \hline
    Before cycling                & 0.83 $\pm$ 0.05 \\ 
    After first 72 cycling run    & 0.79 $\pm$ 0.04 \\
    After second 72 cycling run   & 0.78 $\pm$ 0.05 \\
    After third 72 cycling run    & 0.80 $\pm$ 0.06
    \end{tabular}
    \captionsetup{width=0.9\textwidth}
    \caption{Light transmission test conducted for five optical fibers before and after three thermal cycling runs. Details of the light transmission test procedure is provided in Section \ref{QA/QC}.}
    \label{tab:MultipleThermalCycling}
\end{table}

\subsection{Power Loss at Cryogenic Temperatures}
\label{Sec:loss_power_cryo}
\begin{figure}[h!]
    \centering
    \begin{subfigure}[b]{0.32\textwidth}
        \includegraphics[width=\textwidth]{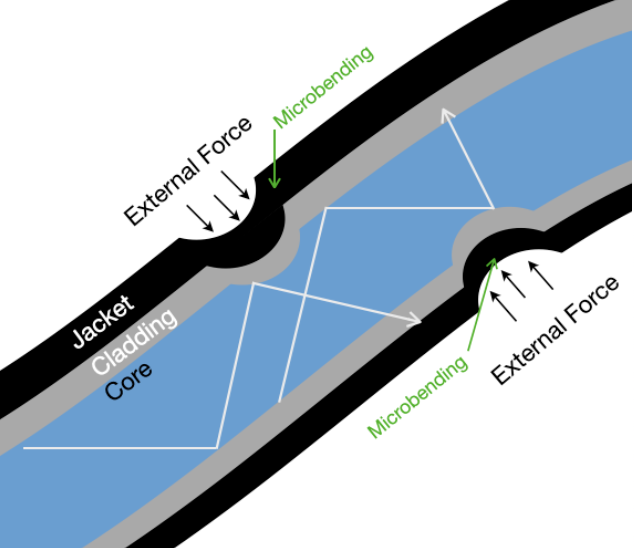}
    \end{subfigure}
    \begin{subfigure}[b]{0.5\textwidth}
        \includegraphics[width=\textwidth]{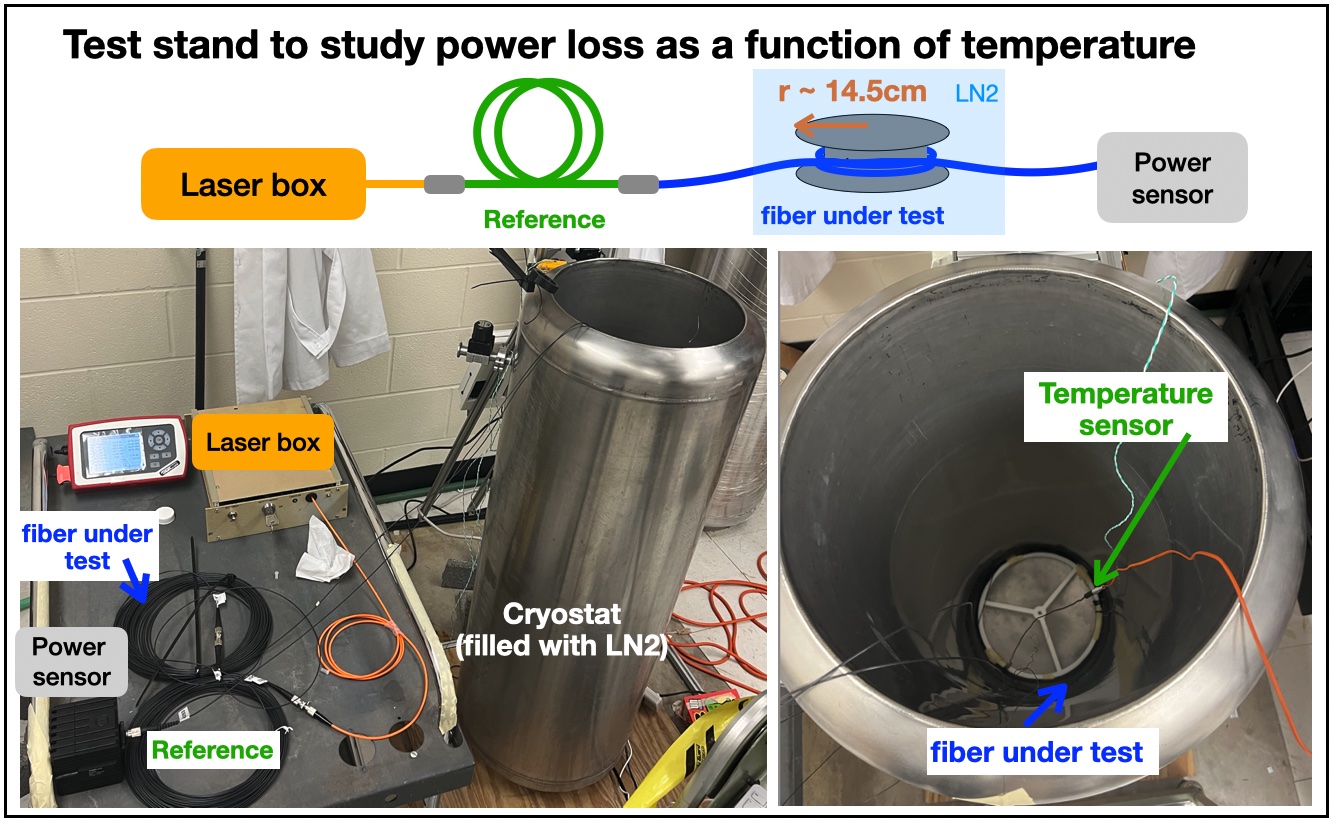}   
    \end{subfigure}
    \captionsetup{width=0.9\textwidth}
    \caption{Left: The schematic view of applying external force to the jacket surface resulting in the deformation of the cladding and core of a fiber. Right: Test stand to study the power loss due to PVDF jacket compression as a function of temperature, one optical fiber with a diameter of 62.5 $\mu$m and a length of 40 m was tested inside a cryostat with $\sim$ 3 L of LN2. The fiber was wrapped around a spool with a radius of $\sim$14.5 cm.}
    \label{fig:PVDFPowervstemp}
\end{figure}

As described in Section~\ref{sec_fiber_jacket}, the optical fibers studied in this paper have a Polyvinylidene fluoride (PVDF) black jacket. The PVDF jacket is a UV light resistant plastic that offers chemical and radiation resistance, a high relative melting point, high mechanical strength, non-flammability, very low thermal conductivity, and physiological inertness~\cite{xijun2022analysis}. The exposure of PVDF to low temperatures~\cite{farboodmanesh2019base} can induce compression forces along the fiber (thermal contraction), contributing to microbending along its surface. Microbending is caused when external forces are applied to the jacket surface of an optical fiber~\cite{deo2021experimental}. The application of external forces to the jacket surface results in deformation of the cladding and core, as shown in Figure \ref{fig:PVDFPowervstemp} (left). While core-cladding deformation may not be visible to the naked eye, it alters the angle of light reflection within the fiber. Consequently, the transmitted light rays are redirected at angles that prevent further internal reflection, causing the light rays to be lost within the cladding. 

\begin{figure}[h!]
    \centering
    \includegraphics[width=0.5\textwidth]{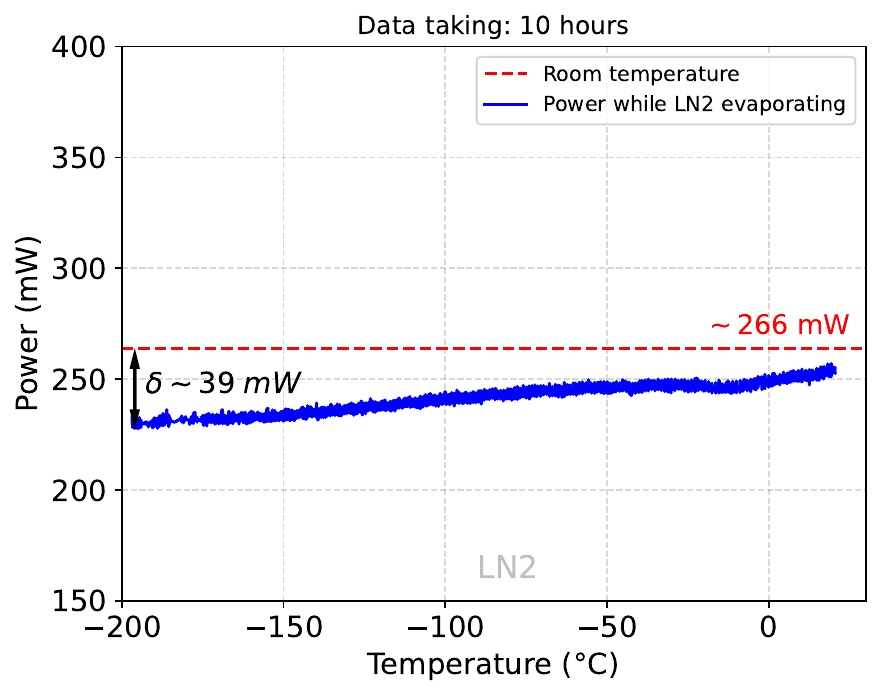} 
    \captionsetup{width=0.9\textwidth}
    \caption{The optical power performance as a function of the temperature. An optical power performance of 266.71 $\pm$ 2.00 mW was recorded at room temperature and is depicted in the plot as a dashed red line. The blue line represents the power performance after submerging the fiber in LN2. At the lowest temperature recorded (-196 $^{\circ}$C), a power loss of $\sim$ 39 mW was measured.}
    \label{fig:PowervsTempStandTest}
\end{figure}

To study the power loss due to PVDF jacket compression as a function of temperature, one optical fiber with a diameter of 62.5 $\mu$m and a length of 40 m was tested inside a cryostat with $\sim$ 3 L of LN2. The fiber was wrapped around a spool with a radius of $\sim$14.5 cm, as illustrated in Figure~\ref{fig:PVDFPowervstemp} (right). Initially, an optical power performance test  was conducted at room temperature. Subsequently, $\sim$ 27 m of the fiber was submerged in LN2 until it completely evaporated, and the optical power was measured as a function of temperature. Throughout the test, temperature variations were monitored using a sensor attached to the fiber. 

Utilizing an input power of 0.3W from a laser box, Figure~\ref{fig:PowervsTempStandTest} illustrates the optical power performance as a function of the temperature. An optical power performance of 266.71 $\pm$ 2.00 mW was recorded at room temperature and is depicted in the plot as a dashed red line. The blue line represents the power performance subsequent to submerging the fiber in LN2. The increase in temperature observed occurred over the 10 hours that it took for the complete evaporation of LN2 in the cryostat. At the lowest temperature recorded (-196 $^{\circ}$C) a power loss of $\sim$ 39 mW was measured. Taking into consideration that $\sim$ 27 m of fiber was submerged in LN2, the optical fiber presented a power loss of $\sim$ 1.44 mW/m.

\subsection{Bending Radius Testing}
\label{bending}
In order to understand the performance of optical fibers used in a PoF system, it is important to characterize the optical power loss as a function of the fiber length for different operating conditions. Two main contributors to optical power loss along the fiber are discussed here for room and cryogenic temperatures.

First, optical power loss arises from cooling the fibers to cryogenic temperatures. This occurs because the fibers are encased in an outermost jacket layer of PVDF plastic, which contracts when exposed to cold temperatures. This contraction around the fragile inner layers can lead to microbending along the fiber, resulting in increased light reflection and absorption during fiber transmission. More details about microbending can be found in Section \ref{Sec:loss_power_cryo}.

The second significant source of power loss occurs due to fiber bending. This loss can be attributed to the operating principle of optical fibers, specifically the concept of total internal reflection. The core and cladding layers of the fiber have different indices of refraction, causing light to be reflected at the fiber wall upon hitting it, a process that repeats along the fiber's length. When the fiber is bent, however, some light strikes the wall at an angle less than the critical angle of the fiber/cladding boundary, allowing it to be transmitted through rather than be reflected. The amount of power loss due to this effect depends on the bending radius and the length of fiber spooled at that radius.

Due to the limited availability of Dewar shapes and volumes in our laboratory, it was not possible for the test presented in this paper to submerge straight lengths of fiber. Instead, the thermal contraction and bending radius power loss are characterized together. For this test, five 3D printed spools of radius 7.5 cm were made to hold the fibers. The five fibers were wrapped around the spools for varying number of loop counts, where these loop counts correspond to different lengths of fiber being bent at the 7.5 cm bending radius. For the 7.5 cm bending radius and each loop count, the power loss measurement is taken at room temperature for the five optical fibers. Later, the five fibers are then submerged in LN2, and again the optical power loss is measured. Finally, the subtraction between the optical power loss at room temperature and in LN2 is calculated, thereby giving the power loss due to thermal contraction and bending of the fibers. 

\begin{figure}[h!]
    \centering
    \includegraphics[width=0.9\textwidth]{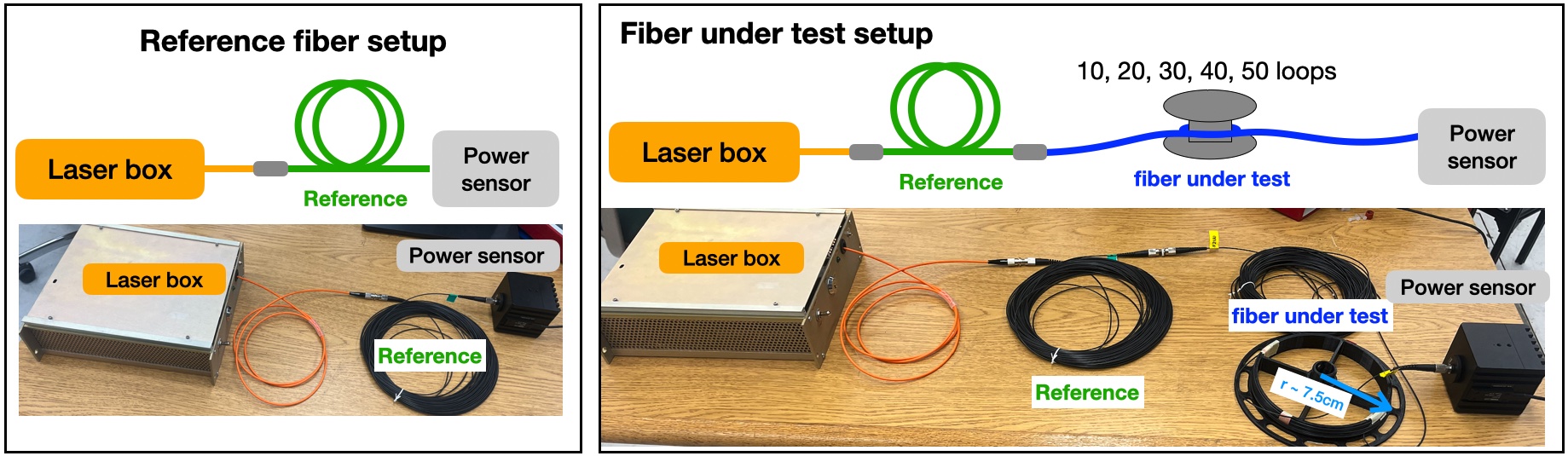}
    \captionsetup{width=0.9\textwidth}
    \caption{Left: The bending radius test setup at room temperature. Right: The laser is connected to a reference fiber, which is also connected to the looped test fiber (10, 20, 30, 40, or 50 loops) at a 7.5 cm radius. \label{fig:bend_room temperature}}
\end{figure}

To obtain the optical power loss measurements, a reference fiber was connected to a laser box (see Figure~\ref{fig:bend_room temperature}, left), then each of the five optical fibers power performance was measured separately using a power sensor (see Figure~\ref{fig:bend_room temperature}, right). Figure~\ref{fig:bend_room temperature} presents the power measurements taken at 10, 20, 30, 40, and 50 loops (corresponding to fiber lengths of 4.71, 9.42, 14.13, 18.84, and 23.55 meters respectively) at room temperature. The power loss at room temperature was calculated by subtracting the power measurements of each fiber from the value obtained with the reference fiber (see Equation \ref{eq:Ploss_room}). The uncertainties for the optical power measurements ($\delta P$) are primarily due to the laser power fluctuations (see Section~\ref{sec_laser}). 

\begin{equation}
    (P_{loss} \pm \delta P)_{room \; temperatures} = P_{\text{tested fiber}}^{\text{room}} - P_{\text{reference fiber}}
    \label{eq:Ploss_room}
\end{equation} \\

\begin{figure}[htbp]
\centering
\includegraphics[width=0.9\textwidth]{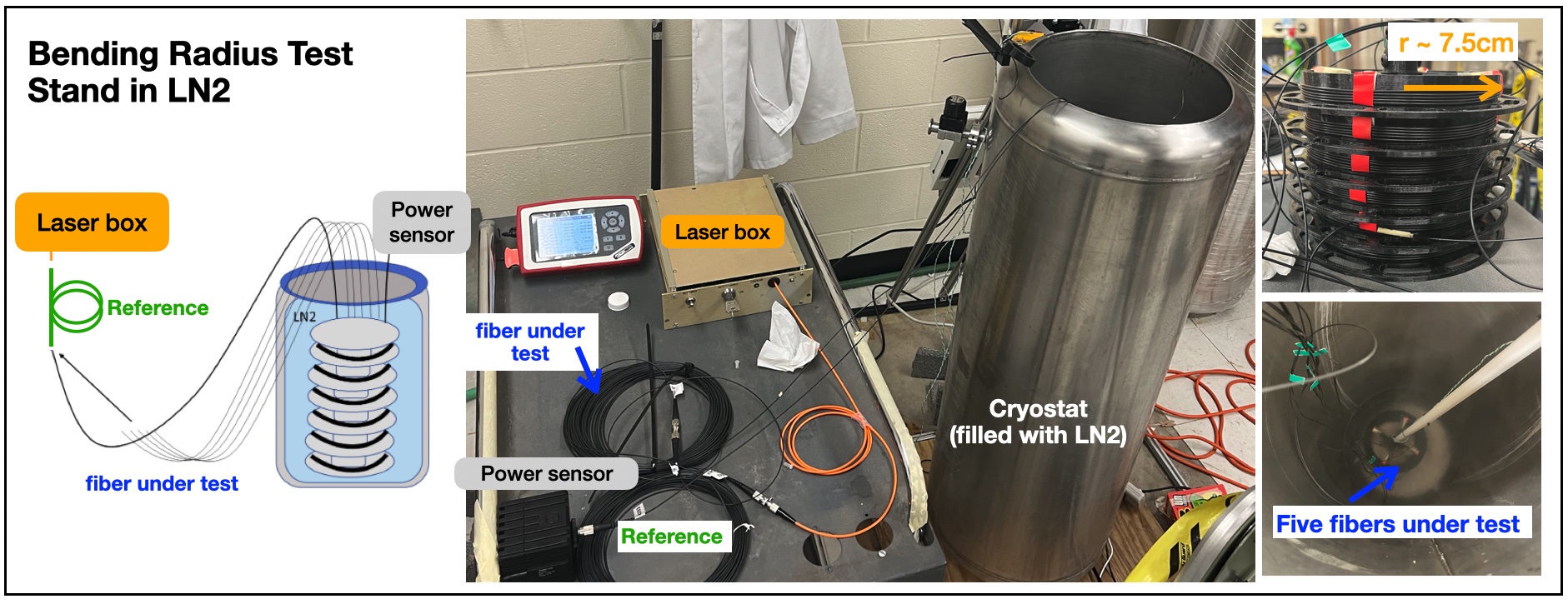}
\captionsetup{width=0.9\textwidth}
\caption{The bending radius test setup in LN2, which follows the same setup as in room temperature, but now the test fiber at the respective loop count is submerged in a dewar filled with LN2.\label{fig:bend_ln2}}
\end{figure}

Subsequently, the fibers were then immersed in LN2, as shown in Figure~\ref{fig:bend_ln2}. The same procedure used for the room temperature measurements was followed for the LN2 measurements, where power measurements were again taken at loop counts of 10, 20, 30, 40, and 50 (corresponding to fiber lengths of 4.71, 9.42, 14.13, 18.84, and 23.55 meters respectively). The power loss in LN2 was calculated by subtracting the power measurements of each test fiber from the value obtained with the respective reference fiber (see Equation \ref{eq:Ploss_LN2}).

\begin{equation}
    (P_{loss} \pm \delta P)_{LN2} = P_{\text{tested fiber}}^{LN2} - P_{\text{reference fiber}}
    \label{eq:Ploss_LN2}
\end{equation}

To find the power loss for the optical fibers due to being exposed to cryogenic temperatures, it is necessary to first establish the baseline power of the fiber at normal operating conditions, which is why the measurements at room temperature were taken as well. In order to calculate the power loss due to thermal contraction, the formula shown in Equation~\ref{eq:bend} was used, where the $P_{loss}$ in LN2 and at room temperatures are subtracted for each power and loop count.

\begin{equation}
    P_{loss} = (P_{loss} \pm \delta P)_{LN2} - (P_{loss} \pm \delta P)_{room \; temperatures}
    \label{eq:bend}
\end{equation}

\begin{table}[htbp]
    \centering
    \smallskip
    \begin{tabular}{c|c|c}
    Input Power  & Power loss per meter  & Power loss per meter \\
    (mW) & (measurement) (mW/m) & (linear fit) (mW/m) \\ \hline
    500 &  1.91 $\pm$ 0.64  & 2.16 $\pm$ 0.40\\
    600 &  2.43 $\pm$ 0.66  & 2.57 $\pm$ 0.39\\
    800 &  3.37 $\pm$ 0.80  & 3.58 $\pm$ 0.57\\
    900 &  3.55 $\pm$ 0.82  & 4.67 $\pm$ 0.59\\
    \end{tabular}
    \captionsetup{width=0.9\textwidth}
    \caption{Power loss per meter measurement for five optical fibers using four input powers provided by the laser box. The five optical fibers were wrapped around a bending radius of $\sim$ 7.5 cm.}
    \label{tab:bending}
\end{table}

\begin{figure}[htbp!]
    \centering
    \includegraphics[width=.48\textwidth]{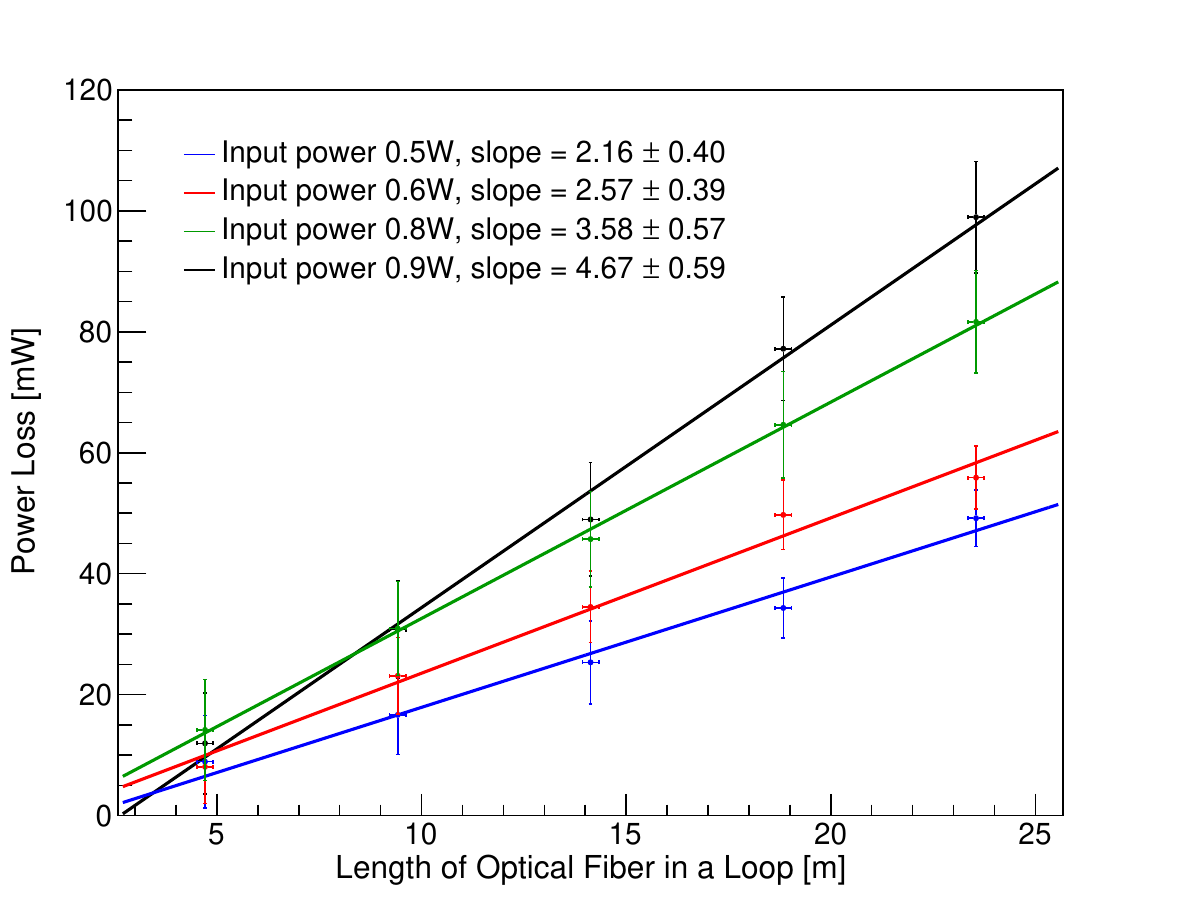}
    \includegraphics[width=.48\textwidth]{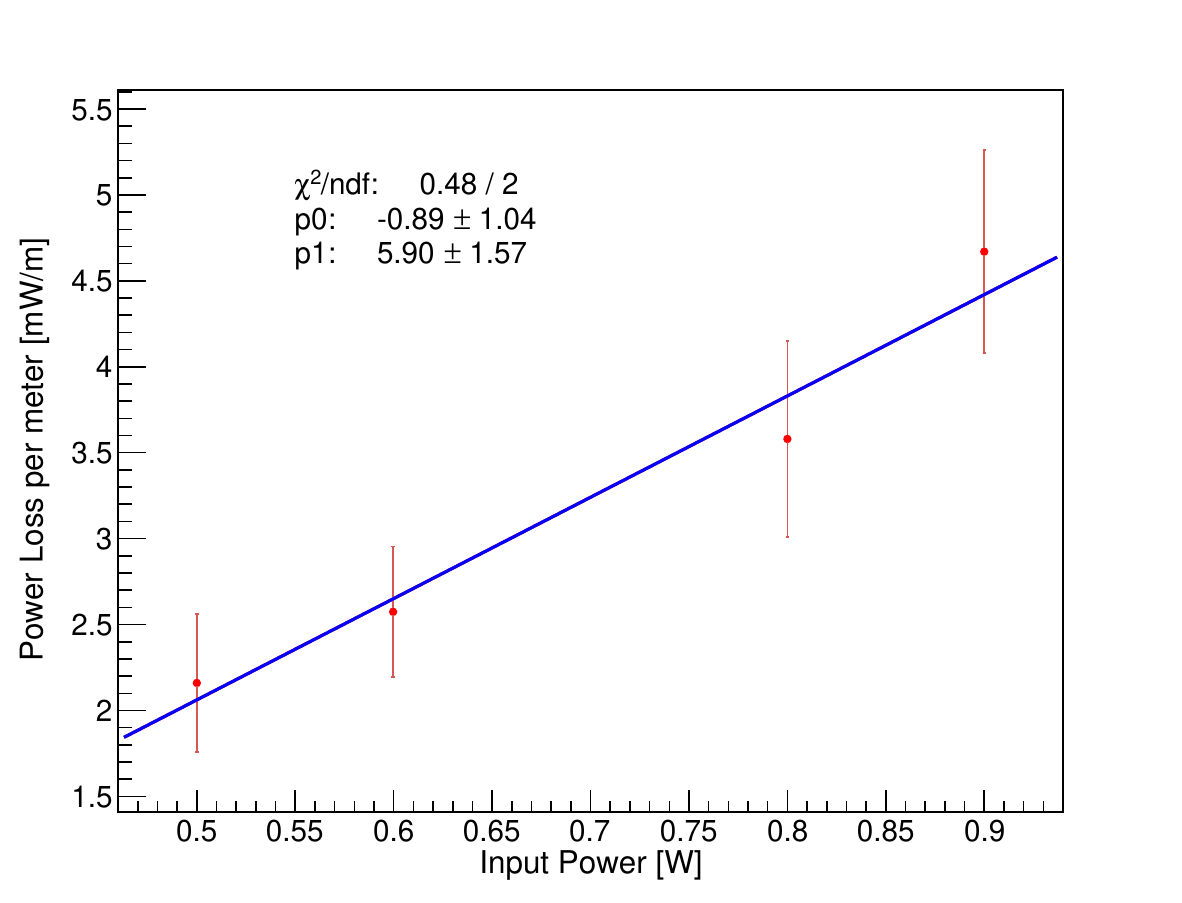}
    \captionsetup{width=0.9\textwidth}
    \caption{Power loss versus length plot for 7.5 cm bending radius for 10, 20, 30, 40 and 50 loops at powers 0.5 W, 0.6 W, 0.8 W, and 0.9 W (left), as well as the results shown as input power vs linear fit slope (right).\label{fig:bend_total}}
\end{figure}

Figure \ref{fig:bend_total} (left) shows the power loss for input powers of 0.5 W, 0.6 W, 0.8 W, and 0.9 W, with the X-axis representing the increasing length of fiber looped at the 7.5 cm bending radius. The error bars in Figure~\ref{fig:bend_total} include the sum in quadrature of the statistical uncertainty (limited by testing of five optical fibers) and the systematic uncertainty associated with the laser power fluctuations. The error on the optical fiber length submerged was $\sim$ 20 cm.

The positive slopes present in the plot indicate that as the loop count increases and more fiber is submerged in LN2, the power loss also increases at a linear rate. In addition, a linear dependence of the optical power loss per meter as a function of the power provided by the laser was measured to be 5.90 $\pm$ 1.57 $m^{-1}$ (see Figure~\ref{fig:bend_total}, right). Table~\ref{tab:bending}, summarizes the measured power loss per meter and the corresponding linear fit parameters obtained for 0.5W, 0.6 W, 0.8 W, and 0.9 W laser powers used in the bending radius test as shown in Figure~\ref{fig:bend_total}.

\subsection{Power Loss Measurement in an Optical Fiber Without Jacket}
\label{sec_fiber_jacket}

\begin{figure}[h!]
    \centering
    \includegraphics[width=0.9\textwidth]{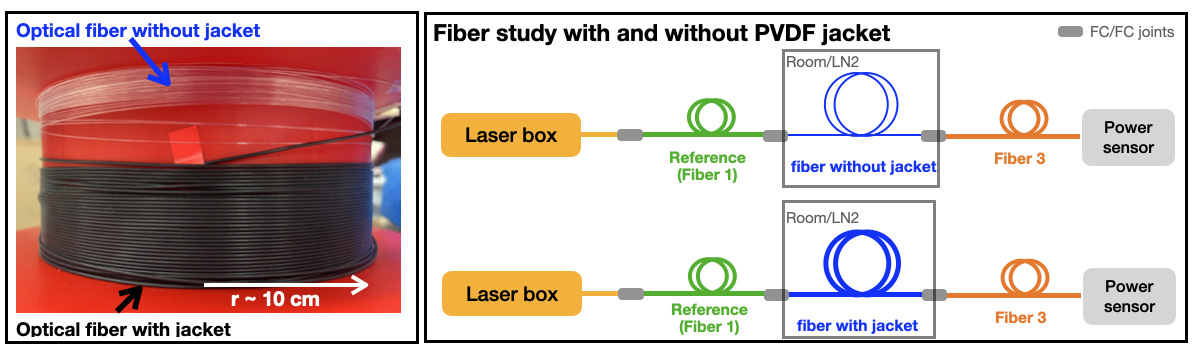}
    \captionsetup{width=0.9\textwidth}
    \caption{Left: Diagram showing the experimental setup for the fiber with and without a jacket wrapped on a spool with a radius of $\sim$ 10 cm. Right: The setup shows fiber 1, fiber with/without jacket (fiber 2) and fiber 3 connected to the laser and power sensor. \label{fig:bare_diagram}}
\end{figure}

As demonstrated in Section~\ref{Sec:loss_power_cryo}, the fiber with a PVDF jacket experiences power loss at cryogenic temperatures due to thermal contraction. A test was carried out to measure the power loss of an optical fiber without a jacket. This test measured the power loss exhibited in LN2 by two optical fibers: one with a PVDF jacket (62.5 $\mu$m core diameter and 40 m length) and the other optical fiber without the jacket (62.5 $\mu$m core diameter and 25 m length). Both optical fibers were wrapped on a spool with radius $\sim$ 10 cm, as shown in Figure~\ref{fig:bare_diagram} (left).

The setup consisted of a reference fiber (Fiber 1) initially connected to a laser box. Next, the fiber being tested with or without jacket (Fiber 2) was connected, and finally, a third fiber (Fiber 3) was introduced in the setup. All interconnections between fibers were established using FC-FC junctions as illustrated in Figure~\ref{fig:bare_diagram} (right). A power performance in room and LN2 temperatures were recorded using a power sensor. For the LN2 measurements, 25 m of the jacketed fiber was introduced into the LN2 to allow for a direct comparison with the 25 m of the fiber without a jacket.

To measure the power loss of the fibers with and without jacket under cryogenic temperatures, Equation~\ref{eq:bend} was used, where the power loss is calculated by subtracting the power loss at LN2 and room temperatures, following a procedure similar to that outlined in Section~\ref{bending}. Table~\ref{tab:bare}, summarizes the power loss per meter calculated for both fibers with and without jackets. The results show that the power loss per meter for fibers with a PVDF jacket is approximately three times higher than that for fibers without a jacket, for both 0.5 W and 0.8 W input powers.

\begin{table}[htbp]
    \centering
    \smallskip
    \begin{tabular}{c|c|c}
    Fiber type & Input Power (mW) & Power loss per meter (mW/m) \\
    \hline
    With Jacket    & \multirow{2}{*}{500} & 1.74 $\pm$ 0.28 \\
    Without Jacket &                      & 0.56 $\pm$ 0.21 \\ \hline
    With Jacket    & \multirow{2}{*}{800} & 3.14 $\pm$ 0.51 \\
    Without Jacket &                      & 1.06 $\pm$ 0.48 \\
    \end{tabular}
    \captionsetup{width=0.9\textwidth}
    \caption{Measurement of power loss per meter for optical fibers with and without jacket in LN2 using 0.5W and 0.8W input powers provided by the laser box.}
    \label{tab:bare}
\end{table}

\subsection{Power Loss due to Polishing} 

\begin{figure}[h!]
    \centering
    \includegraphics[width=0.45\textwidth]{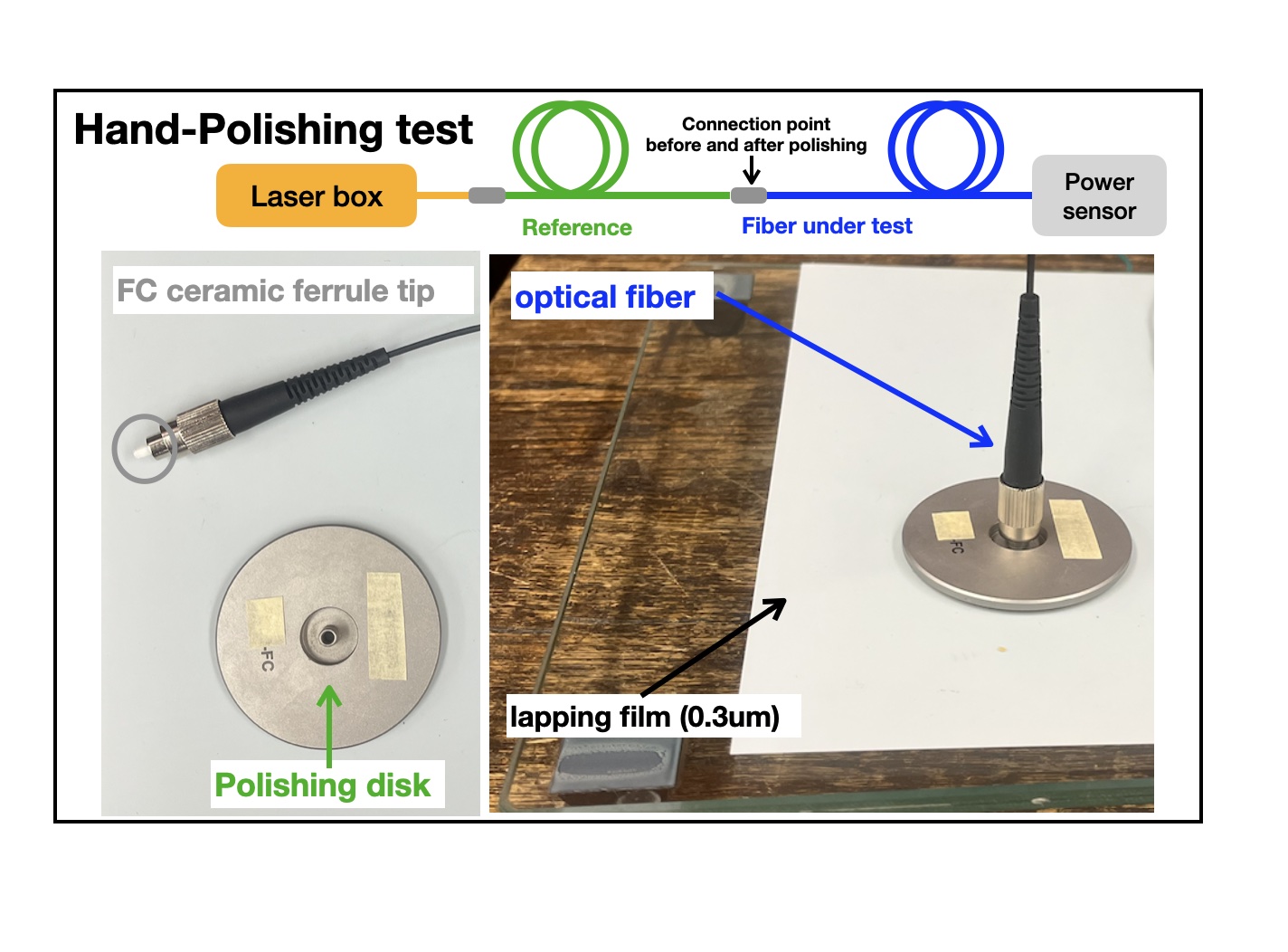}
    \includegraphics[width=0.4\textwidth]{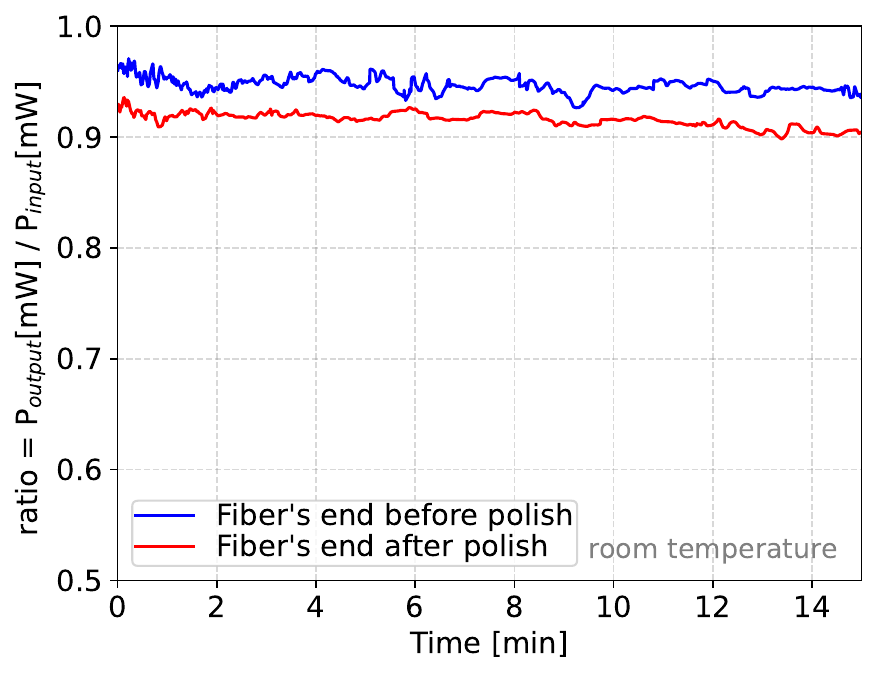}
    \captionsetup{width=0.9\textwidth}
    \caption{Left: Optical fiber end polishing was performed by hand using a polishing disk and lapping film with 0.3 $\mu$m. Right: The ratio between power performance measurements before and after hand-polishing. \label{fig:polishing}}
\end{figure}

The removal of imperfections such as epoxy and scratches from the optical connector's ferrule tip using lapping (polishing) film can have an impact in the quality of fiber end surface. Poor polishing or overpolishing of the optical fiber tip could result in a higher optical power loss. A preliminary test to study the effect of hand-polishing on the power performance at room temperature for a new fiber with a 62.5 $\mu$m core, 40 meters in length, and with ceramic ferrule tips is performed. First, one end of the fiber being tested (new fiber) is connected to a reference fiber using a FC-FC joint, while the other end is connected to a power sensor (see Figure \ref{fig:polishing}, left). 

Afterward, the fiber was disconnected from the reference fiber, hand-polished with a 0.3 $\mu$m lapping film for less than one minute, and then connected again. Finally, a power transmission test was performed before and after hand-polishing. Figure~\ref{fig:polishing} (right) shows the ratio between power performance measurements before and after hand-polishing. Based on the power ratio, a noticeable decrease in performance is observed after hand-polishing the optical fibers. It is important to mention that hand-polishing can be affected by the uniform pressure, speed, and time applied by the person performing the polishing.

\subsection{Optical Fiber Tensile Strength Test}
\label{sec_tensil_strength}

\begin{figure}[h!]
    \centering
    \includegraphics[width=0.4\textwidth]{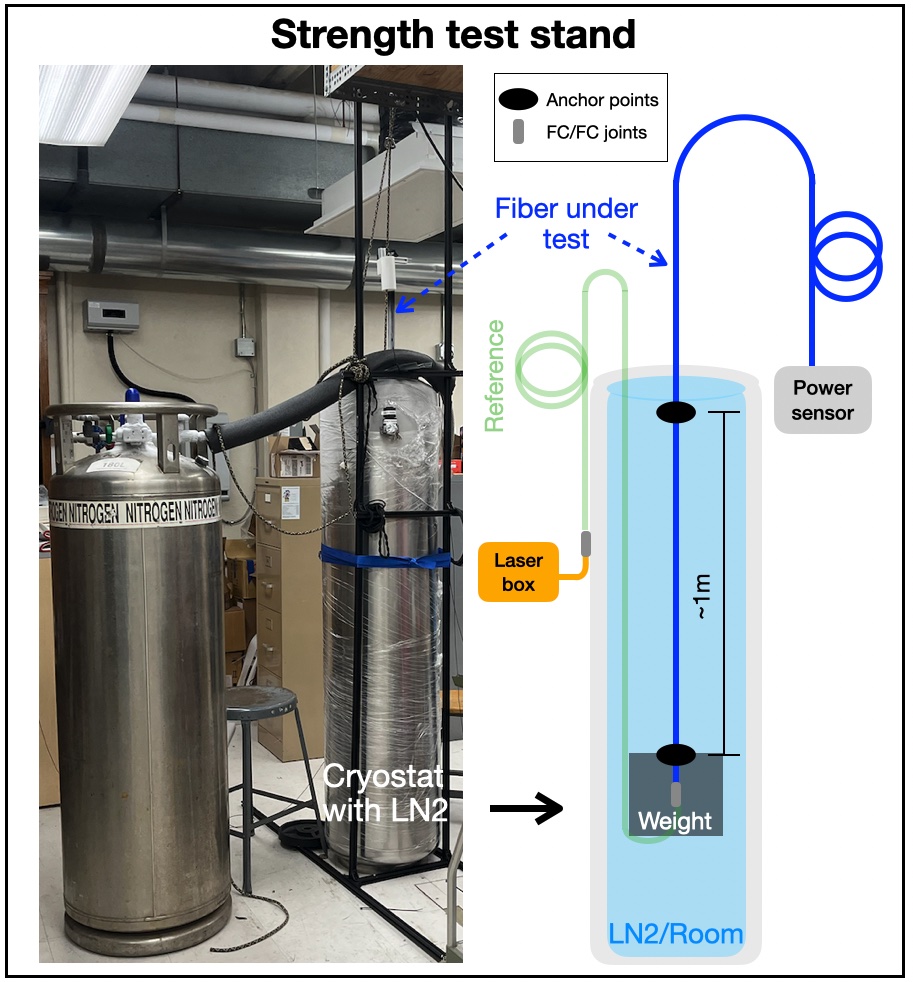}
    \captionsetup{width=0.9\textwidth}
    \caption{Tensile strength test setup for room and LN2 conditions. The fiber under test is secured with two anchor points to ensure $\sim$ 1 m length during the strength test. Power sensor measure the optical power for varying tensile strengths.}
    \label{fig:Strength}
\end{figure}

Tensile strength refers to the maximum stress that a material can bear before breaking when it is allowed to be stretched or pulled. Figure \ref{fig:Strength} show a test setup built to assess the tensile strength of an optical fiber under varying environmental conditions, including room and LN2 temperatures. The fiber was suspended at a length of one meter using two anchor points, as shown in Figure \ref{fig:Strength}. The two anchor points were secured in a customized support designed and built for this test. The choice of one meter of optical fiber was made to minimize the impact of power losses associated with exposure to LN2 temperatures, as described in Section \ref{sec_fiber_jacket}.

\begin{figure}[h!]
    \centering
    \includegraphics[width=0.56\textwidth]{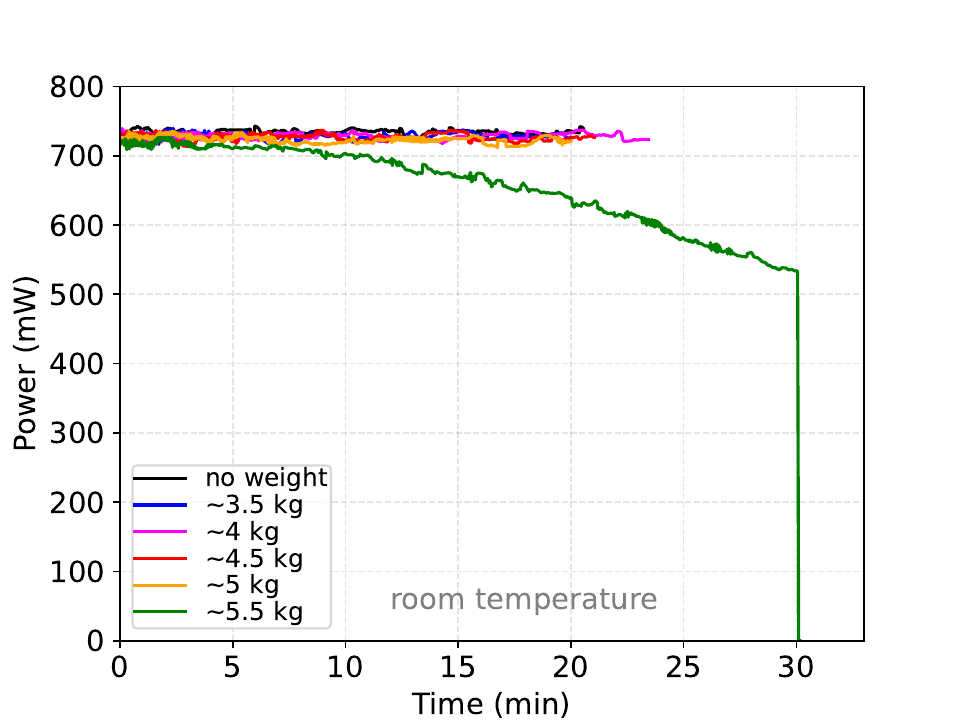}
    \includegraphics[width=0.32\textwidth]{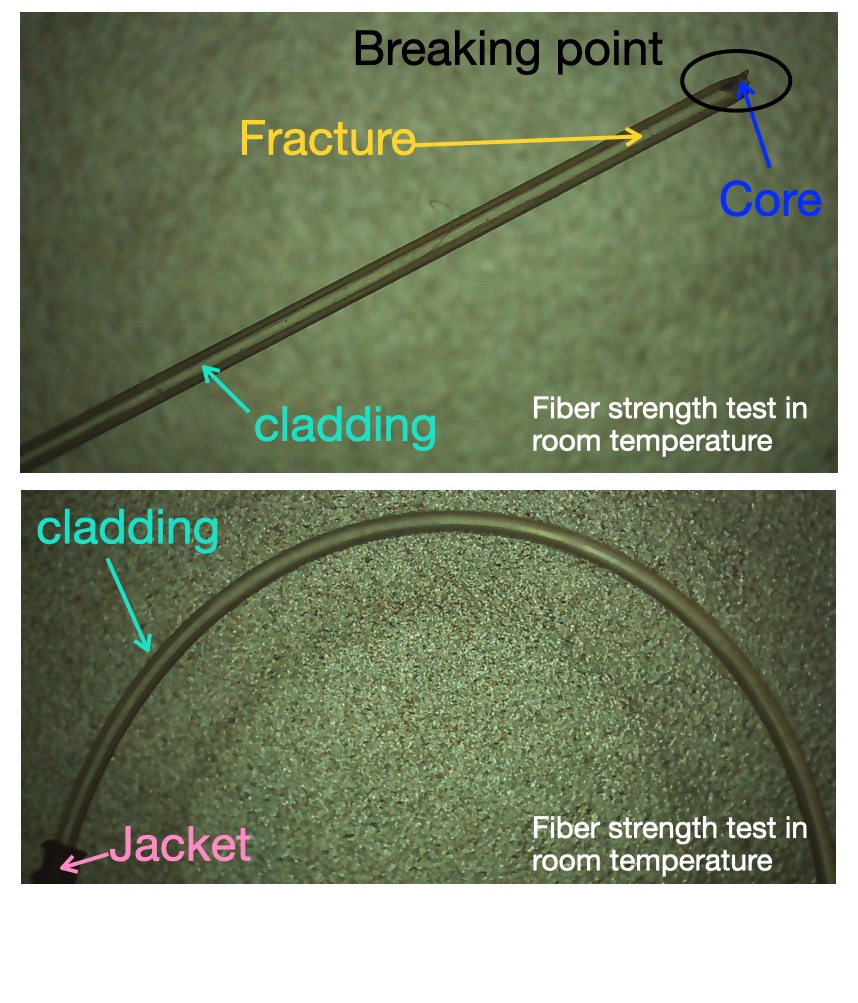}
    \captionsetup{width=0.9\textwidth}
    \caption{Left: Tensile strength test results conducted at room temperature with an input power of 0.8W using fibers with a 62.5 $\mu$m core. The colored lines correspond to the power performance of a fiber during strength testing with 3.5 kg, 4 kg, 4.5 kg, 5 kg and 5.5 kg compared with applying no weights. Right: Fiber inspection after strength tests at room temperature.}
    \label{fig:Strength_room_tempr}
\end{figure}

One anchor point of the fiber is secured at the top of the structure, and the second anchor point is attached to a weight support. During the test multiple weights were added while monitoring the optical power performance until the optical fiber reaches its breaking point. Figure \ref{fig:Strength_room_tempr} (left) shows the results of increasing weight on the optical power performance at room temperature. The optical fiber breaking point was found to be approximately 5.5 kg for room temperature. The fiber was inspected after the strength test with a digital microscope, as shown in Figure \ref{fig:Strength_room_tempr} (right), revealing fractures in both the cladding and core of the optical fiber.

\begin{figure}[ht!]
    \centering
    \includegraphics[width=.47\textwidth]{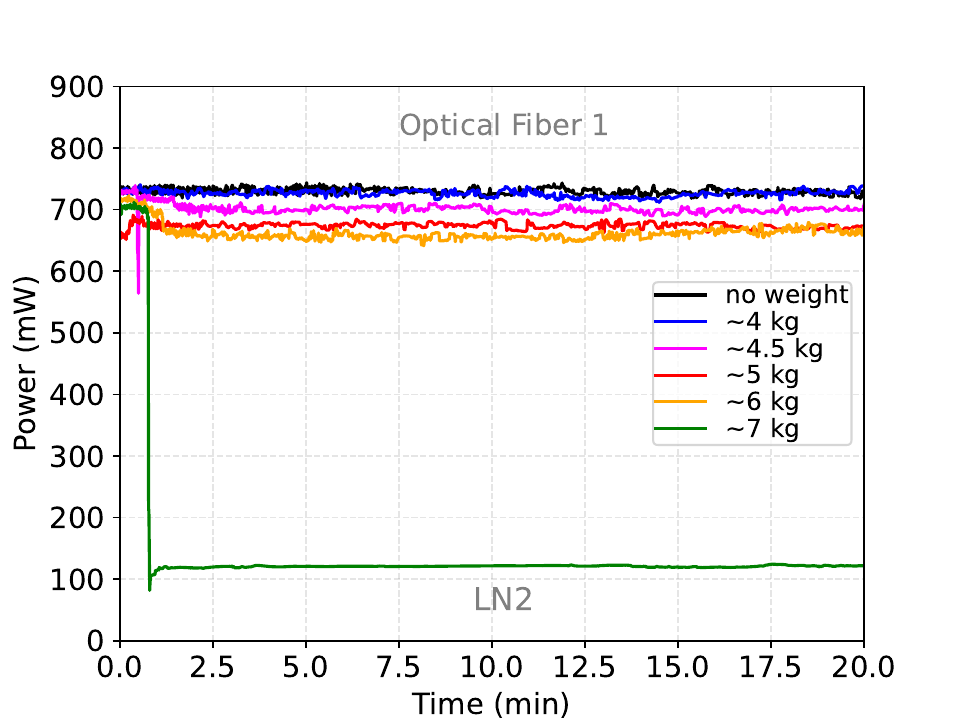}
     \includegraphics[width=.47\textwidth]{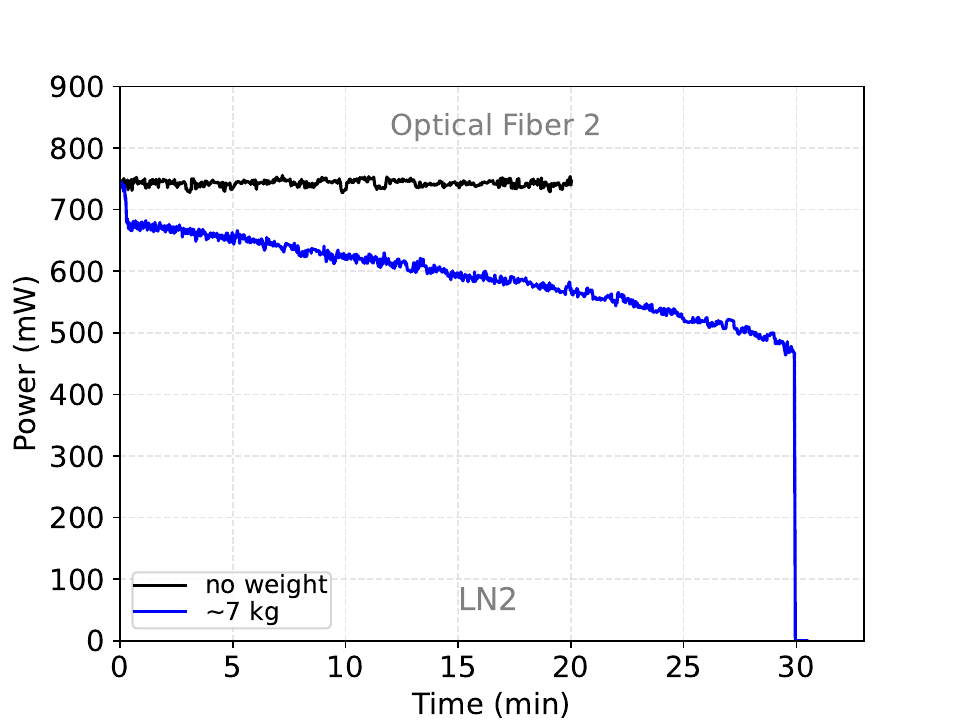}\\
    \includegraphics[width=.48\textwidth]{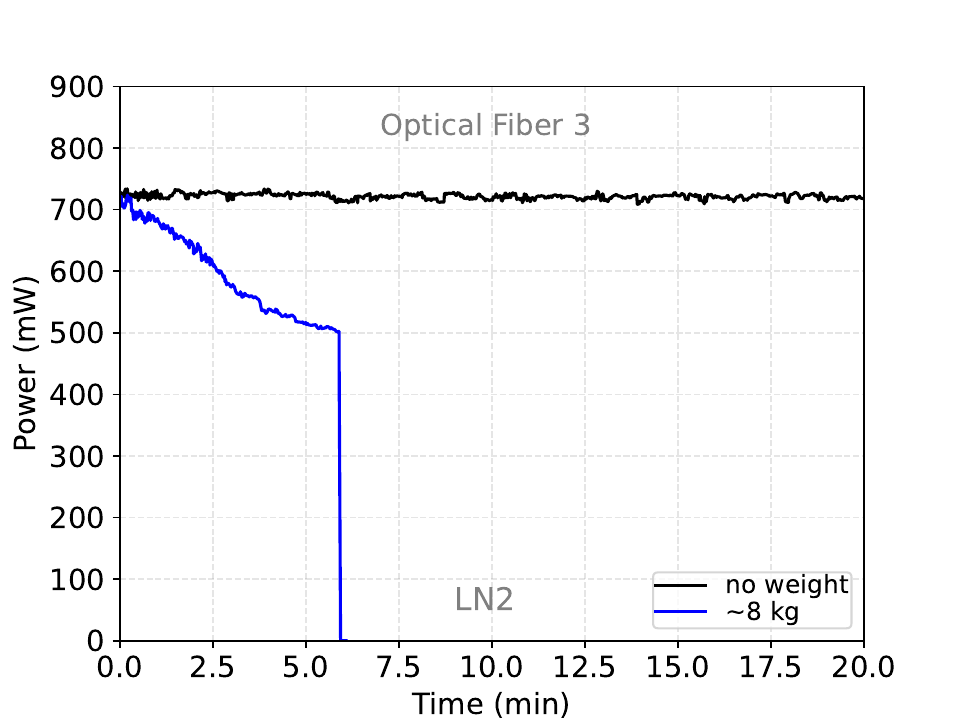}
    \includegraphics[width=.43\textwidth]{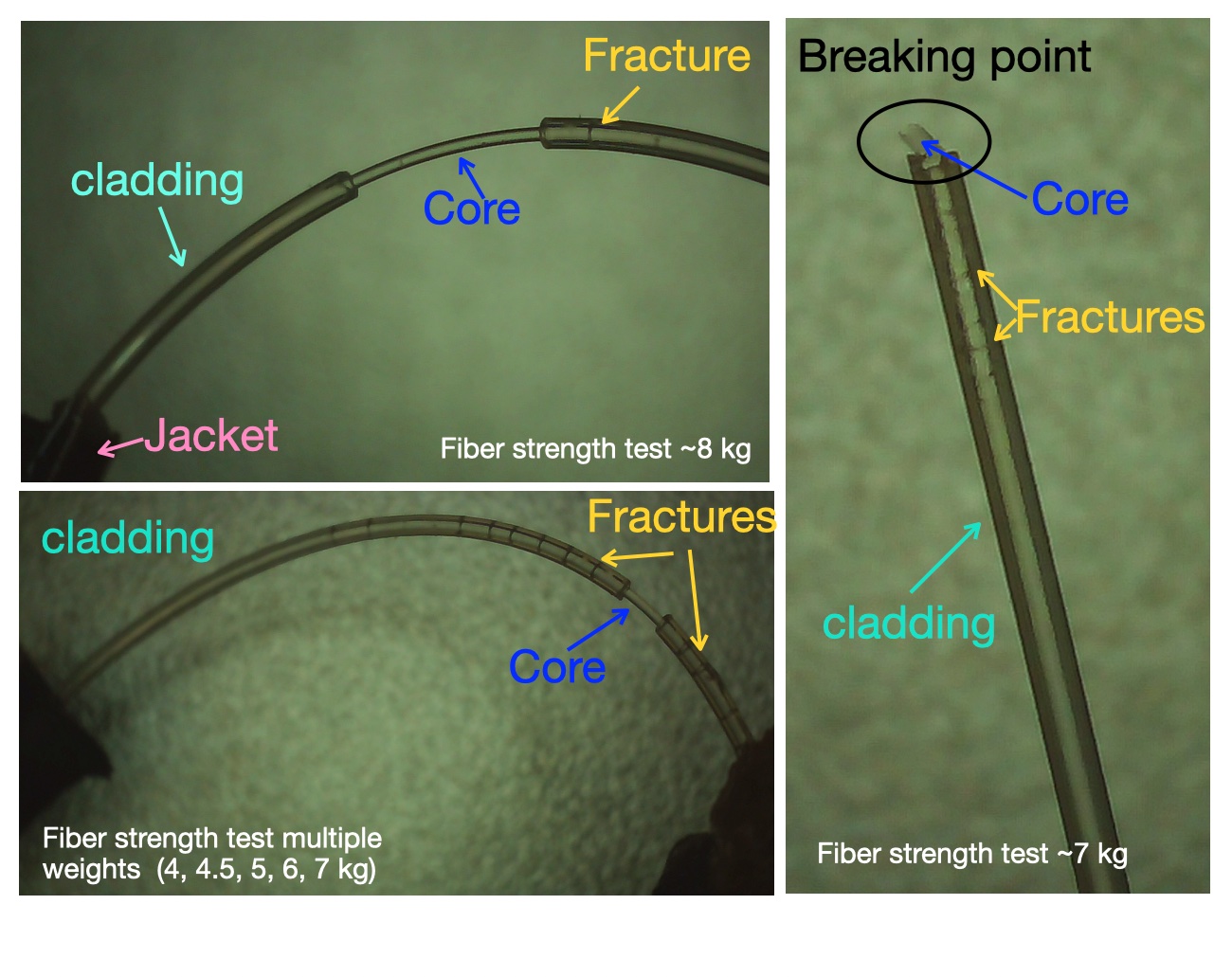}
    \captionsetup{width=0.9\textwidth}
    \caption{Tensile strength test conducted in LN2 with an input power of 0.8W using fibers with a 62.5 $\mu$m core. Upper left: Power performance a fiber during strength testing with six different weights. Upper right: Power performance of a fiber during strength testing with a $\sim$ 7 kg. Lower left: Power performance of a fiber during strength testing with a $\sim$ 8 kg. Lower right: Fiber inspection after strength tests in LN2.}
    \label{fig:Strength_LN2}
\end{figure}

A similar procedure to that conducted at room temperature was carried out for LN2. The optical fiber was submerged to a depth of $\sim$ 1 m, as shown in Figure \ref{fig:Strength}. Figure \ref{fig:Strength_LN2} (upper left) shows the optical power performance of a fiber when subject to tensile strengths using weights of 4 kg, 4.5 kg, 5 kg, 6 kg, and 7 kg and an input power of 0.8 W provided by the laser box. At 7 kg, it was discovered that the jacket and cladding broke, and the core had a fracture that reduced the power transmission (green line). A second test is conducted using a different fiber and placing $\sim$ 7 kg to verify the fiber fracture breaking point found in the first test. The optical power performance is shown in Figure \ref{fig:Strength_LN2} (upper right) where the blue line represents the power decreases over time until the fiber reached its breaking point at 30 min. A third test was conducted using $\sim$ 8 kg and the fiber broke after $\sim$ 6 min as shown in Figure \ref{fig:Strength_LN2} (lower left). 

The fibers used in each test were inspected with a digital microscope, as shown in Figure \ref{fig:Strength_LN2} (lower right), revealing fractures in both the cladding and core of the optical fiber. The volume of the LN2 displaced by the $\sim$ 7 kg was $\sim$ 1.5 L, corresponding to approximately 1.2 kg of LN2. Considering the buoyancy force, the apparent weight is equivalent to $\sim$ 5.8 kg.

\section{PoF Laser}
\label{sec_laser}
\begin{figure}[htbp]
    \centering
    \includegraphics[width=.6\textwidth]{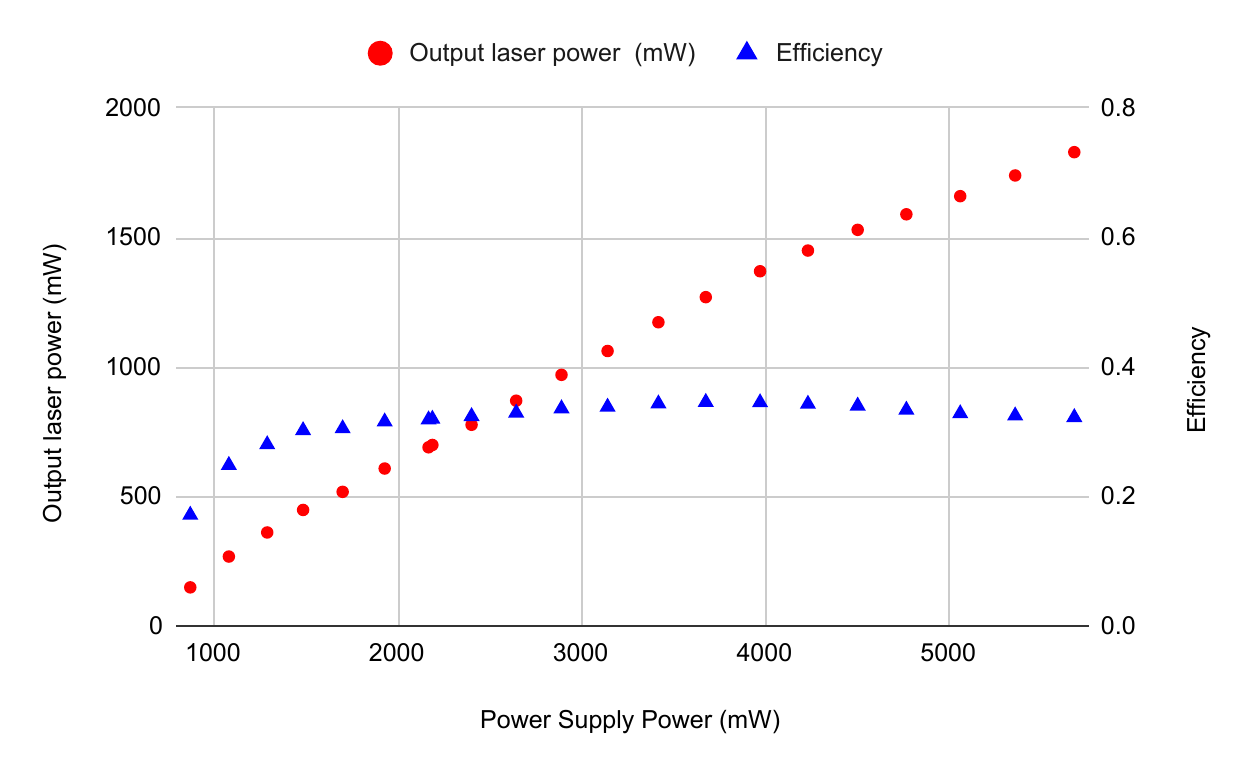}
    \caption{Lasing efficiency vs Supplied Power. }
    \label{fig:laser_diagram_eff}
\end{figure}

The Broadcom laser~\cite{LaserBroadcom} used for the measurements reported in this paper operates at 808 nm and can be directly powered from the power supply. The maximum lasing efficiency is plateaued at 35\% as shown in Figure~\ref{fig:laser_diagram_eff}. A commercial regulating circuit is used in prototypes for fast turnaround. A custom-designed control circuit is planned to be used for DUNE FD2 to simplify the circuitry and meet DUNE safety requirements.   

The PoF laser box contains a gallium arsenide (GaAs) 808-nm laser transmitter module that transmits optical power through an optical fiber to the OPCs. This high-power laser module has a 62.5$\mu$m pigtail optical fiber and FC/PC ceramic ferrules. The laser-transmitter was designed for PoF applications by Broadcom~\cite{LaserBroadcom} and provides up to 2 W of power at room temperature. The laser box is assembled using a commercial Broadcom laser controller. Laser transmitters are widely used in multiple applications such as telecommunications, space, imaging, and more~\cite{LaserBroadcom}. Fluctuations in the output power provided by the lasers needed to be studied in order to understand if these variations could affect the stability of the OPC performance. 

\begin{figure}[h!]
    \centering
    \qquad
    \includegraphics[width=0.47\textwidth]{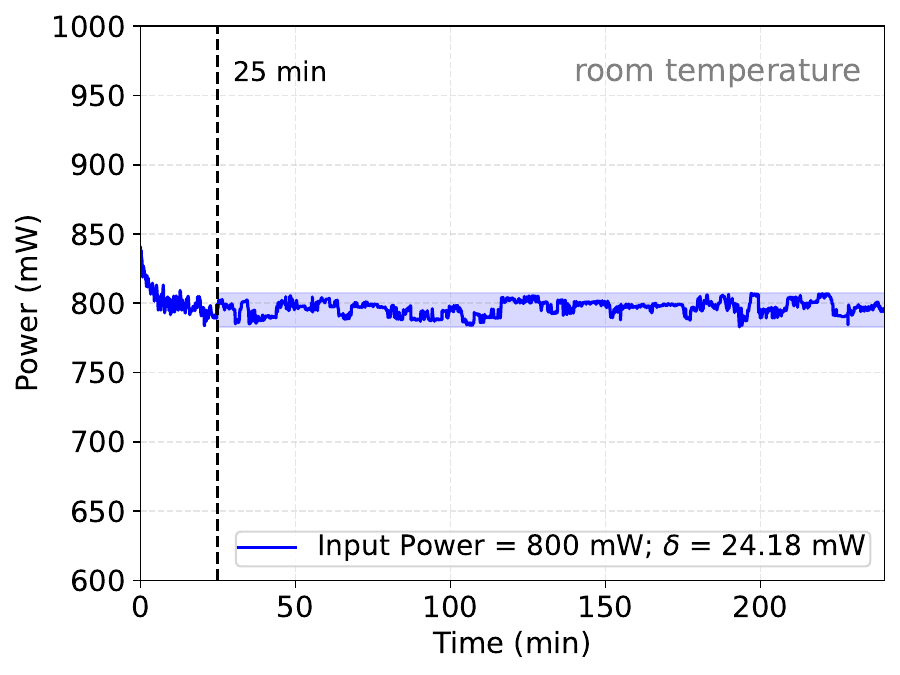}
    \includegraphics[width=0.47\textwidth]{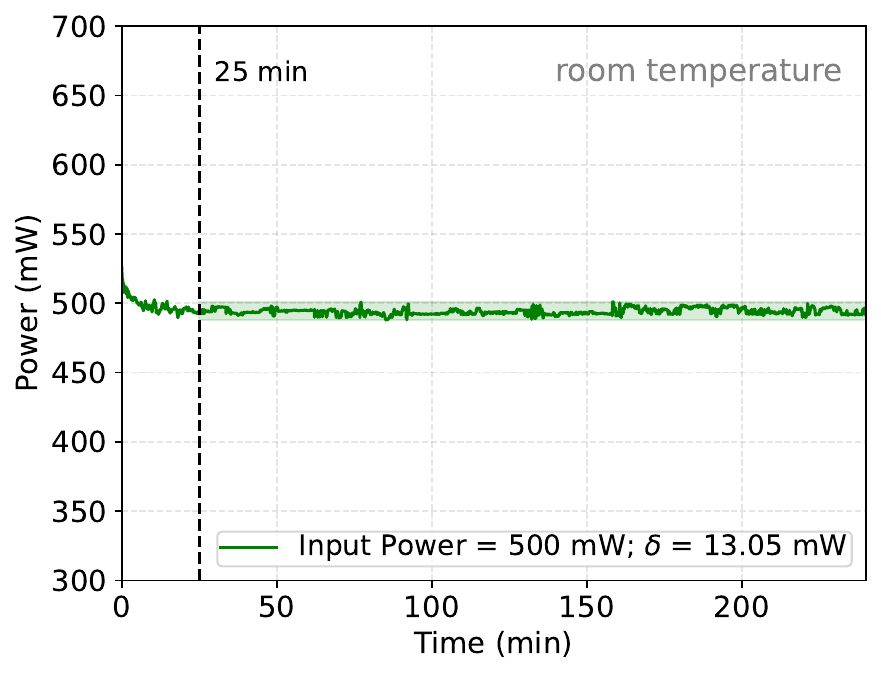}
    \captionsetup{width=0.9\textwidth}
    \caption{Laser power performance is depicted for a laser transmitter with a 62.5 $\mu$m core and an input power of 0.8 W (left) and 0.5 W (right) over a duration of 240 minutes, with power fluctuations of about 24 mW (left) and 13 mW (right) observed at room temperature.\label{fig:StabilityPlots}}
\end{figure}

\begin{figure}[h!]
    \centering
    \includegraphics[width=.4\textwidth]{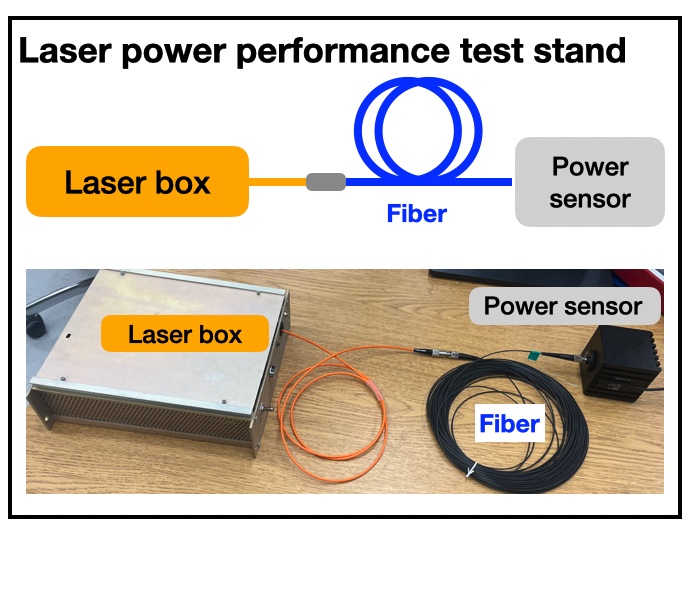}
    \includegraphics[width=.5\textwidth]{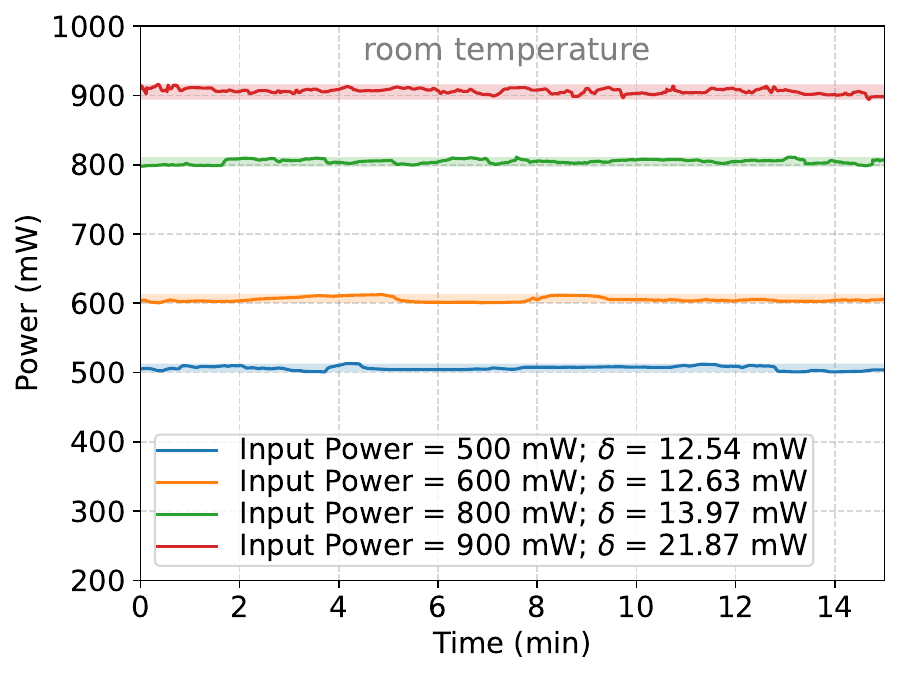}
    \captionsetup{width=0.9\textwidth}
    \caption{Left: The laser power performance test stand includes a laser box with a 62.5 $\mu$m pigtail connected to a reference fiber, which is then connected to the power sensor. Right: The power fluctuations due to regular laser operation as a function of different laser-transmitter powers with colored lines representing different laser powers, where $\delta$ defines the difference between upper and lower limit of power.}
    \label{fig:LaserMultiplePowers}
\end{figure}

In order to study the laser-transmitter fluctuations, a test was conducted to monitor the power performance of the laser over time using a sphere photo diode sensor. Figure~\ref{fig:LaserMultiplePowers} (left) show the setup used to measure the laser power performance at room temperature. The setup includes a laser box with a 62.5 $\mu$m pigtail connected to a reference fiber, which is then connected to the power sensor. 
Figure \ref{fig:StabilityPlots} (left) and (right) displays the power performance of a single GaAs laser transmitter during 240 minutes of operation, with input powers of 0.8W and 0.5W, respectively. In both cases, the laser transmitter stabilized after 25 minutes of operation.

\begin{figure}[!h]
    \centering
    \includegraphics[width=.47\textwidth]{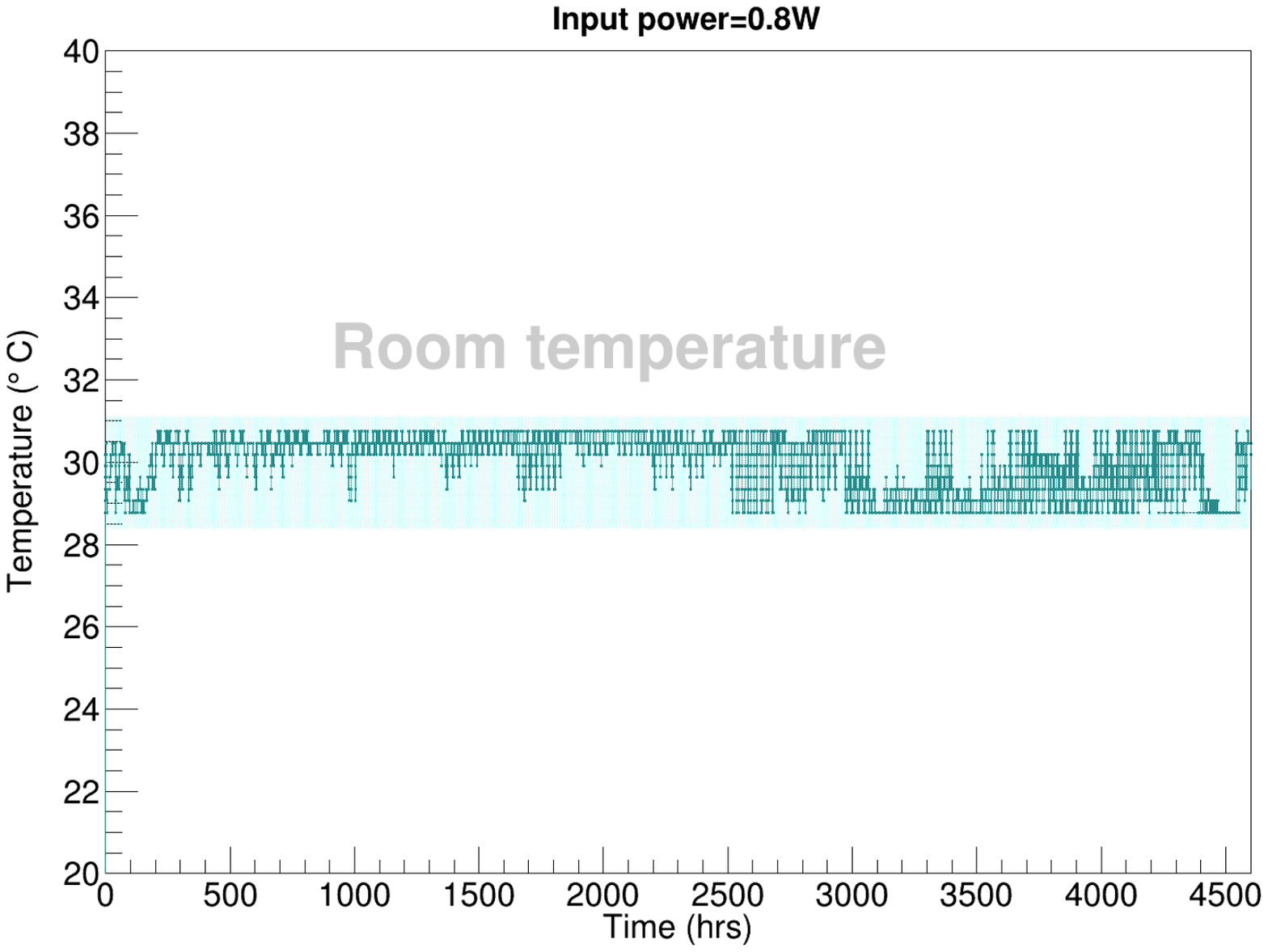}
    \includegraphics[width=.47\textwidth]{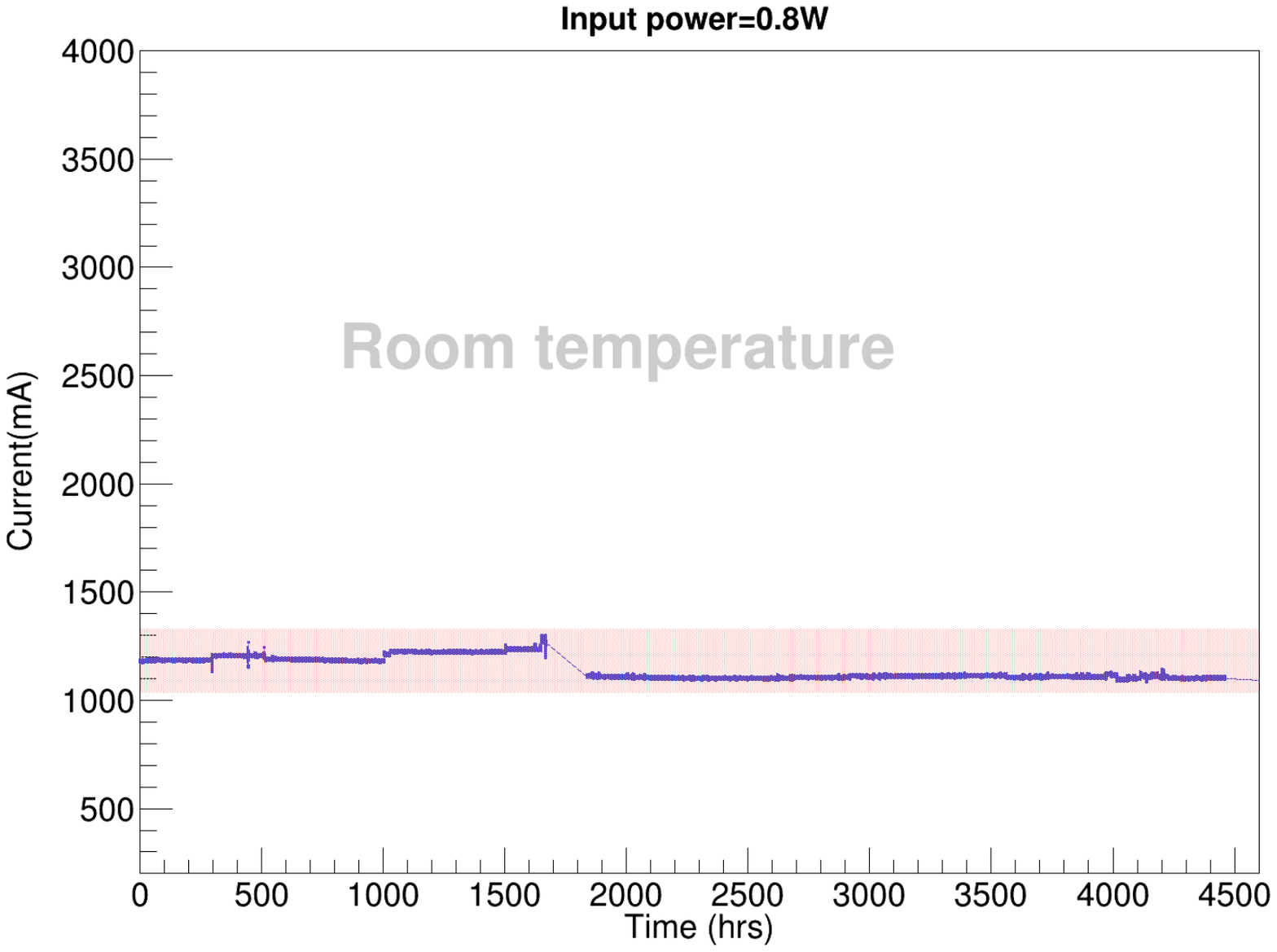}
    \captionsetup{width=0.9\textwidth}
    \caption{Long term measurements of laser temperature (left) and laser current (right) over a period of 6.1 months. The laser temperature (green color band) and laser current (pink color band) variations remained below 2.5°C and 225 mA, respectively.} 
    \label{fig:long_term_laser}
\end{figure}

Figure~\ref{fig:LaserMultiplePowers} (right) shows the power fluctuations due to standard laser operation as a function of different laser-transmitter powers, with colored lines representing different laser powers. Other factors can contribute to fluctuations in the laser output power, such as variations in laser temperature and instability of the laser current. Section~\ref{subsec_longterm} describes the setup built to monitor the operation of a single GaAs laser-transmitter. Figure~\ref{fig:long_term_laser} presents continuous measurements of laser temperature (left) and laser current (right) over a period of 6.1 months. 

\section{Electrical Power Delivery}
\label{sec_pd}

PoF technology enables electrical power supply to devices in harsh environments such as high voltage and cryogenic conditions. 
Sections \ref{sec_opc} and \ref{sec_fiber} present measurements of the optical-to-electrical power conversion efficiency of the OPCs and  power loss in optical fibers. These measurements could make it possible to approximately estimate the maximum electrical power delivered by an OPC in a PoF system when using a specific length of optical fibers (see Table \ref{tab:FiberType}) and a certain number of OPCs, depending on the application's requirement for supplying power to electronic devices operating at room and LN2 temperatures.

\begin{figure}[h!]
    \centering
    \includegraphics[width=.47\textwidth]{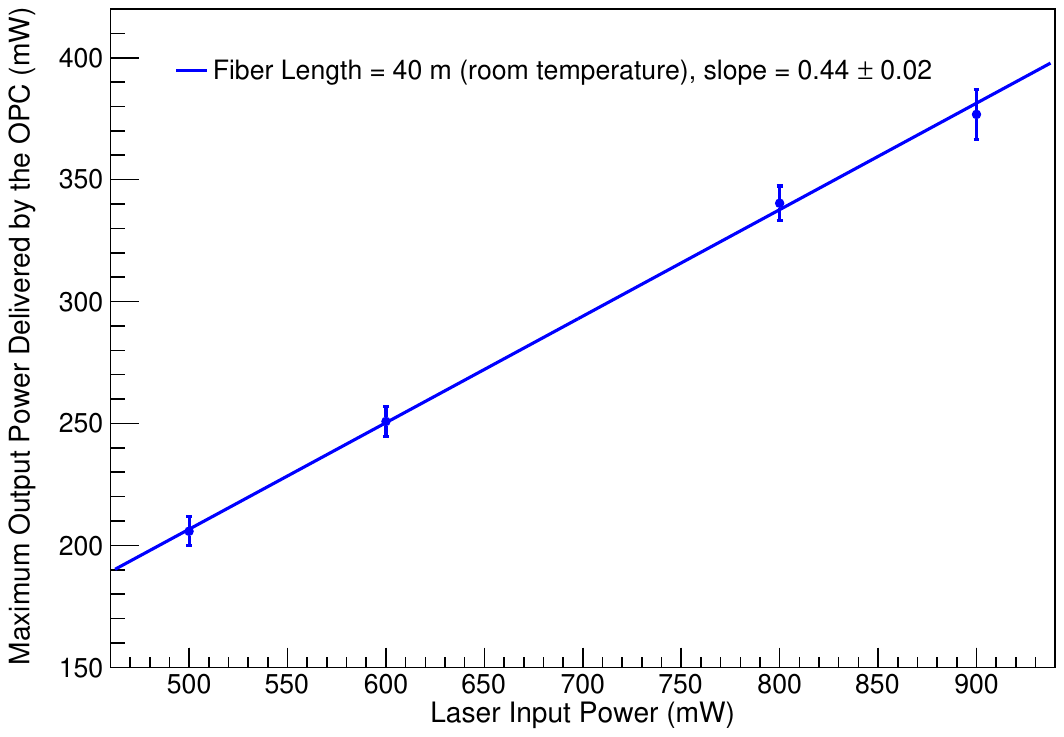}
    \includegraphics[width=.47\textwidth]{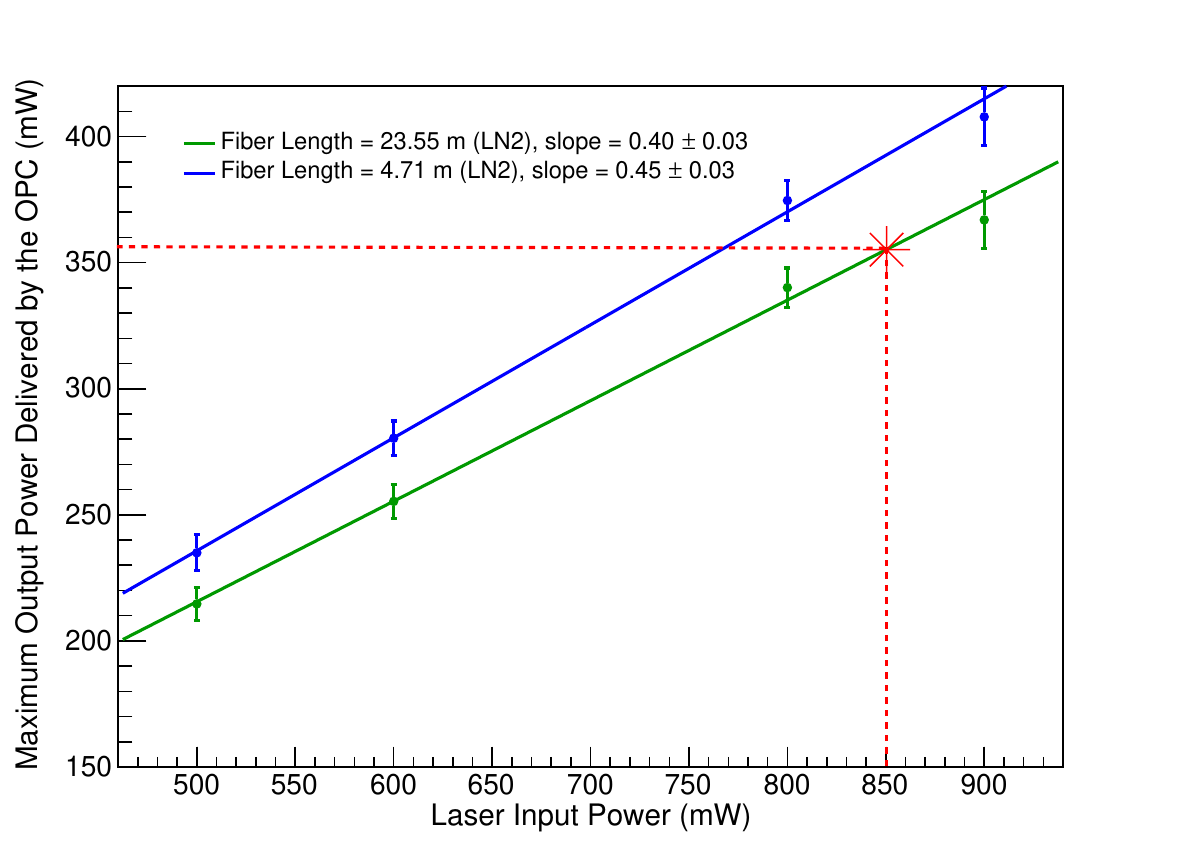}
    \captionsetup{width=0.9\textwidth}
    \caption{The measurement of the maximum output power delivered by an OPC (mW) for varying laser inputs at both room temperature and LN2 conditions, shown in the left and right figures, respectively. Additionally, a linear fit model is applied to the data.}
    \label{fig:OutvsInPConversion}
\end{figure}

\begin{table}[h!]
    \centering
    \smallskip
    \begin{tabular}{c|c|c|c}
    \multirow{3}{*}{Input Power (mW)}  & Environment &  \multirow{3}{*}{Length (m)} & Maximum Output  \\
      & Condition   &                              & Power Delivered \\ 
      &             &                              & by the OPC (mW) \\ \hline
    \multirow{3}{*}{500 $\pm$ 12.54} & room temperature & 40.00 $\pm$ 0.2 &  205.78 $\pm$ 5.98 \\ 
                                     & LN2              & 4.71  $\pm$ 0.2 &  234.94 $\pm$ 7.11 \\
                                     & LN2              & 23.55 $\pm$ 0.2 &  214.67 $\pm$ 6.65 \\ \hline
    \multirow{3}{*}{600 $\pm$ 12.63} & room temperature & 40.00 $\pm$ 0.2 &  250.82 $\pm$ 6.07 \\
                                     & LN2              & 4.71  $\pm$ 0.2 &  280.37 $\pm$ 6.86 \\
                                     & LN2              & 23.55 $\pm$ 0.2 &  255.33 $\pm$ 6.78 \\ \hline
    \multirow{3}{*}{800 $\pm$ 13.97} & room temperature & 40.00 $\pm$ 0.2 &  340.32 $\pm$ 7.05 \\
                                     & LN2              & 4.71  $\pm$ 0.2 &  374.64 $\pm$ 7.79 \\
                                     & LN2              & 23.55 $\pm$ 0.2 &  340.11 $\pm$ 7.76 \\ \hline
    \multirow{3}{*}{900 $\pm$ 21.87} & room temperature & 40.00 $\pm$ 0.2 &  376.75 $\pm$ 10.32 \\
                                     & LN2              & 4.71  $\pm$ 0.2 &  407.81 $\pm$ 11.46 \\
                                     & LN2              & 23.55 $\pm$ 0.2 &  366.95 $\pm$ 11.43 \\ 
    \end{tabular}
    \captionsetup{width=0.9\textwidth}
    \caption{Summary of the maximum output power delivered by the OPC at both room and LN2 temperatures for four laser input powers provided by the laser box.}
    \label{tab:MaxPoFDelivery}
\end{table}

Figure \ref{fig:OutvsInPConversion} (left) and (right) show the maximum electrical output power delivered by the OPC as a function of laser input powers of 0.5W, 0.6W, 0.8W, and 0.9W at room and LN2 temperatures, respectively. For the LN2 measurements, the maximum output power delivered by the OPC is measured for shorter (4.71 m) and longer (23.55 m) optical fiber lengths submerged in LN2. For example, in a scenario where a new PoF application in a cryogenic environment demands $\sim$ 350 mW of electrical power (see dashed red line in Figure \ref{fig:OutvsInPConversion}, right) one option could be to require a laser input power of $\sim$ 850 mW with an optical fiber length of $\sim$23 m. Table \ref{tab:MaxPoFDelivery} summarizes the maximum output power delivered by the OPC at both room and LN2 temperatures for a laser input power of 0.5 W, 0.6 W, 0.8 W, and 0.9 W and different optical fiber lengths. 

\section{Mitigation of Light Leakage in PoF Systems for Photodetection Applications in HEP}
\label{sec_dune}

\begin{figure}[h!]
    \centering
    \includegraphics[width=.44\textwidth]{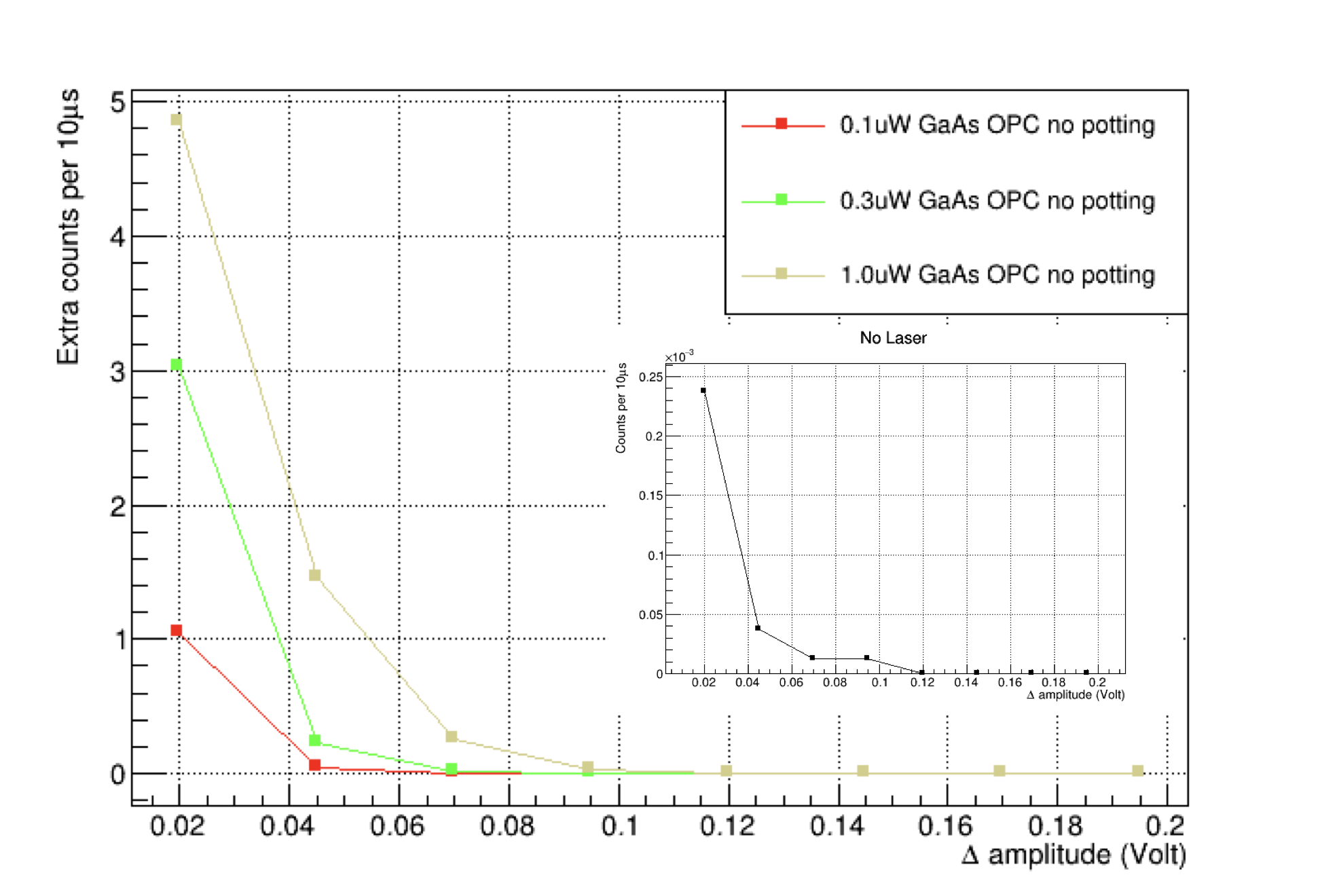}
    \qquad
    \includegraphics[width=.44\textwidth]{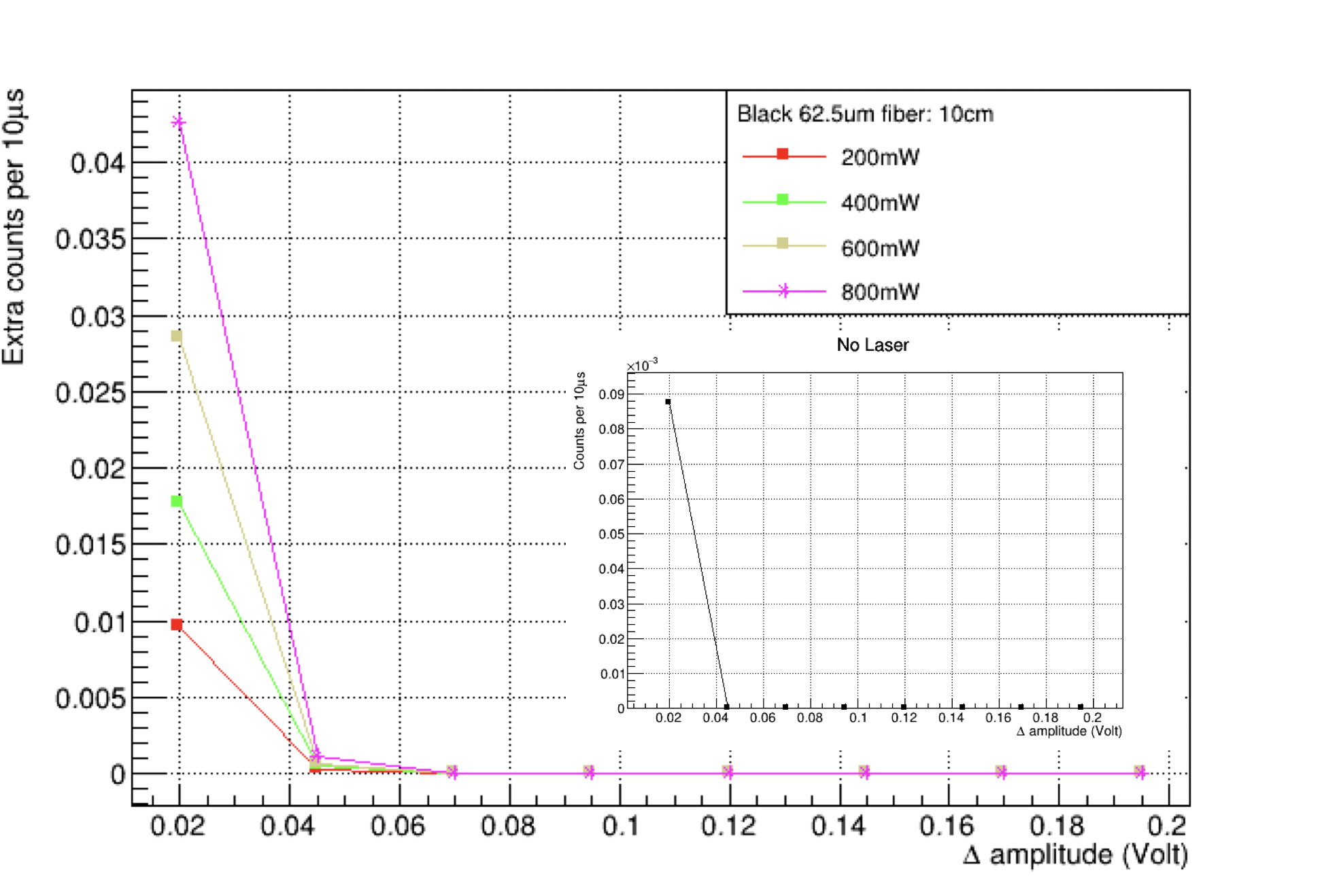}
    \includegraphics[width=.44\textwidth]{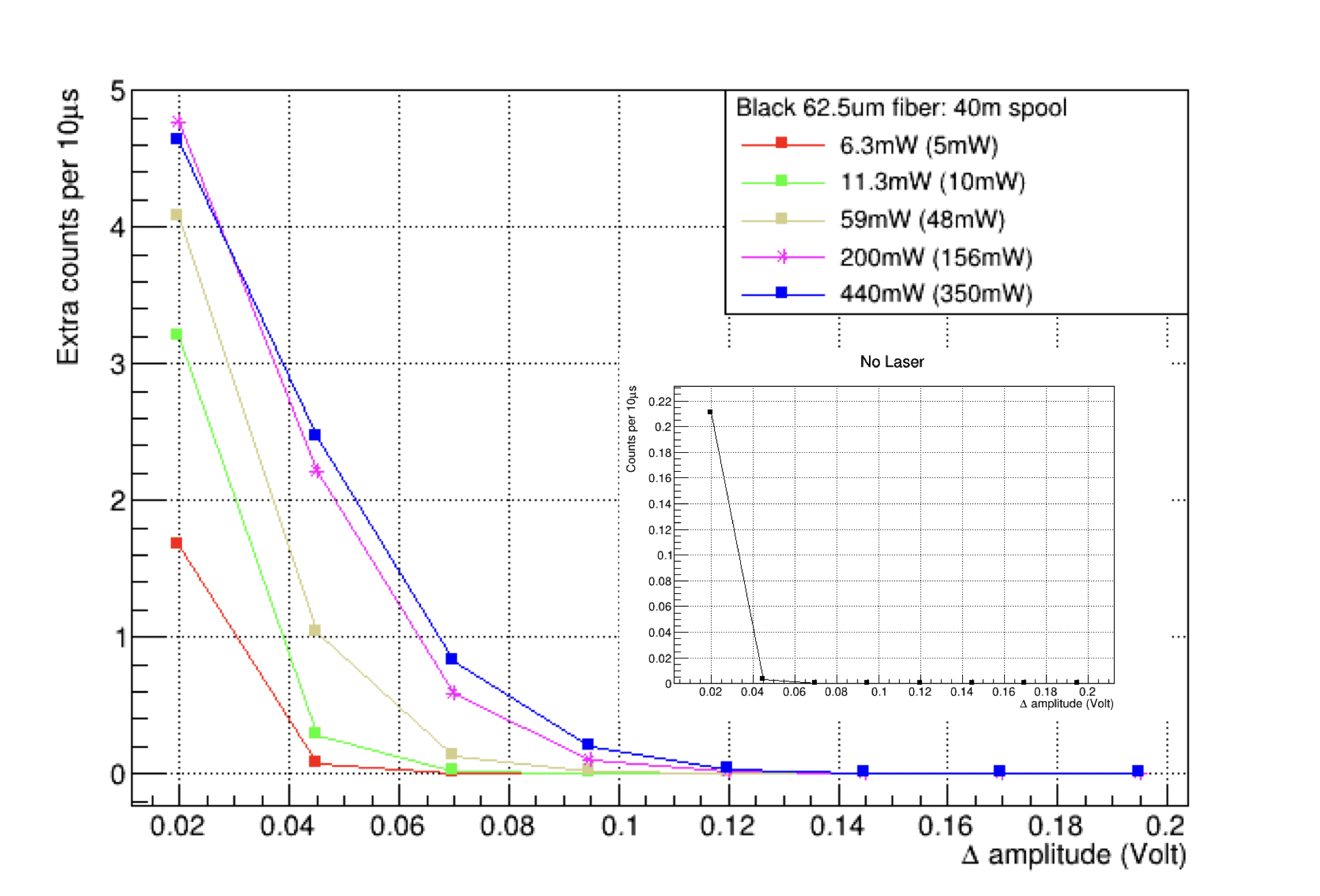}
    \qquad
    \includegraphics[width=.44\textwidth]{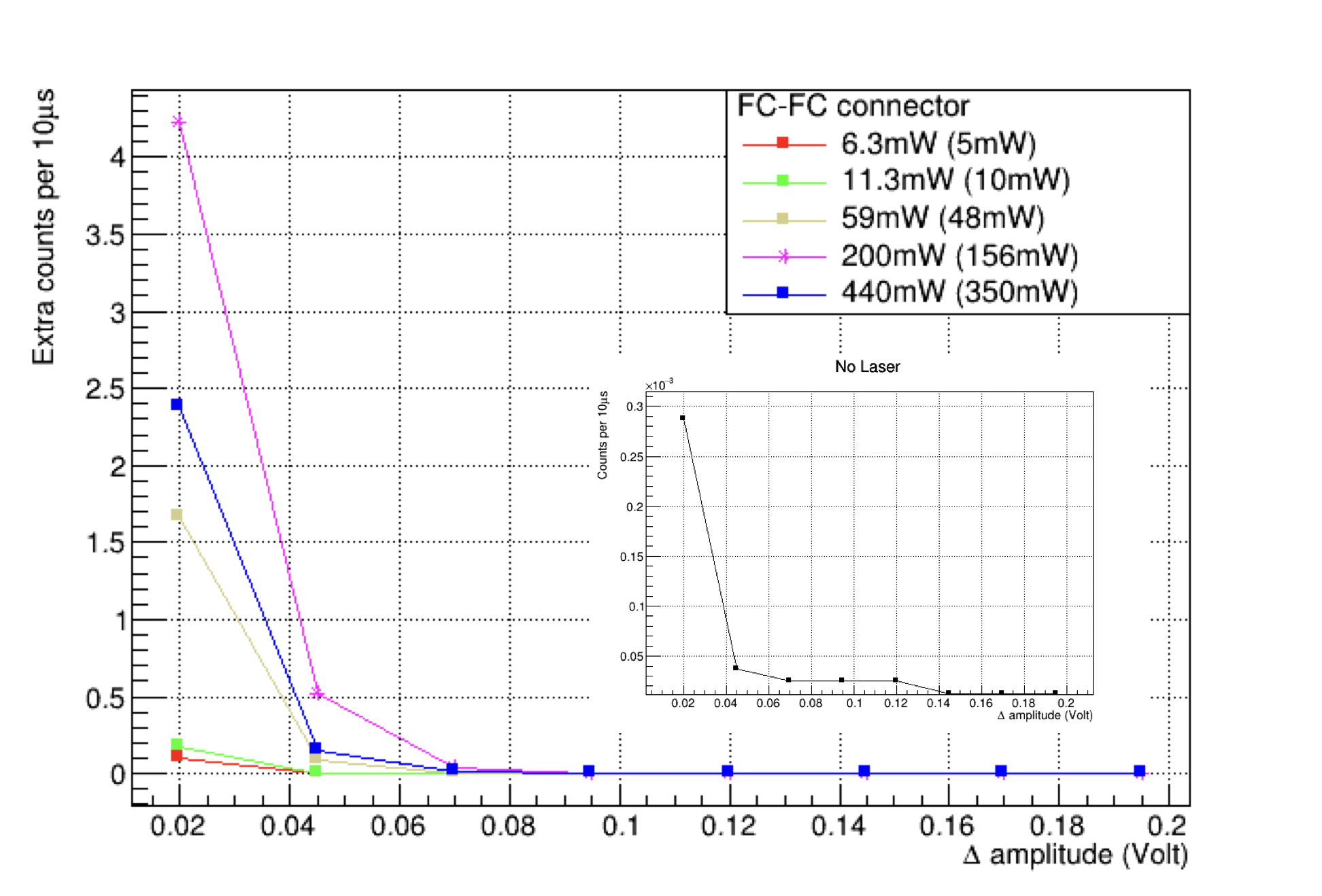}
    \captionsetup{width=0.9\textwidth}
    \caption{OPC (top left) light noise reaches a few photo-electrons per 10 $\mu$s when sub-$\mu$W laser power is delivered. The single PE amplitude is defined as 20 mV above the waveform baseline as displayed in the horizontal axis. A 10 cm long fiber (top right) light noise level is low. A 40m-long spool of fiber (bottom left) light noise level is more than a few PE when PoF is running at normal operation power. An FC-FC connector (bottom right) has a few PE per 10$\mu$s noise level. For the bottom two plots, a 40m-long 62.5$\mu$m fiber and an extra FC-FC connector are added to the measurement setup. In the legends, the powers specified outside (in) the brackets are the actual measured power before (after) the FC-FC connector and the 40m-long black fiber. Providing these two measured powers is for readers' reference since there is non-negligible power loss for a 40m-long fiber in LAr as described in previous sections. In the bottom left plot, when measuring the 40m-long fiber spool, the SiPM board needs to be taken outside the sealed black box because the fiber can not fit inside. As a result, the background noise is higher than in this test as shown in the inset plot. Nevertheless, the metric in the main plot is extra photo-electrons counts on top of this background noise from the environment.}
    \label{fig:PoFlightleakCharacterization}
\end{figure}

The 808 nm IR light escaping from the PoF system could change the SiPM signal waveform baseline if there is no shielding protection in applications related to photodetectors such as those in DUNE~\cite{DUNE:2023nqi}. The PoF IR light noise sources include the OPC, the optical fibers as well as the commonly used FC-FC fiber connectors. The light leakage reduces the signal-to-noise ratio of the photon detection system and the overall data quality. It is therefore critical to understand the noise sources and develop mitigation solutions at the lab (see Figure \ref{fig:PoFlightleakCharacterization}). 

The light noise level is defined as the average number of single photoelectron (PE) signals detected per 10 $\mu$s. As an example, the expected background contribution from the Argon-39 (dominant) and other radiological background is 200 kHz in the DUNE photodetector application~\cite{DUNE:2023nqi}. This translates to 2 PE/10$\mu$s and also sets the target noise level. Any mitigation solution needs to reach a noise level below 2 PE per 10$\mu$s so that it will not be more dominant than the radiological background. The noise level of each component from the PoF system is subsequently characterized in the lab. A Hamamatsu 4$\times$5-array SiPM with a breakdown voltage of 36V in LAr is used to detect noise IR photons from the PoF system components. The SiPM signal is read out at room temperature by a custom-designed analog readout board. A Keysight oscilloscope is used for data acquisition at 250 Msps (4ns time resolution), with a 60 MHz low pass filter applied. The SNR of this readout system reaches 5.7 and is good for the characterization of light noise. 

\begin{figure}[htbp]
    \centering
    \includegraphics[width=.4\textwidth]{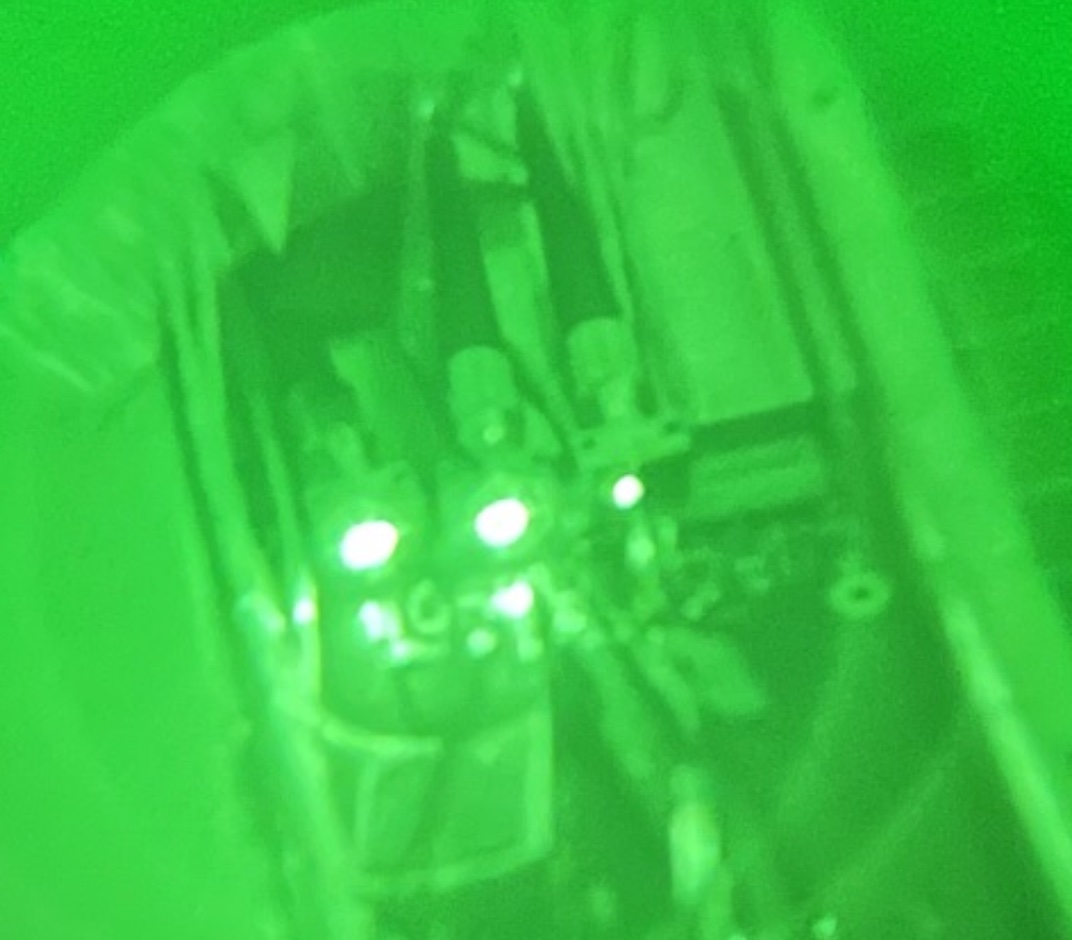}
    \captionsetup{width=0.9\textwidth}
    \caption{Light leakage (bright spots) from the pins of OPCs seen by an infrared viewer when 400 mW of laser power is delivered using optical fibers.}
    \label{fig:leakopchundredmW}
\end{figure}

To measure the light leak from OPCs, a commercial GaAs OPC is fixed at 7 cm away from the center of the 4$\times$5-array SiPM, with the OPC pins pointing perpendicularly to the SiPM array. This distance is smaller than that in any prototypes or DUNE FD2 where the OPCs are much further away from the photosensors. The OPC-SiPM system is then placed in a well-sealed black box that shields environmental light noise before being put in a Dewar filled with LAr. The SiPM detects a few PE per 10$\mu$s noise from OPC with PoF operating at sub-$\mu$W (red and green) as shown in Figure~\ref{fig:PoFlightleakCharacterization}. At 1 $\mu$W laser power, light leaks already start to produce too much current for SiPM. At normal PoF operating power of hundreds of mW, an OPC light leak (Figure~\ref{fig:leakopchundredmW}) produces too much current in SiPM that the SiPM signal waveform baseline oscillates. The SiPMs bias voltage dropped below breakdown and is not able to detect SPEs. In this case, counting photoelectrons is not meaningful. 

The light noise from a 10 cm-long 62.5 $\mu$m black fiber is also studied. A short straight fiber is run parallel to the 4$\times$5-array SiPM at a distance of 3 cm, again smaller than the distance in any realistic prototypes where fibers are much further away from photosensors. The fiber-SiPM system is similarly placed in the black box inside a Dewar of LAr. The light noise from this 10 cm long fiber is 10$^{-2}$ PE with PoF running at its normal operating power of hundreds of mW as shown in Figure~\ref{fig:PoFlightleakCharacterization}. A further test of a 40m-long spool of the same 62.5 $\mu$m black jacket fiber shows a much higher light noise level as expected. The noise reaches a few PE already when PoF is running at tens of mW (red and green curves in Figure~\ref{fig:PoFlightleakCharacterization}). Starting at 59 mW PoF power, light noise is too much for SiPM to work normally and the count of PEs is not meaningful anymore. 

\begin{figure}[h!]
    \centering
    \includegraphics[width=0.6\textwidth]{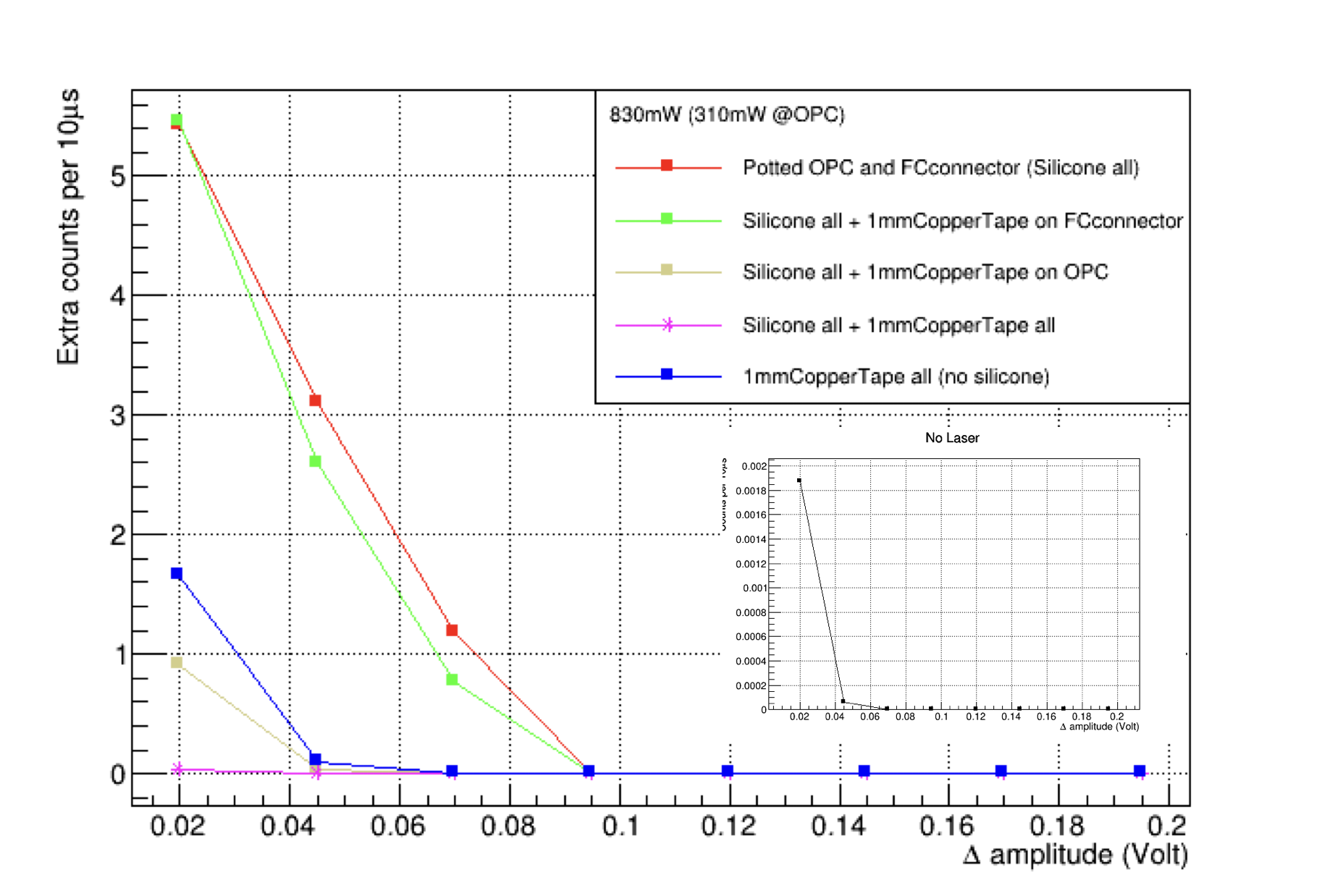}
    \captionsetup{width=0.9\textwidth}
    \caption{Several investigated solutions to mitigate the light noise from the PoF system.}
    \label{fig:PoFNoiseSolution}
\end{figure}

A similar light noise test is performed on a typical FC-FC connector when the PoF system is on. The FC-FC connector runs parallel to the SiPM array at a distance of 3 cm inside the black box. Light noise from a typical rectangular FC-FC connector is at a few PE per 10$\mu$s level as shown in Figure~\ref{fig:PoFlightleakCharacterization}. Such connectors only appear in small prototypes and will not be used in DUNE FD2.

In summary, the most critical light noise, beyond the single photoelectron level, comes from the OPC output pins and the FC-FC connector that interfaces the optical fiber and the OPCs. Investigations of several solutions (Figure~\ref{fig:PoFNoiseSolution}) show that covering both OPC and the OPC FC-FC connector with adequate electronic-grade silicone paste and a minimum of 1mm-thick copper tape can easily bring light noise down to the level of $10^{-2}$ SPE per 10 $\mu$s, fulfilling the DUNE photodetector application requirement. This measure could be further strengthened with a much thicker metallic box to house the entire readout including silicone potted PoF OPCs and FC-FC connectors. In addition, all fibers can be shielded within black PTFE tubes to further reduce the light noise. These measures are demonstrated to be effective in the DUNE application.

\section{Application of PoF Technology at CERN Cold Box Test Facility}
\label{sec_dune_application} 

\begin{figure}[h!]
    \centering
    \includegraphics[width=0.45\textwidth]{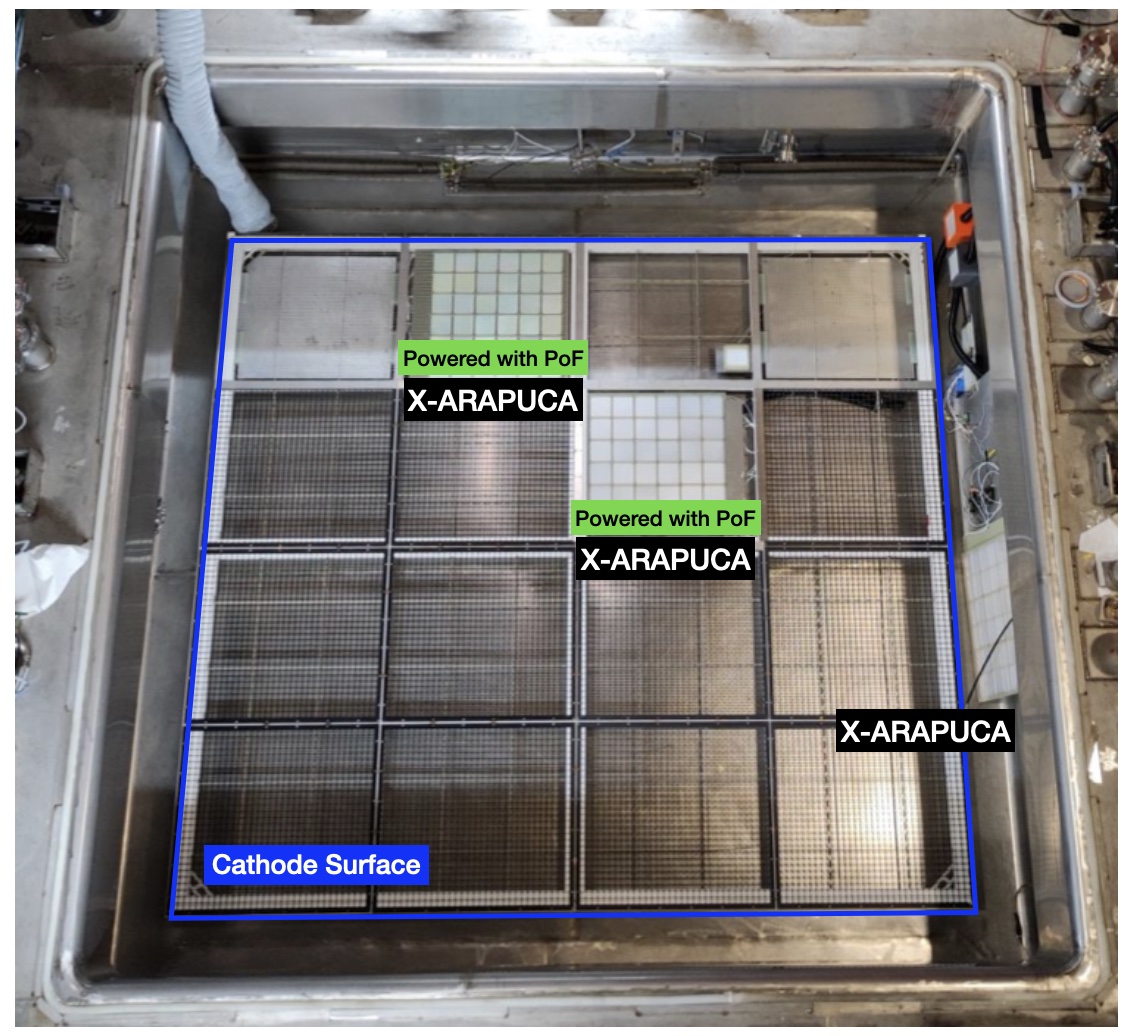}
    \includegraphics[width=0.45\textwidth]{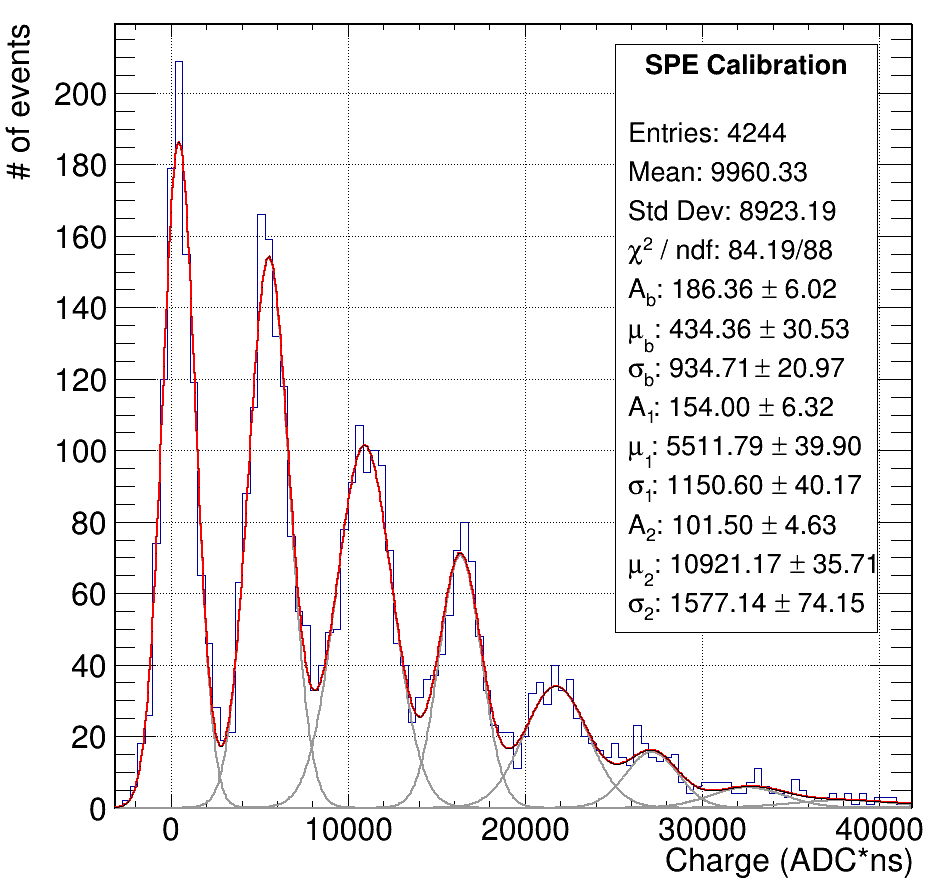}
    \captionsetup{width=0.9\textwidth}
    \caption{Left: Top view of the cold box facility at CERN, showing two X-ARAPUCA photodetector prototypes installed on the cathode surface and powered by PoF technology. Right: Photo-electron charge distribution for a X-ARAPUCA photodetector located at the cathode surface and powered by PoF technology. The distribution is fitted by a multi-Gaussian function, the parameters presented in the figure are from the three first Gaussian. The first set of parameters (A$_{b}$, $\mu_{b}$, and $\sigma_{b}$) corresponds to the baseline Gaussian fit, while the second (A$_{1}$, $\mu_{1}$, and $\sigma_{1}$) and third (A$_{2}$, $\mu_{2}$, and $\sigma_{2}$) sets correspond to the distribution of one and two photo-electrons. The figures are obtained and adapted from the DUNE FD2 TDR \cite{DUNE:2023nqi}.}
    \label{fig:TDRPoF}
\end{figure}

The CERN Neutrino Platform's cold box test facility ~\cite{CERNcoldbox} enables the development and prototyping of next-generation neutrino detectors, such as the DUNE prototypes (e.g., ProtoDUNE Vertical Drift~\cite{DUNE:2023nqi}). The cold box test facility at CERN has served as a platform for the successful integration of PoF technology to power X-ARAPUCAs photodetectors located in high-voltage and cryogenic environments, providing validation prior to their ProtoDUNE VD deployment. The cold box cryostat has dimensions of (3 $\times$ 3 $\times$ 1) m$^3$ and is filled with LAr, where the photodetectors are situated on a cathode surface (see Figure \ref{fig:TDRPoF}). The distance between the anode and cathode plane is $\sim$ 25 cm, with high voltage (10 kV) applied \cite{DUNE:2023nqi}, allowing for vertical drift. The cold box tests used SiPM based X-ARAPUCAs photodetector modules read-out by DCEM boards. The DCEM readout board is powered by PoF while the SiPM electrical signals are converted to optical signals and transmitted through a Signal-over-Fiber (SoF) laser diode through an optical fiber.

The PoF configuration of the X-ARAPUCA photodetector electronics utilizes two GaAs OPCs similar to the ones described in Section \ref{sec_opc}, each supplying $\sim$ 6.5V when operated in LAr. These units receive power through a 62.5$\mu$m core optical fiber (see Table \ref{tab:FiberType}) and uses an 808 nm, 2-Watt laser transmitter (Section \ref{sec_laser}). The analog signal is also transmitted using optical fibers. Figure \ref{fig:TDRPoF} (right) (taken from \cite{DUNE:2023nqi} ) shows the charge spectrum using LED pulses from the calibration system, with a Signal-to-Noise ratio (SNR) of 5.9 \cite{DUNE:2023nqi}. This demonstration marked the first successful instance of powering photodetectors at high voltage and in a cryogenic environment using PoF technology in HEP.

\section{Conclusions} 
\label{sec_con}

In this paper, the first demonstration of PoF technology for cryogenic use and its first ever application in a HEP detector system is presented. Detailed characterization of the components of a PoF system, such as optical fibers and OPCs, and their operation at both room and LN2 temperatures, is also provided. A study of optical power loss at cryogenic temperatures for multiple optical fiber lengths submerged in LN2 showed a linear dependence on the input power and the amount of fiber submerged. It is also demonstrated that optical fibers with jackets have about three times greater power loss than fibers without jackets when submerged in LN2. The OPC's optical-to-electrical power conversion efficiency was measured to be about 51\% in LN2 and 45\% at room temperature for a range of optical input powers ranging from 0.5 - 1 W. A tensile strength test made it possible to estimate an approximate breaking point of $\sim$ 5.5 kg for the studied PVDF optical fiber at room temperature and $\sim$ 7 kg in LN2. A long-term test of the OPC operating in LN2 showed that during $\sim$ 6.1 months of operation, the efficiency variation from conversion of optical-to-electrical power is less than 2.6\%. Laser fluctuations has been studied under different input powers, and it was observed that the laser transmitters power stabilizes after $\sim$ 25 min of operation. 
During $\sim$ 6.1 months of the long term test operation, the laser temperature and laser current variations remained below 2.5°C and 225 mA, respectively.  The results presented in this paper serve as a foundation for future endeavors towards the use of PoF technology in other systems operating in HV environments and cryogenic conditions, such as HEP dark matter experiments, space exploration technologies, and other applications where low noise, superior isolation, and optimal efficiency are needed. Further investigations are underway to continue improving the PoF technology.

\section{Acknowledgment}
This manuscript has been authored by Fermi Research Alliance, LLC under Contract No. DE-AC02-07CH11359 with the U.S. Department of Energy, Office of Science, Office of High Energy Physics. Dr. Shi would like to thank the Visiting Scholars Award Program of the Universities Research Association for supporting the light leakage study at Fermilab. The authors also thank Broadcom and MH GoPower Company Limited for their participation in the R$\&$D process to customize OPCs and fibers for operation in cryogenics. Additionally, the authors would like thanks to Bora Kim and Minjoo Lawrence Lee at UIUC for their fruitful discussions and studies aimed at improving the efficiency of OPCs. This material is based upon work supported by the U.S. Department of Energy, Office of Science, Office of High Energy Physics under Award Number DE-SC0024450. Moreover, this work is supported by subcontract No. 687391 between Fermi Research Alliance, LLC and South Dakota School of Mines and Technology.

\printbibliography[title=References]

\end{document}